\documentclass[longauth]{aa}

\usepackage[varg]{txfonts}

\usepackage{graphicx}
\usepackage{natbib}
\bibpunct{(}{)}{;}{a}{}{,} 
\usepackage{color}
\usepackage{float}

\usepackage{hyperref}
\usepackage{longtable}
\usepackage{rotating}
\usepackage{lscape}
\usepackage{soul}
\usepackage{subcaption}
\usepackage{adjustbox}

\usepackage{xcolor}

\newcommand{\sbl}{\mbox{erg s$^{-1}$ cm$^{-2}$ arcsec$^{-2}$}}
\newcommand{\kms}{\mbox{km s$^{-1}$}}

\newcommand{\oii}{\mbox{[O\,{\scshape ii}}]}
\newcommand{\fesc}{\mbox{$f_{\rm esc}^{\rm LyC}$}}

\begin{document}

\title{Linking \ion{Mg}{ii} and \oii\ spatial distribution to ionizing photon escape in confirmed LyC leakers and non-leakers \thanks{Based on observations obtained with the Hobby-Eberly Telescope (HET), which is a joint project of the University of Texas at Austin, the Pennsylvania State University, Ludwig-Maximilians-Universität München, and Georg-August-Universität Göttingen. The HET is named in honor of its principal benefactors, William P. Hobby and Robert E. Eberly.} \thanks{Some of the data presented herein were obtained at the W. M. Keck Observatory, which is operated as a scientific partnership among the California Institute of Technology, the University of California and the National Aeronautics and Space Administration. The Observatory was made possible by the generous financial support of the W. M. Keck Foundation.}}\titlerunning{Linking gas spatial distribution and LyC photon escape}

\author{Floriane~Leclercq\inst{\ref{austin}\thanks{e-mail: floriane.leclercq@austin.utexas.edu}}
\and John~Chisholm\inst{\ref{austin}}
\and Wichahpi~King\inst{\ref{austin}}
\and Greg Zeimann\inst{\ref{austin}}
\and Anne~E.~Jaskot\inst{\ref{williams}} 
\and Alaina~Henry\inst{\ref{stsci}}
\and Matthew~Hayes\inst{\ref{stockholm}}
\and Sophia~R.~Flury\inst{\ref{umass}} 
\and Yuri~Izotov\inst{\ref{ukraine}}
\and Xavier~J.~Prochaska\inst{\ref{calif}} 
\and Anne~Verhamme\inst{\ref{geneva}}
\and Ricardo~O.~Amorín\inst{\ref{laserena}}
\and Hakim~Atek\inst{\ref{iap}}
\and Omkar~Bait\inst{\ref{geneva}}
\and Jérémy~Blaizot\inst{\ref{cral}}
\and Cody~Carr\inst{\ref{china1}}\inst{\ref{china2}} 
\and Zhiyuan~Ji\inst{\ref{arizona}}
\and Alexandra~Le~Reste\inst{\ref{stockholm}} 
\and Harry~C.~Ferguson\inst{\ref{stsci}} 
\and Simon~Gazagnes\inst{\ref{austin}} 
\and Timothy~Heckman\inst{\ref{hopkins}}
\and Lena~Komarova\inst{\ref{michigan}}
\and Rui~Marques-Chaves\inst{\ref{geneva}}
%\and Valentin~Mauerhofer\inst{\ref{groningen}}
\and Göran~Östlin\inst{\ref{stockholm}}
\and Alberto~Saldana-Lopez\inst{\ref{geneva}} 
\and Claudia~Scarlata\inst{\ref{minnesota}}
\and Daniel~Schaerer\inst{\ref{geneva}} 
\and Trinh~X.~Thuan\inst{\ref{virginia}} 
\and Maxime~Trebitsch\inst{\ref{groningen}}
\and Gábor~Worseck\inst{\ref{postdam}} 
\and Bingjie~Wang\inst{\ref{penstate1}}\inst{\ref{penstate2}}\inst{\ref{penstate3}} 
\and Xinfeng~Xu\inst{\ref{northwest1}}\inst{\ref{northwest2}}
}

\institute{Department of Astronomy, The University of Texas at Austin, 2515 Speedway, Stop C1400, Austin, TX 78712-1205, USA \label{austin}
\and Astronomy Department, Williams College, Williamstown, MA 01267, USA \label{williams}
\and Space Telescope Science Institute, 3700 San Martin Drive, Baltimore, MD 21218, USA \label{stsci}
\and The Oskar Klein Centre, Department of Astronomy, Stockholm University; AlbaNova, SE- 10691 Stockholm, Sweden \label{stockholm}
\and Department of Astronomy, University of Massachusetts, Amherst, MA 01003, USA \label{umass}
\and Bogolyubov Institute for Theoretical Physics, National Academy of Sciences of Ukraine, 14-b Metrolohichna str., Kyiv, 03143, Ukraine \label{ukraine}
\and Department of Astronomy and Astrophysics, UCO/Lick Observatory, University of California, 1156 High Street, Santa Cruz, CA 95064, USA\label{calif}
\and Observatoire de Genève, Université de Genève, Chemin Pegasi 51, 1290 Versoix, Switzerland \label{geneva}
\and Instituto de Investigación Multidisciplinar en Ciencia y Tecnología, Departamento de Física y Astronomía, Universidad de La Serena, Avda. Juan Cisternas 1200, La Serena, Chile \label{laserena}
\and Institut d’astrophysique de Paris, CNRS UMR7095, Sorbonne Université, 98bis Boulevard Arago, F-75014 Paris, France \label{iap}
\and CNRS, Centre de Recherche Astrophysique de Lyon UMR5574, Univ Lyon, Univ Lyon1, Ens de Lyon, F-69230 Saint-Genis-Laval, France\label{cral}
\and Center for Cosmology and Computational Astrophysics, Institute for Advanced Study in Physics, Zhejiang University, Hangzhou 310058,  China \label{china1}
\and Institute of Astronomy, School of Physics, Zhejiang University, Hangzhou 310058, China \label{china2}
\and Steward Observatory, University of Arizona, 933 N. Cherry Avenue, Tucson, AZ 85721, USA\label{arizona}
\and Department of Physics and Astronomy, Johns Hopkins University, Baltimore, MD 21218, USA \label{hopkins}
\and Astronomy Department, University of Michigan, Ann Arbor, MI, 48103, USA \label{michigan}
\and Kapteyn Astronomical Institute, University of Groningen, PO Box 800, 9700 AV Groningen, The Netherlands \label{groningen}
\and Minnesota Institute for Astrophysics, School of Physics and Astronomy, University of Minnesota, 316 Church St. SE, Minneapolis, MN 55455, USA \label{minnesota}
\and Astronomy Department, University of Virginia, PO Box 400325, Charlottesville, VA 22904-4325, USA\label{virginia}
\and Institut fur Physik und Astronomie, Universitat Potsdam, Karl-Liebknecht-Str. 24/25, D-14476 Potsdam, Germany \label{postdam}
\and Department of Astronomy \& Astrophysics, The Pennsylvania State University, University Park, PA 16802, USA;\label{penstate1}
\and Institute for Computational \& Data Sciences, The Pennsylvania State University, University Park, PA 16802, USA\label{penstate2}
\and Institute for Gravitation and the Cosmos, The Pennsylvania State University, University Park, PA 16802, USA\label{penstate3}
\and Department of Physics and Astronomy, Northwestern University, 2145 Sheridan Road, Evanston, IL, 60208, USA\label{northwest1}
\and Center for Interdisciplinary Exploration and Research in Astrophysics (CIERA), Northwestern University, 1800 Sherman Avenue, Evanston, IL, 60201, USA\label{northwest2}
}

%%%%%%%%%%%%%%%%%%%%%%%%%%%%%%%%%%%%%%%%%%%%%%%%%%%%%%%%%%%%%%%%%%%%%%%%%%%%%%%%

\abstract{
The geometry of the neutral gas in and around galaxies is a key regulator of the escape of ionizing photons. We present the first statistical study aiming at linking the neutral and ionized gas distributions to the Lyman continuum (LyC) escape fraction (\fesc) in a sample of 22 confirmed LyC leakers and non-leakers at $z\approx0.35$ using the Keck Cosmic Web Imager (Keck/KCWI) and the Low Resolution Spectrograph 2 (HET/LRS2). Our integral field unit data enable the detection of neutral and low-ionization gas, as traced by \ion{Mg}{ii}, and ionized gas, as traced by [\ion{O}{ii}], extending beyond the stellar continuum for 7 and 10 objects, respectively.
All but one object with extended \ion{Mg}{ii} emission also shows extended [\ion{O}{ii}] emission; in this case, \ion{Mg}{ii} emission is always more extended than [\ion{O}{ii}] by a factor 1.3 on average. 
Most of the galaxies with extended emission are non or weak LyC leakers (\fesc < 5\%), but we find a large diversity of neutral gas configurations around these weakly LyC-emitting galaxies. Conversely, the strongest leakers (\fesc > 10\%) appear uniformly compact in both \ion{Mg}{ii} and [\ion{O}{ii}] with exponential scale lengths $\lesssim$1 kpc. Most are unresolved at the resolution of our data. We also find a trend between \fesc\ and the spatial offsets of the nebular gas and the stellar continuum emission. 
Moreover, we find significant anti-correlations between the spatial extent of the neutral gas and the [\ion{O}{iii}]/[\ion{O}{ii}] ratio, and H$\beta$ equivalent width, as well as positive correlations with metallicity and UV size, suggesting that galaxies with more compact neutral gas sizes are more highly ionized.
The observations suggest that strong LyC emitters do not have extended neutral gas halos and ionizing photons may be emitted in many directions.
Combined with high ionization diagnostics, we propose the \ion{Mg}{ii}, and potentially [\ion{O}{ii}], spatial compactness are indirect indicators of LyC emitting galaxies at high-redshift. 
}

\keywords{galaxies: formation – galaxies: evolution - galaxies: halos - intergalactic medium}

\maketitle

\section{Introduction}
\label{sec:1}

Cosmic reionization marks a crucial time in the history of the universe at $z > 5$ when the predominantly neutral intergalactic medium (IGM) was ionized by the first luminous sources \citep[see e.g., the recent review of][]{Robertson2022}. Observations have established roughly when cosmic reionization ended, but much remains unclear about how this process occurred \citep[e.g.,][]{Becker2001, Fan2006, Schroeder2013, Ouchi2018, Inoue2018, Banados2018, Kulkarni2019a, Bosman2022}. Competing theories debate whether accreting black holes or massive stars in galaxies produced the requisite ionizing photons \citep[e.g.,][]{Madau2015, Lewis2020, Trebitsch2021}, but the observations are insufficient to resolve which source dominated cosmic reionization \citep[e.g.,][]{Finkelstein2015, Kulkarni2019b, Grazian2022, Grazian2023}.

Before the advent of the James Webb Space Telescope (JWST), the reionization was typically solved by assuming that most of the ionizing photons come from the faintest galaxies \citep[e.g.,][]{Finkelstein2019, Maseda2020, Mascia2023}. JWST is however reshaping our understanding of cosmic reionization by revealing much more active galactic nuclei \cite[AGN, e.g.,][]{Goulding2023}, as well as more UV-bright galaxies \cite[e.g.,][]{Naidu2022-jwst,Bouwens2023,Casey23} than expected at $z>6$. Besides, some UV-bright galaxies have recently been found to emit much more ionizing photons than previously expected \citep{Marques-Chaves2021, Marques-Chaves2022}. These recent discoveries significantly revitalize the exploration of the origins of cosmic reionization. 
According to theoretical work, young and massive stars are promising candidates, but only if $10-20$\% of their ionizing radiation (or Lyman continuum, $\lambda<$ 912 \AA, hereafter LyC) can escape from the interstellar medium (ISM) and circum-galactic medium (CGM; e.g. \citealt{Ouchi2009, Robertson2015, R18}). However, the fraction of the total ionizing radiation that escapes distant galaxies (\fesc), reaching and re-ionizing the IGM, is not directly observable during the epoch of reionization (EoR) since the partially neutral IGM is optically thick to LyC photons \citep{I14, Garel2021}. Indirect observables tracing LyC escape, also called LyC tracers, are thus required to determine the sources of cosmic reionization, and need to be tested on low redshift galaxies for which we can detect LyC photons.

Recently, the Cosmic Origins Spectrograph (HST/COS) has led a revolution in the detection of ionizing continua from galaxies at $z\sim0.3$ \citep{Leitet2013, Borthakur14, Leitherer2016, Izotov16a, Izotov16b, Izotov18a, Izotov18b, Wang19,Izotov21, Izotov22, Flury22}. The Low-redshift Lyman Continuum Survey (LzLCS) is a 134 orbit Cycle-27 Hubble Space Telescope program (PID: 15626) targeting 66 star-forming galaxies at $z\sim0.3$ \citep{Flury22}.
This survey reported 35 new $>2\sigma$ significance LyC detections. The LzLCS survey is therefore the first statistical sample of both LyC emitters and non-emitters that can investigate indirect LyC escape methods.
However, the resultant correlations between directly observed LyC escape fractions and some of the most easily observable indirect counterparts of the galaxies in the EoR, including [\ion{O}{iii}]/[\ion{O}{ii}] flux ratio (O$_{32}$) and H$\beta$ equivalent widths, have over an order of magnitude scatter and are therefore insufficient at inferring the LyC escape fraction. Some of the scatter is physical and can be reduced using a multivariate analysis (Jaskot et al. submitted).

Simulations suggest that this scatter largely arises due to the neutral gas geometry \citep{Katz2020,Choustikov2024}. For this reason, indirect indicators that probe the line-of-sight neutral \ion{H}{i} gas distribution correlate the best with the observed LyC escape. These indicators include resonant transitions like the Lyman $\alpha$ (Ly$\alpha$) line, the \ion{Mg}{ii} emission line, and the FUV absorption lines \citep[e.g.,][]{Steidel2018, Gazagnes2020, Izotov21, Naidu2022}.
Emission lines can inform on the global gas geometry and may therefore be more closely related to the global \fesc\ rather than the line-of-sight \fesc\ traced by the "pencil beam" probes from absorption lines.
Since the Ly$\alpha$ profile is impacted by the neutral IGM at high redshift and the FUV absorption lines require detecting the faint stellar continuum at high signal-to-noise, the \ion{Mg}{ii} emission lines appear to be the leading candidate for indirect LyC tracer \citep{He18, C20}. Recent studies indeed suggest that \ion{Mg}{ii} may be fundamental for JWST to uncover the sources of cosmic reionization \citep[e.g.,][]{Izotov22, XU2022, Xu23}. \ion{Mg}{ii} emission (which traces species with ionization potential between 7.6--15~eV) coincides with \ion{H}{i} gas and, similar to Ly$\alpha$, the \ion{Mg}{ii} doublet lines are resonant transitions, implying that they probe the column density of the Mg$^+$ gas in the ground state, and by extension the \ion{H}{i} column density (by assuming a gas-phase metallicity, \citealt{C20}). 
Moreover, \ion{Mg}{ii} is optically thin at column densities where \ion{H}{i} becomes optically thin to the LyC. 
While \ion{Mg}{ii} could partly co-exist with ionized hydrogen gas, \cite{XU2022} demonstrated that the \ion{Mg}{ii} escape fraction scales effectively with \fesc (their Fig.~5), thereby strengthening the use of \ion{Mg}{ii} as an indirect LyC probe.
\ion{Mg}{ii} can therefore be used to map the neutral and low-ionization gas properties and thus characterize the scattering medium within which the LyC photons travel. This is required to understand how the ionizing radiation escapes the ISM and CGM of star forming galaxies.

Historically, the exploration of the CGM was undertaken using absorption line techniques \citep[see][for a review]{T17}. These methods reveal that large gas reservoirs surround star forming galaxies at any redshift \citep[e.g.,][]{Bouche06, Bouche07, Bouche12, Schroetter16, Schroetter19, Ho17, P17, Ra18, Zabl19, Du20, Lundgren21}. The limitations of these methods is that they only provide a pencil beam information and thus do not probe the global gas geometry.
Imaging the gas directly in emission in and around star-forming galaxies has become common with the advent of sensitive IFUs like the Multi-Unit Spectroscopic Explorer (VLT/MUSE, \citealt{Ba10}) and the Keck Cosmic Web Imager (Keck/KCWI, \citealt{Martin10kcwi}). Neutral gas has been mapped up to tens of kpc around low \citep[e.g.,][]{O14, H13, Zabl2021, Burchett2021, Leclercq2022, Runnholm2023, Dutta2023, Guo2023Natur} and high redshift \citep[e.g.,][]{S11,W16,L17,W18,L20,Claeyssens2022} star-forming galaxies using \ion{Mg}{ii} and Ly$\alpha$. These studies revealed that galaxies are embedded in large gas reservoirs with diverse configurations in terms of density, extent, shape and kinematics. 
This neutral gas interface must play a role in the escape of ionizing photons. Simulations indeed show that gas geometry along the line of sight is a critical parameter \citep{Katz2020,CenKimm2015}. Nevertheless, it still remains to be determined observationally whether LyC escape is connected to gas distribution. 

In this paper, we explore for the first time the connection between the ionizing photons leakage and the spatial distributions of neutral (as traced by \ion{Mg}{ii}) and ionized (as traced by [\ion{O}{ii}]) gas in a statistical sample of confirmed LyC leakers and non-leakers selected from the combined LzLCS sample \citep{Izotov16a, Izotov16b, Izotov18a, Izotov18b, Wang19, Izotov21, Flury22}, referred to as the LzLCS+ sample, and \cite{Izotov22}. We characterize the \ion{Mg}{ii} and [\ion{O}{ii}] spatial properties of 22 galaxies using KCWI and the Low Resolution Spectrograph 2 (HET/LRS2, \citealt{Chonis_2016}), and connect the nebular spatial distribution to their LyC properties. 
The paper is organized as follows: we describe the data acquisition and reduction, as well as the sample selection and properties in Sect.~\ref{sec:2}. Sect.~\ref{sec:3} describes our image extraction and modeling procedure as well as resulting measurements of the \ion{Mg}{ii} and [\ion{O}{ii}] emission. In Sect.~\ref{sec:4}, we connect the spatial properties of the neutral and ionized gas to the LyC leakage properties of our targets, both individually and in stacks. Finally, we discuss our results and present our summary and conclusions in Sects.~\ref{sec:5} and \ref{sec:6}, respectively.\\

Throughout the paper, all magnitudes are expressed in the AB system and distances are in physical units that are not comoving. We assume a flat $\Lambda$CDM cosmology with $\Omega_{\rm m}$ = 0.315 and $\rm H_0$ = 67.4 $\kms$ Mpc$^{-1}$ \citep{Planck20}; in this framework, a 1\arcsec\ angular separation corresponds to 5.1~kpc proper at the median redshift of our sources ($z\approx$ 0.35).

\section{Data and sample}
\label{sec:2}

\subsection{Galaxy sample and global properties}
\label{sec:21}

\begin{figure}
\centering
   \resizebox{\hsize}{!}{\includegraphics{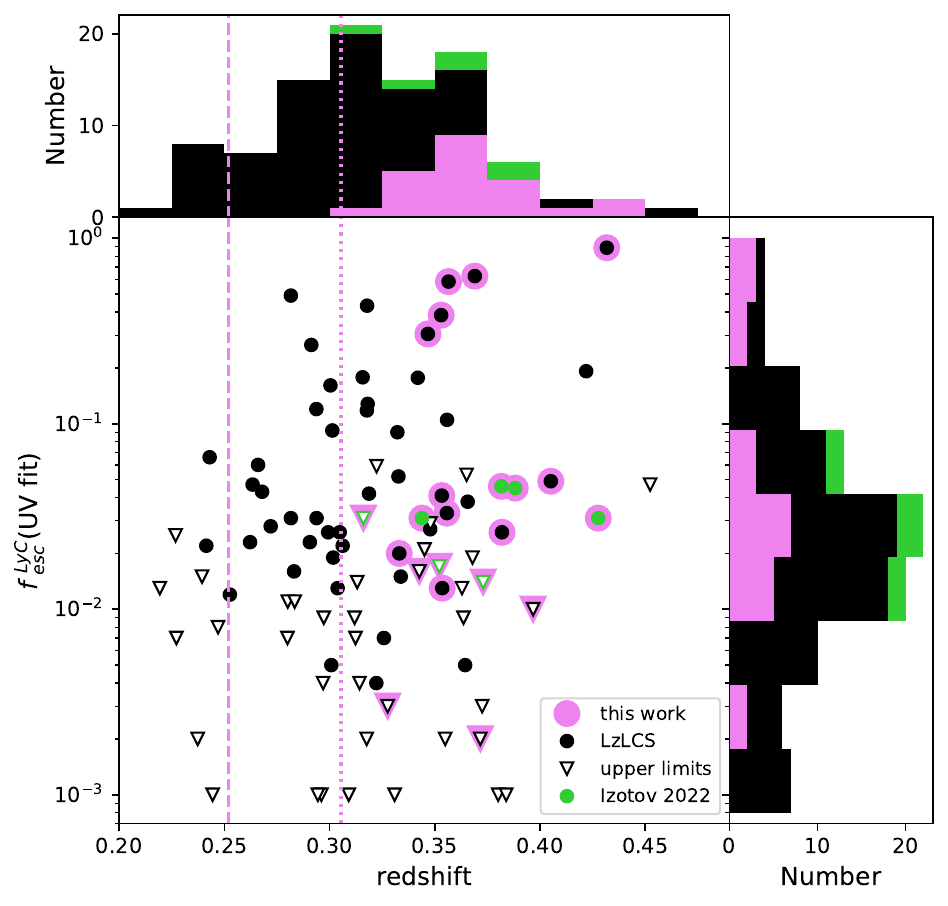}}
    \caption{Selection of our diverse sample of LyC leakers (pink) from the \cite{Izotov22} (green) and LzLCS+ \citep[black;][]{Izotov16a,Izotov16b,Izotov18a,Izotov18b,Wang19,Izotov21,Flury22} samples. The x- and y- axis show the redshift and LyC escape fraction derived in the literature from fitting the COS UV spectrum, respectively (see Sect.~\ref{sec:21}). The triangle symbols show the \fesc\ upper limits for the different samples. The dashed and dotted lines indicate the redshift corresponding to the wavelength cutoff of KCWI ($\lambda=3500$ \AA) and LRS2 ($\lambda=3650$ \AA), respectively. We show the redshift and \fesc\ distributions at the top and right of the figure, respectively.}
    \label{fig:sample}
\end{figure}

Our sample consists of 22 galaxies taken from the \cite{Izotov22} and LzLCS+ \citep{Izotov16a, Izotov16b, Izotov18a, Izotov18b,Wang19,Izotov21,Flury22} samples. While we observed all of the \cite{Izotov22} targets with KCWI, we selected 15 sources from the LzLCS+ sample. 
Of these, five objects were chosen as strong LyC leakers and were observed with KCWI.
The remaining 10 objects were observed with LRS2 and were selected to have coordinates that did not overlap with the Hobby-Eberly Telescope Dark Energy Experiment survey \citep[HETDEX,][]{Gebhardt2021} which consumes the majority of the dark time on HET.
Moreover, our targets are all at $z>0.3$ because the transmission in the blue channel of the ground-based instruments used for this work (where the \ion{Mg}{ii} 2800 \AA\ doublet is detected) is significantly reduced at lower redshift ($\lambda<3600$ \AA). 
Given these constraints, our sample spans a wide range of \fesc\ values with strong (4 objects with \fesc > 20\%) and weaker LyC leakers (11 objects with 1\% < \fesc < 5\%), as well as non-leakers (7 objects with stringent upper-limits on \fesc).
Figure~\ref{fig:sample} illustrates our sample selection. 
Their LyC escape fractions were derived in the literature from fitting the COS UV spectrum \citep{Flury22, Saldana-Lopez2022, Izotov22}. They range from stringent upper limits (\fesc < 0.01) to \fesc $\sim$ 0.9.
Our sample spans a redshift range from 0.3161 to 0.4317 and a stellar mass range from $\approx$10$^{7.5}$ to 10$^{10}$ M$_\odot$. They are star-forming galaxies with star formation rates varying from $\approx$4 to 40~M$_\odot$ y$^{-1}$. 
These properties were measured in \citet[see their Sects.~6.1 and 6.2]{Flury22} and \citet[see their Sect.~7]{Izotov22} and are reported in Table~\ref{tab:gal_prop}.

\begin{table*}
\centering
\captionof{table}{Integrated properties of the galaxy sample.}
\def\arraystretch{1.5}
\begin{tabular}{lcccccccccc}
\hline
ID & $z$ & log$_{10}~M_*$ & SFR & 12+log$_{10}$($\frac{\rm O}{\rm H}$) & $\beta^{1550}_{\rm obs}$ & $E$(B$-$V) & EW(H$\beta$) & O$_{32}$ & $r_{50}^{\rm UV}$ & $f_{\rm esc}^{\rm LyC}$(UV)  \\
   &     & [M$_\odot$]    & [M$_\odot$ yr$^{-1}$] & & & [mag]  & [\AA] & & [kpc] & \\
\hline
\hline
J0047+0154       & 0.354 & 9.2±0.4  & 23.1±1.1  & 8.29±0.04 & $-$2.25  & 0.129  & 62±1 & 4.5±1.1  & 0.6±0.2 & 0.013$_{-0.003}^{+0.021}$ \\
J0130-0014       & 0.316 & 8.63     & 3.7       & 7.97±0.02 & $-$1.52  & 0.202  & 200 & 7.4±0.4   & 0.1       & <0.031 \\
J0141-0304       & 0.382 & 9.99     & 36.0      & 8.06±0.02 & $-$0.83  & 0.319  & 220 & 5.6±0.2   & 0.2      & 0.046$_{-0.006}^{+0.006}$ \\
J0804+4726$^{*}$ & 0.357 & 8.5±0.4  & 1.6±1.1   & 7.46±0.06 & $-$2.25  & 0.078  & 287±27 & 34.6±1.2 & 0.4±0.1 & 0.584$_{-0.37}^{+0.416}$  \\
J0811+4141       & 0.333 & 8.4±0.4  & 31.6±1.1  & 7.94±0.09 & $-$1.75  & 0.163  & 151±7 & 69.2±1.2 & 0.7±0.2 & 0.02$_{-0.005}^{+0.012}$  \\
J0834+4805       & 0.343 & 9.1±0.4  & 9.7±1.1   & 8.17±0.03 & $-$1.88  & 0.141  & 184±5 & 4.5±1.1  & 1.2±0.2 & <0.016     \\
J0844+5312       & 0.428 & 8.2      & 25.3      & 8.02±0.02 & $-$0.66  & 0.347  & 196 & 4.9±0.2   & 0.2      & 0.031$_{-0.006}^{+0.006}$ \\
J0912+5050       & 0.328 & 8.8±0.4  & 15.4±1.0  & 8.18±0.04 & $-$2.00  & 0.119  & 91±2 & 3.9±1.1  & 1.2±0.1 & <0.003     \\
J0919+4906       & 0.405 & 7.5±0.1  & 7.3±1.1   & 7.91±0.06 & $-$1.78  & 0.194  & 223±13 & 16.9±1.1 & 0.4±0.2 & 0.049$_{-0.027}^{+0.183}$ \\
J0940+5932       & 0.372 & 9.4±0.4  & 35.7±1.1  & 8.38±0.07 & $-$1.33  & 0.234  & 31±1 & 1.8±1.1  & 0.8±0.2 & <0.002     \\
J1014+5501       & 0.373 & 7.5      & 8.6       & 7.96±0.02 & $-$1.35  & 0.230  & 240 & 6.8±0.4   & --        & <0.014 \\
J1033+6353$^{*}$ & 0.347 & 9.1±0.4  & 15.8±1.0  & 8.24±0.04 & $-$2.16  & 0.092  & 81±2 & 4.6±1.1  & 0.5±0.1 & 0.305$_{-0.116}^{+0.154}$ \\
J1046+5827       & 0.397 & 7.9±0.1  & 4.2±1.1   & 7.95±0.05 & $-$2.54  & 0.028  & 148±7 & 6.0±1.1  & --        & <0.01      \\
J1137+3605       & 0.344 & 9.2      & 21.1      & 7.81±0.01 & $-$0.44  & 0.385  & 280 & 7.4±0.3   & 0.2      & 0.031$_{-0.009}^{+0.008}$ \\
J1154+2443$^{*}$ & 0.369 & 8.2±0.1  & 2.7±1.1   & 7.71±0.04 & $-$2.44  & 0.046  & 165±9 & 14.±1.1  & 0.6±0.2 & 0.625$_{-0.241}^{+0.375}$ \\
J1157+5801       & 0.352 & 9.3      & 13.0      & 7.81±0.01 & $-$0.77  & 0.329  & 263 & 9.0±0.5   & 0.2      & <0.017 \\
J1243+4646$^{*}$ & 0.432 & 7.8±0.1  & 5.0±1.1   & 7.84±0.05 & $-$2.57  & 0.024  & 241±11 & 18.4±1.1 & 0.4±0.2 & 0.889$_{-0.263}^{+0.111}$ \\
J1256+4509$^{*}$ & 0.353 & 8.8±0.1  & 2.6±1.0   & 7.91±0.04 & $-$2.44  & 0.045  & 203±7 & 24.4±1.1 & 0.3±0.1 & 0.385$_{-0.093}^{+0.249}$ \\
J1352+5617       & 0.388 & 9.4      & 12.6      & 8.05±0.03 & $-$0.87  & 0.312  & 172 & 3.8±0.2   & --        & 0.045$_{-0.011}^{+0.011}$ \\
J1503+3644       & 0.356 & 8.2±0.1  & 15.6±1.1  & 8.38±0.03 & $-$1.76  & 0.161  & 180±5 & 6.9±1.1  & 0.4±0.1 & 0.033$_{-0.006}^{+0.019}$ \\
J1517+3705       & 0.353 & 9.0±0.4  & 39.8±1.1  & 8.32±0.03 & $-$1.70  & 0.197  & 107±3 & 3.2±1.1  & 0.4±0.1 & 0.041$_{-0.023}^{+0.08}$  \\
J1648+4957       & 0.382 & 8.5±0.4  & 27.5±1.1  & 8.25±0.10 & $-$1.53  & 0.164  & 72±3 & 3.6±1.2  & 1.4±0.2 & 0.026$_{-0.014}^{+0.051}$ \\
\hline
\end{tabular}
\smallskip\\
\small{\textbf{Notes.} Source identifier (ID), redshift ($z$), stellar mass ($M_*$), star formation rate (SFR), oxygen abundance (12+log$_{10}$($\frac{\rm O}{\rm H}$)), H$\beta$ equivalent width (EW(H$\beta$)), [\ion{O}{iii}]/[\ion{O}{ii}] ratio (O$_{32}$), NUV half light radius ($r_{50}^{\rm UV}$) and LyC escape fraction \fesc(UV) measurements are from \cite{Flury22} and \cite{Izotov22}. The UV slope at 1550 \AA\ ($\beta^{1550}_{\rm obs}$) and the dust extinction $E$(B$-$V) are measured in \cite{Saldana-Lopez2022} using SED fitting. $^{*}$: objects identified as strong LyC leakers following the criteria of \cite{Flury22b}, i.e., $>5\sigma$ detection and \fesc > 5\%.}
\label{tab:gal_prop}
\end{table*}

\subsection{Observations and data reduction}
\label{sec:22}
Both integral field units HET/LRS2 and Keck/KCWI data have been used to characterize the spatial distribution of our 22 galaxies (Sect.~\ref{sec:21}). Ten galaxies were observed with the blue configuration of LRS2 (refereed to as LRS2-B), and 12 with KCWI (see Table~\ref{tab:obs} for details). 

\begin{table*}
\centering
\captionof{table}{Integral field spectrograph observations.}
\def\arraystretch{1.5}
\begin{tabular}{lccllcc}
\hline
ID & R.A. & Decl. & Instr. & Config. & Exp. time & Obs. date \\
   & [deg]  & [deg] &   &   &  [s] & [mm/jj/yyyy]   \\
\hline
\hline
J0047+0154       & 00:47:42.84 & $+$01:54:39.91 & HET/LRS2   & LRS2-B  &  3600    &  12/16/2022  \\
J0130-0014       & 01:30:32.37 & $-$00:14:32.52 & Keck/KCWI  & BL-small &  3600    &  10/01/2019  \\
J0141-0304       & 01:41:42.85 & $-$03:04:51.12 & Keck/KCWI  & BL-small &  3600    &  09/30/2019  \\
J0804+4726       & 08:04:25.11 & $+$47:26:06.63 & HET/LRS2   & LRS2-B  &  17400   &  12/30/2021  \\
J0811+4141       & 08:11:12.04 & $+$41:41:45.92 & HET/LRS2   & LRS2-B  &  5400    &  01/05/2022  \\
J0834+4805       & 08:34:40.06 & $+$48:05:40.91 & HET/LRS2   & LRS2-B  &  8400    &  12/30/2021  \\
J0844+5312       & 08:44:57.90 & $-$53:12:30.11 & Keck/KCWI  & BL-small  &  3600    &  01/28/2020  \\
J0912+5050       & 09:12:08.07 & $+$50:50:08.63 & HET/LRS2   & LRS2-B  &  7200    &  01/06/2022  \\
J0919+4906       & 09:19:55.78 & $+$49:06:08.75 & HET/LRS2   & LRS2-B  &  5400    &  01/08/2021  \\
J0940+5932       & 09:40:00.66 & $+$59:32:44.42 & HET/LRS2   & LRS2-B  &  5400    &  01/24/2022  \\
J1014+5501       & 10:14:23.78 & $+$55:01:43.82 & Keck/KCWI  & BL-medium  &  4500    &  01/14/2021  \\
J1033+6353       & 10 33 44.05 & $+$63:53:17.20 & HET/LRS2   & LRS2-B  &  5400    &  04/03/2022  \\
J1046+5827       & 10:46:01.98 & $+$58:27:56.95 & Keck/KCWI  & BL-small  &  3600    &  01/29/2020  \\ 
J1137+3605       & 11:37:47.77 & $+$36:05:04.62 & Keck/KCWI  & BL-small  &  3600    &  01/29/2020  \\
J1154+2443       & 11:54:48.85 & $+$24:43:33.03 & Keck/KCWI  & BL-small  &  3600    &  01/27/2022  \\ 
J1157+5801       & 11:57:44.80 & $+$58:01:42.69 & Keck/KCWI  & BL-small  &  3000    &  01/28/2020  \\ 
J1243+4646       & 12:43:00.63 & $+$46:46:50.40 & Keck/KCWI  & BL-small  &  9600    &  04/12/2021  \\ 
J1256+4509       & 12:56:44.15 & $+$45:09:17.01 & Keck/KCWI  & BL-small  &  5400    &  01/27/2022  \\ 
J1352+5617       & 13:52:35.80 & $+$56:17:01.41 & Keck/KCWI  & BL-small  &  3600    &  01/29/2020  \\ 
J1503+3644       & 15:03:42.83 & $+$36:44:50.75 & Keck/KCWI  & BM-small  &  3600    &  01/30/2019  \\
J1517+3705       & 15:17:07.40 & $+$37:05:12.27 & HET/LRS2   & LRS2-B  &  6300    &  07/22/2022  \\
J1648+4957       & 16:48:49.35 & $+$49:57:50.85 & HET/LRS2   & LRS2-B  &  5400    &  05/27/2022  \\
\hline
\end{tabular}
\smallskip\\
\small{\textbf{Notes.} ID: source identifier in \cite{Flury22} or \cite{Izotov22}. R.A.: right ascension in degree (J2000). Decl.: declinaison of the source in degree (J2000). Instr.: instrument used to observe the target. Config.: configuration of the instrument. LRS2-B corresponds to the blue arm of LRS2 (see Sect.~\ref{sec:221}). For KCWI, gratings and slicers are indicated using the formatting grating-slicer. Exp. time: total science exposure time in seconds. Obs. date: first night of the observations.}
\label{tab:obs}
\end{table*}

\subsubsection{HET/LRS2}
\label{sec:221}
Our LRS2 \citep{Chonis_2016} observations were taken as part of the UT21-1-019, UT22-1-011, UT22-2-016 (PI Chisholm), UT22-2-021 (PI Leclercq) and UT22-3-011 (PI Endsley) programs between January 2021 and December 2022 (see Table~\ref{tab:obs}). LRS2 is installed on the 10m Hobby-Eberly Telescope \citep{Ramsey98, Hill21} at the McDonald Observatory.
LRS2 comprises two spectrographs separated by 100\arcsec\ on sky: LRS2-B (with wavelength coverage of 3650 \AA\ -- 6950 \AA) and LRS2-R (with wavelength coverage of 6450 \AA\ -- 10500 \AA). There are two channels for each spectrograph: UV and orange for LRS2-B and red and far red for LRS2-R. Each spectrograph has 280 fibers, each with a diameter of 0.59$\arcsec$, covering  6\arcsec\ $\times~$12\arcsec\ with nearly unity fill factor \citep{Chonis_2016}. The LRS2 spectral sampling is 0.7 \AA\ per pixel and the spectral resolution depends on the LRS2 spectrograph arm (UV: 1.63 \AA, Orange: 4.44 \AA, Red: 3.03 \AA, Far red: 3.78 \AA). We obtained LRS2-B observations for 10 sources with total exposure time varying between 3600 and 17400 seconds (see Table~\ref{tab:obs}). We also obtained LRS2-R observations that were combined to the LRS2-B datacubes during data reduction (see below), but did not use the LRS2-R data in this study.

We performed the LRS2 initial reductions using the Panacea\footnote{https://github.com/grzeimann/Panacea} pipeline including: fiber extraction, wavelength calibration, astrometry, and flux calibration. On each exposure, we combined fiber spectra from the two channels into a single data cube accounting for differential atmospheric refraction (DAR). 
The DAR is taken into account by correcting the spatial shift channel by channel, using an empirical model for each channel built on tens of standard stars.
We then identified the target galaxy in each observation by collapsing the cubes and by fitting the resulting white light images with a 2D Gaussian model. We finally rectified the data cubes to a common sky coordinate grid with the target at the center. 
To normalize each cube, we measured H$\beta$ in both the LRS2-B and LRS2-R IFUs at the observed wavelength of $\approx$6564 \AA\ at $z\sim0.35$. After normalization of blue and red cubes, we stacked the individual exposures together using a variance weighted mean.
Our LRS2 data cubes have a spatial scale of 0\farcs4 by 0\farcs4 spatial pixels (spaxels) and are seeing-limited, with a median resolution of $\approx$1\farcs74 ($\sim$9 kpc at $z=0.35$). The LRS2 point spread function (PSF) of our observations was characterized by fitting standard stars observations using a Moffat function \citep{M69} to account for the PSF wings (see Appendix~\ref{ap:1a}). Throughout the paper, the LRS2 PSF is illustrated on figures with a blue circle whose size corresponds to the PSF full width at half maximum (FWHM).

For sky subtraction, we took the biweight spectrum of all spaxels at radius larger than 4$\arcsec$ from the target to minimize self-subtraction. 
Our galaxies are compact in SDSS \citep{Flury22}, ensuring that all the flux is included in the 4$\arcsec$ aperture.
Besides, this 4$\arcsec$ aperture is larger than the curve of growth radius of the white light, \ion{Mg}{ii} and [\ion{O}{ii}] narrow band (NB) images for all the objects. The curve of growth radius ($R_{\rm CoG}$) is defined as the radius at which the averaged flux in a 1-pixel wide annulus reaches zero (see \citealt{L17}).
Given that the sky varies from fiber to fiber over the 6$\arcsec\times$12$\arcsec$ field of view of LRS2 \citep{Chonis_2016}, we performed an additional residual sky subtraction. We modeled the sky residuals by masking in 2$\arcsec$ regions around the center of the galaxy as well as the emission lines of the galaxy (from \ion{Mg}{ii} to H$\alpha$) and by smoothing the data with a 2\farcs5 Gaussian kernel. 
This masked and smoothed sky residuals model is then subtracted from the data to obtain the final data cube.

\subsubsection{Keck/KCWI}
\label{sec:222}

The KCWI observations were taken between 2019 and 2022 (PIs Chisholm and Prochaska). The KCWI IFU offers several configurations with different gratings impacting the spectral resolution and wavelength range, and beam-slicers impacting the spatial resolution and field of view. For most of our observations, we choose a configuration consisting of the small image slicer and the BL grating with central wavelength of 4600 \AA\ and a 8$\farcs$4 $\times$ 20$\farcs$4 field of view with a 1$\times$1 pixel binning.
This combination allowed us to obtain a wavelength coverage of 3330 -- 5937 \AA\ (2467 -- 4398 \AA\ rest frame at $z\sim0.35$), a spectral resolution of R = 3600 (83 \kms) and a spatial sampling of 0\farcs35 $\times$ 0\farcs147. We note that two objects have been observed with different configurations. J1503+3644 has a better spectral resolution than the rest of the sample (R = 8000 or 37~\kms) resulting in a shorter wavelength coverage (2700 -- 3300 \AA\ rest frame) that does not cover the [\ion{O}{ii}]~3727~\AA\ doublet. J1014+5501 was observed with the medium slicer because of sub-optimal transparency and thus has a spectral resolution of R = 1800 (166~\kms). We refer to Table~\ref{tab:obs} for the configuration details of each target.

We used the KCWI KDERP pipeline Version 1.2.1\footnote{https://github.com/Keck-DataReductionPipelines/KcwiDRP} \citep{Morrissey2018} to reduce our KCWI observations. Details about our KCWI KDERP data reduction steps and the CWITools pipeline can be found in King et al. (in prep). Briefly, the major steps are the following: (1) bias and overscan subtraction, gain correction, cosmic rays removal; (2) dark and scattered light subtraction; (3) geometric transformation and wavelength calibration; (4) illumination correction and flat fielding; (5) standard sky subtraction; (6) data and variance cubes are produced in air wavelengths using the maps from (3); (7) differential atmospheric refraction correction using the observed airmass, the orientation of the image slicer, and the wavelengths of the exposures; (8) flux calibration using an inverse sensitivity curve made from standard star observations. After performing these eight steps, the final cubes were run through the CWITools pipeline \citep{OSullivan20} in order to coadd the individual exposures and propagate the errors accordingly. The contributions from the individual input cubes were weighted by exposure time and projected on a common coadd grid with 0\farcs14 $\times$ 0\farcs14 pixel size to optimize the spatial sampling (see Fig.~6 of \citealt{OSullivan20}), except for the J1014+5501 exposures which was observed with the medium slicer and projected on a common pixel grid of 0\farcs29 $\times$ 0\farcs29.
Every exposure was inspected and any exposure with poor seeing or cirrus absorption were discarded, leading to the total exposure time reported in Table~\ref{tab:obs} with a minimum of 50 minutes on target.

The KCWI point spread function of our observations was characterized by modeling standards star observations using a Moffat function to account for the PSF wings (see Appendix~\ref{ap:1a} for details). The median spatial resolution of the KCWI sample is 1$\arcsec$ or $\sim$5 kpc at the median redshift of the sample (Table~\ref{tab:psf}). Throughout the paper, the KCWI PSF is shown on figures with a blue ellipse whose size corresponds to the PSF FWHMs and position to the PSF rotation angle.

\section{\ion{Mg}{ii} and [\ion{O}{ii}] spatial distributions} 
\label{sec:3}

We now use our IFU observations of our sample of LyC leakers and non-leakers to characterize and compare the spatial extent of the neutral and low-ionization gas, using Mg~II~2796, 2803~\AA\ line doublet (which traces species with ionization potential between 7.6--15~eV), and the ionized gas, using [\ion{O}{ii}]~3727, 3729~\AA\ line doublet (which traces species with ionization potential between 13.6--35.1] eV). 
We note that none of our galaxies show \ion{Mg}{i}~2852~\AA\ absorption (ionization potential of 7.6 eV), indicating that we do not detect gas in lower ionization state than \ion{Mg}{ii}.

We start by extracting \ion{Mg}{ii} and [\ion{O}{ii}] flux optimized NB images from the datacubes, as well as continuum images in Sects.~\ref{sec:31} and \ref{sec:32}, respectively. We then perform a 2D modeling procedure (Sect.~\ref{sec:33}) to characterize and compare the spatial extents (Sect.~\ref{sec:34}) and offsets (Sect.~\ref{sec:35}) of the different gas phases.

\subsection{Narrow band image construction}
\label{sec:31}

We constructed \ion{Mg}{ii} and [\ion{O}{ii}] narrow band images of 12$\arcsec\times$12$\arcsec$ and 8$\arcsec\times$8$\arcsec$ from the LRS2 and KCWI datacubes, respectively. The LRS2 NB images are larger than the KCWI ones because they have a lower spatial resolution (see table~\ref{tab:obs}); a larger spatial aperture is thus needed to encompass all the flux. These large spatial apertures ensure that all the detectable flux is included and that enough background is available to estimate the limiting surface brightness level (see below).

\begin{figure*}
\centering
\begin{minipage}{\textwidth}
    \begin{subfigure}[t]{\linewidth}
    \vspace{0.1cm}
        \includegraphics[width=\textwidth]{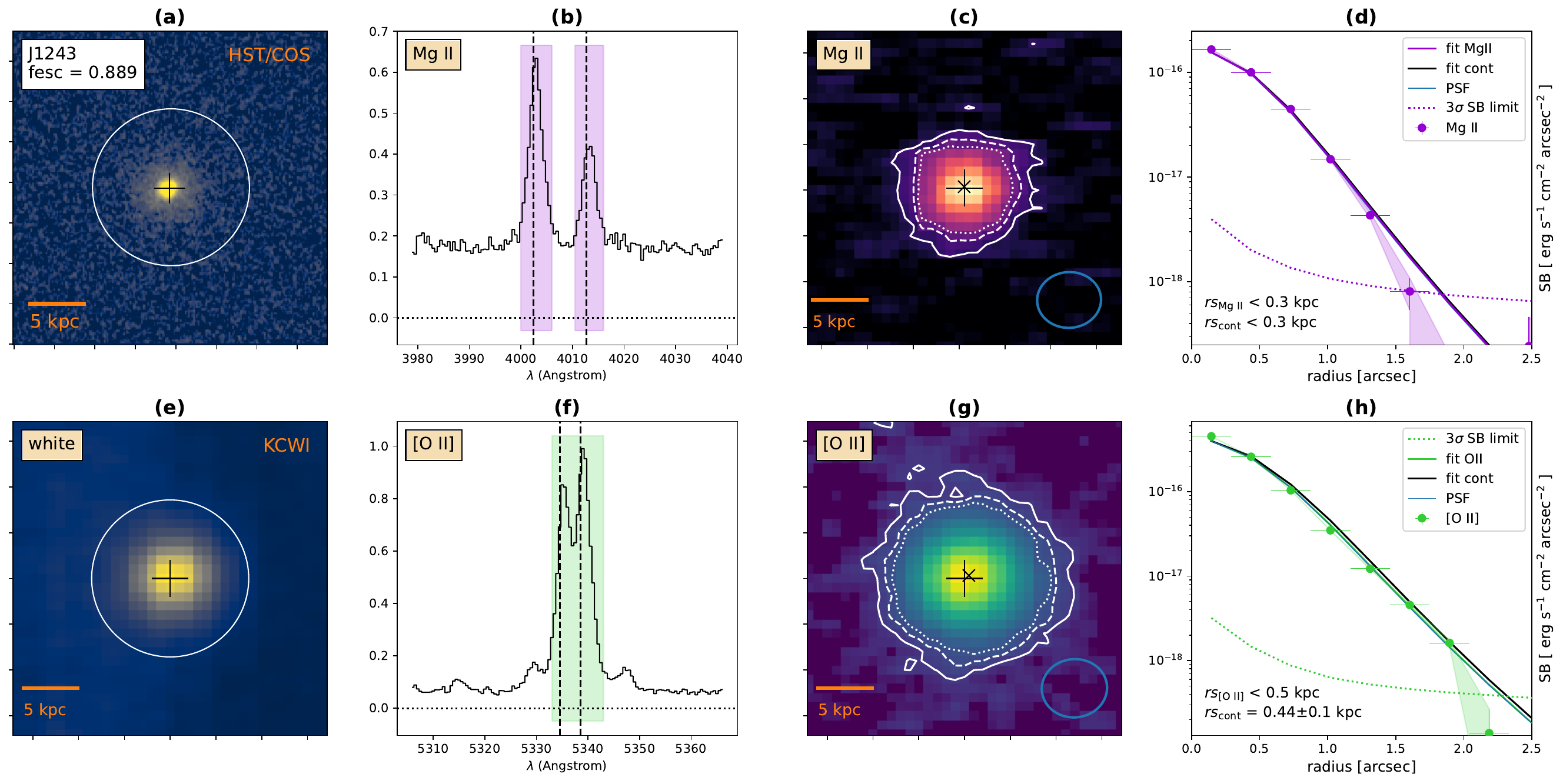}
    \end{subfigure} \\
    \begin{subfigure}[b]{\linewidth}
        \vspace{0.6cm}
        \includegraphics[width=\textwidth]{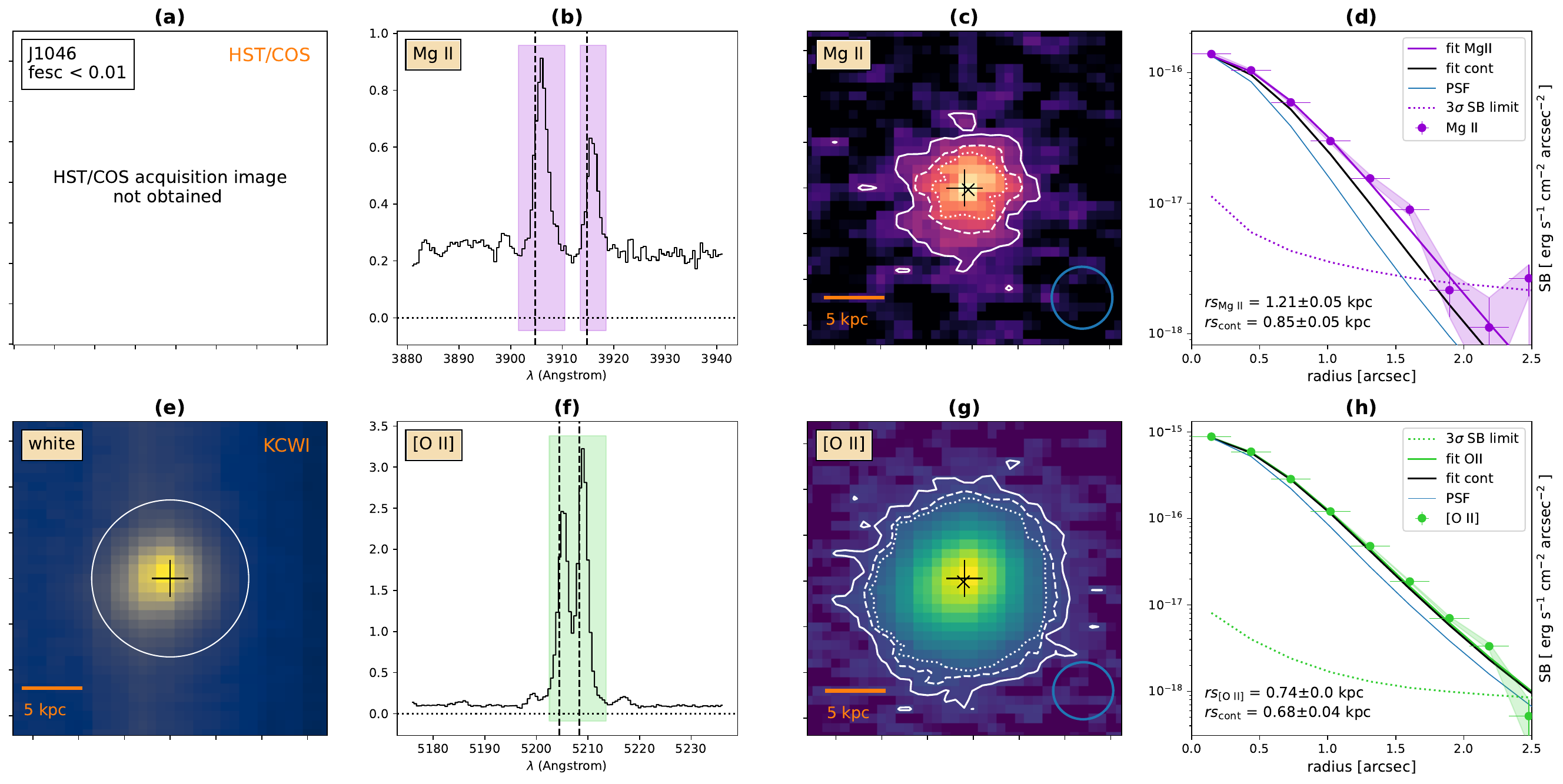}
    \end{subfigure} 
\end{minipage}
\caption{Examples of a LyC leaking and non-leaking sources from our sample observed with KCWI. The first two rows show J1243+4646, our strongest LyC leaker (\fesc = 0.889), and the last two rows show J1046+5827, a non-leaker (\fesc< 0.01). \textit{(a)}: HST/COS acquisition image (when available) plotted with a power-law stretch. The circle indicates the 2$\farcs$5 COS spectroscopic aperture (diameter). This aperture is also shown on the white light image (e) and was used to integrate the spectrum shown in panels (b) and (f) for comparison. \textit{(b)}: Observed \ion{Mg}{ii} line doublet extracted in a 2$\farcs$5 aperture. The purple areas show the narrow band (NB) image spectral coverage. The vertical lines indicate the systemic redshift from \cite{Flury22} and \cite{Izotov22}. \textit{(c)}: \ion{Mg}{ii} NB image plotted with a power-law stretch. The white contours correspond to \ion{Mg}{ii} significance levels of 3, 6 and 9$\sigma$ (solid, dashed,
and dotted, respectively). The blue circle shows the PSF FWHM estimated at \ion{Mg}{ii} wavelengths. The "+" and "x" symbols show the best-fit centroids of the emission line and nearby continuum, respectively. If neighboring sources were masked for the analysis, the mask is shown with an orange contour. \textit{(d)}: Radial surface brightness (SB) profiles of the observed and modeled \ion{Mg}{ii} emission (purple dots and line, resp.) compared to the modeled continuum (black) and PSF (blue) estimated at \ion{Mg}{ii} wavelengths. If the source is unresolved, the blue and black lines are under the purple or green line.
The 3$\sigma$ SB limit is shown as a dotted line. The shaded area represents the 1$\sigma$ uncertainties from bootstrapping analysis. The horizontal error bars corresponds to the annulus width used to build the profiles. \textit{(e)}: IFU white light image. \textit{(f)}: [\ion{O}{ii}] line doublet. \textit{(g)}: Same as (c) for [\ion{O}{ii}]. \textit{(h)}: Same as (d) for [\ion{O}{ii}].}
\label{fig:profiles}
\end{figure*}

\begin{figure*}
\centering
\begin{minipage}{\textwidth}
    \begin{subfigure}[t]{\linewidth}
    \vspace{0.1cm}
        \includegraphics[width=\textwidth]{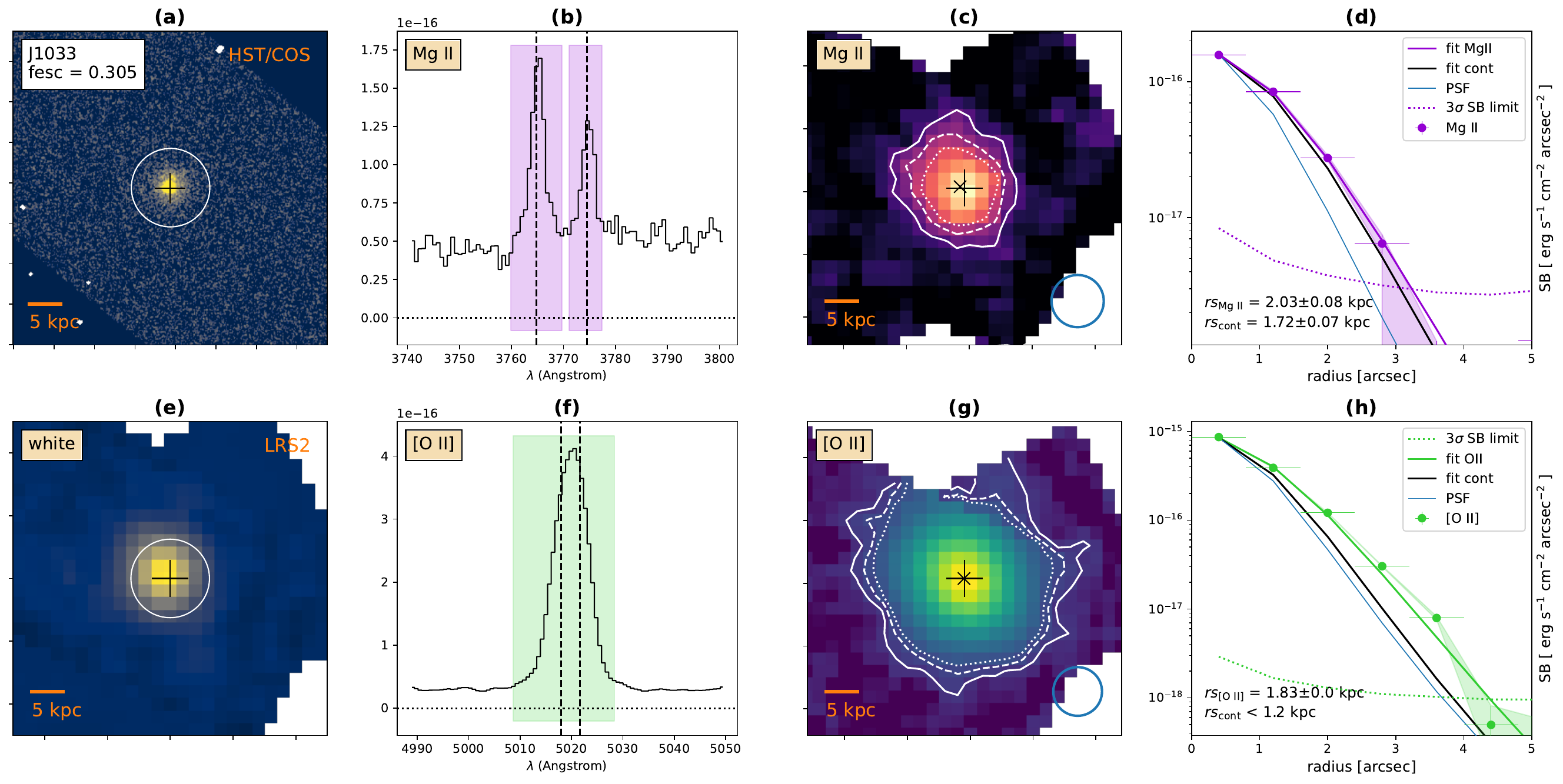}
    \end{subfigure} \\
    \begin{subfigure}[b]{\linewidth}
        \vspace{0.6cm}
        \includegraphics[width=\textwidth]{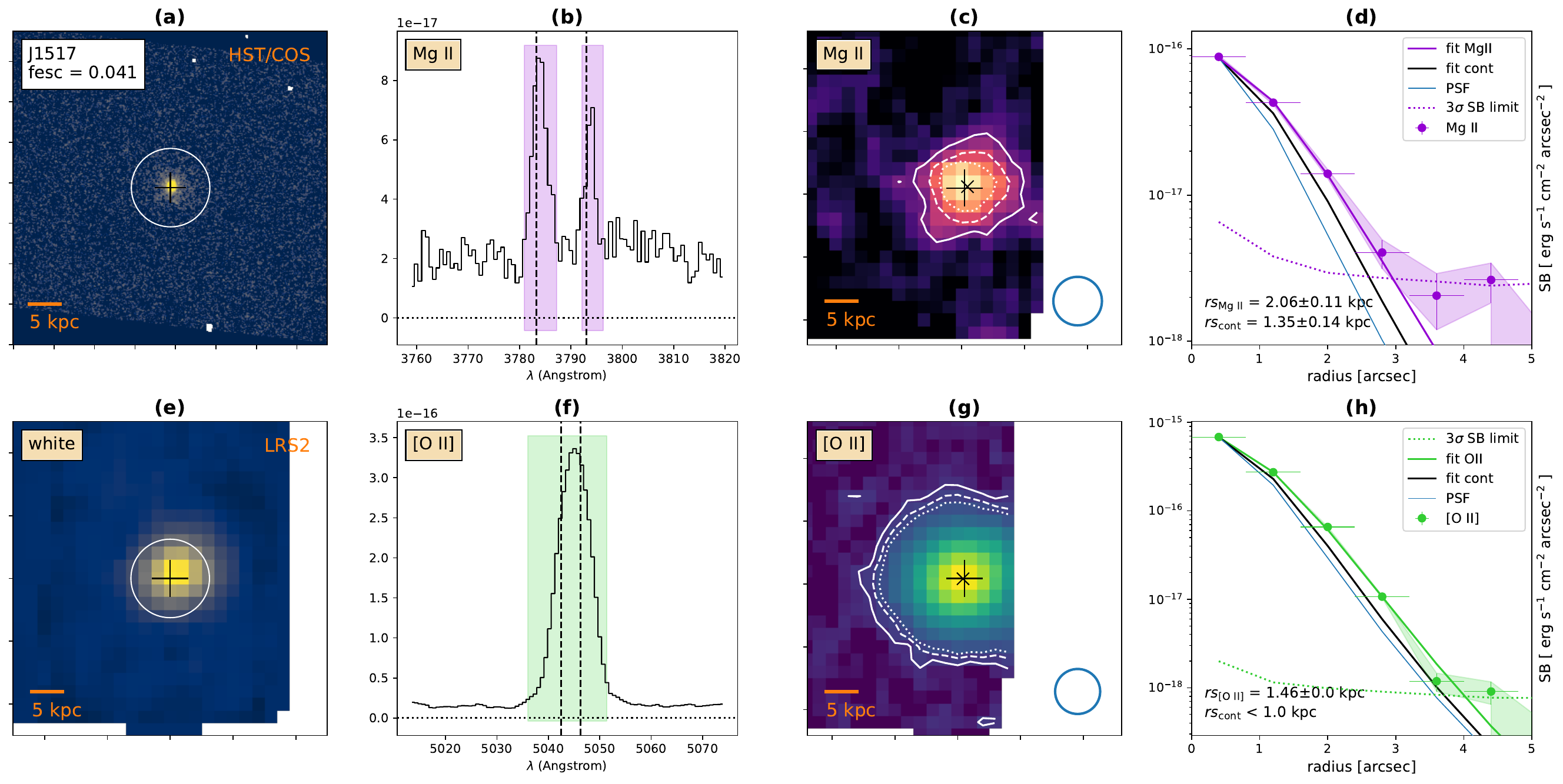}
    \end{subfigure} 
\end{minipage}
\caption{Same as Fig.~\ref{fig:profiles} but showing two objects observed with LRS2. The first two rows show the LyC leaker J1033+6353 (\fesc = 0.305), and the last two rows show the weaker leaker J1517+3705 (\fesc = 0.04). The white areas are regions on the sky uncovered by our detector.}
\label{fig:profiles_2}
\end{figure*}

For each source, we constructed a continuum-only cube by performing a spectral median filtering on the data cube using a wide spectral window of 200 spectral pixels (see \citealt{HerenzWisotzki2017} for the validation of this continuum subtraction method on data cubes).
After subtracting this continuum-only cube from the original one, we obtained an emission line only cube from which we optimally created the emission line NB images as follows: (i) we determine the spatial aperture that maximizes the integrated line flux by increasing the aperture until the integrated flux decreases because of the addition of noise, (ii) the line is extracted in the optimal aperture set in the previous step and its borders are determined by wavelengths for which the continuum subtracted flux density reaches zero, (iii) the NB image is created by summing the continuum-subtracted cube over the wavelength range delimiting the line. This procedure is applied to both the 2796~\AA\ and 2803~\AA\ lines of the \ion{Mg}{ii} doublet. The total NB image of the \ion{Mg}{ii} emission is finally obtained by adding the 2796~\AA\ and 2803~\AA\ NB images. We extracted the [\ion{O}{ii}] NB images from the data cubes using the same procedure. The [\ion{O}{ii}] lines can be blended with the high-order Balmer lines H$_{13}$ 3722\AA\ and H$_{14}$ 3734\AA, so their borders are set when the flux increases again or manually when needed. 
We note that these Balmer lines are much fainter than [\ion{O}{ii}], so any contamination does not impact our results. We also note that with this procedure we miss the wings of the broader [\ion{O}{ii}] component, which represent a small percentage of the total flux.
The resulting spectral windows used to create the \ion{Mg}{ii}~2796~\AA, \ion{Mg}{ii}~2803~\AA, and [\ion{O}{ii}] NB images are on average 5.5 \AA, 5.0 \AA\ and 11 \AA\ wide, corresponding to 436~$\kms$, 396~$\kms$, and 655~$\kms$ at z $\approx$ 0.35, respectively.
The resulting \ion{Mg}{ii} and [\ion{O}{ii}] NB images, as well as their corresponding spectral windows, are shown in the middle panels of Figs.~\ref{fig:profiles} and ~\ref{fig:profiles_2} for one strong LyC emitter (\fesc\ > 30\%) and one weaker leaker (\fesc $\leq$ 4\%) observed with KCWI and LRS2, respectively. Note that the spectra shown in these figures were extracted in a different aperture (COS-like aperture of 2\farcs5) than the line used to determine the spectral width of the NB images (aperture maximizing the integrated flux). The figures for the whole sample can be found in the Appendix~\ref{ap:2}.

To determine the 1$\sigma$ limiting flux value of the LRS2 NB images, we took the standard deviation of the pixels located outside a radius of 4$\arcsec$. This radius is larger than the curve of growth radius of the \ion{Mg}{ii} and [\ion{O}{ii}] emission line NB images, ensuring that we only select pixels with noise. The background pixels ($r$ > 4") were clipped (3$\sigma$) to eliminate the very noisy pixels (e.g., at the edges of the cubes). The LRS2 significance maps were then obtained by dividing the NB images by the corresponding 1$\sigma$ flux limit value.

For the KCWI NB images, we used the variance from the data cubes to estimate the 1$\sigma$ limiting flux. We checked that the errors from the KCWI error cubes were consistent with the values measured on the data using the same method as for the LRS2 data. We chose to use the values from the error cubes because they contain the pixel-per-pixel error information and therefore limit the impact of very noisy pixels. The KCWI significance maps were obtained by dividing the NB images by the square root of the corresponding variance images.

The contours in Figs.~\ref{fig:profiles} and \ref{fig:profiles_2} show the 3, 6 and 9$\sigma$ significance levels for the KCWI and LRS2 data. The resulting 1$\sigma$ \ion{Mg}{ii} and [\ion{O}{ii}] limiting surface brightness (SB) values reach on average 8$\times$10$^{-18}$ and 3$\times$10$^{-18}$~\sbl, respectively. This is comparable to the previous KCWI studies \citep[e.g.,][]{Burchett2021}.

\subsection{Continuum image construction}
\label{sec:32}

To compare the extent of the \ion{Mg}{ii} and [\ion{O}{ii}] emission to the stellar and nebular continuum and determine whether the metal line emission is more spatially extended than the continuum emission, we also generated continuum images from the data cubes. 
For \ion{Mg}{ii}, we averaged over two spectral windows located at $\pm$1200~\kms\ around the line doublet extremities, and that have the same velocity width as the corresponding NB image (on average 800 \kms\ at $z=0.35$ corresponding to the median redshift of the sample). These windows are close enough to the lines of interest and avoid the \ion{Mg}{i}~2852~\AA\ absorption line. 
For [\ion{O}{ii}], we averaged over one spectral window located at $-$2000~\kms\ of the [\ion{O}{ii}] line doublet, and that has the same velocity width as the NB image ($\approx$655~\kms). We refrain from using a window redder than the [\ion{O}{ii}] lines because of the numerous emission lines detected in this area. We have checked that using continuum windows at different velocities and total velocity widths do not affect the results. 
We note that we did not attempt to disentangle the stellar and nebular continuum emission. The nebular continuum can contribute up to 20\% to the observed continuum in highly ionized compact galaxies \citep{Amorin2012}. Previous studies found that the nebular continuum is often more extended than the stellar continuum \citep[e.g.,][]{PapaderosOstlin2012}. Including the nebular continuum could thus lead to an overestimation of the size of the stellar continuum, however, this fact actually reinforces our conclusions (see Sect.~\ref{sec:34}).

\subsection{Two-dimensional exponential modeling}
\label{sec:33}

To characterize the spatial distribution of our sources, we fitted the NB images with a two-dimensional exponential distribution following previous studies of extended resonant emission lines \citep[e.g., Ly$\alpha$,][]{S11}. 
We used the python module {\sc LMFIT} \citep{Newville2014} and the following equation :
\begin{equation}
    C(r)= [I_c~\exp(- r / rs)] \otimes {\rm PSF}_{\rm 2D,instr}(\lambda_{line}),
\end{equation}
with $I_c$ and $rs$ the central intensity and exponential scale length, respectively, and $r=\sqrt{(y-y_0)^2+(x-x_0)^2}$ with [$x_0$, $y_0$] the center coordinates.
All parameters (center, scale length, and amplitude) are free to vary. The fit takes into account the PSF by convolving the model with the PSF kernel of the instrument (PSF$_{instr}$), which depends on the wavelength of interest ($\lambda_{line}$, see Appendix~\ref{ap:1}). This approach holds the advantage to allow the direct comparison of the resulting parameters across different datasets acquired under varying conditions and utilizing different instruments. This particularly applies to our analysis, which involves separate datasets from two distinct instruments (Sect.~\ref{sec:2}). The neighboring galaxies visible in the NB images were masked to avoid contamination (J0130-0014 and J1256+4509).

We estimated the uncertainties associated with the best-fit parameters using a bootstrap Monte Carlo technique. We generated 100 instances of both the \ion{Mg}{ii} and [\ion{O}{ii}] NB images where each spaxel was randomly drawn from a normal distribution centered on the initial spaxel value and with standard deviation derived from the estimated median noise value for LRS2 data and from the variance image for the KCWI data (Sect.~\ref{sec:31}). Each realization of a given NB image set was fitted as described above. The final best-fit values and associated errors were determined from the median and standard deviation, respectively, of the resulting parameter distributions.

The accuracy of size measurements for compact objects is impacted by the resolution limit due to the PSF. To establish the threshold scale length below which our measurements become unreliable, we ran our modeling procedure for a range of simulated flux distributions convolved with the appropriate PSF and combined with random realizations of the noise.  
For every modeled source, we incrementally decreased the exponential scale length until we could no longer recover the input value. The scale length limit is a function of wavelength due to the PSF dependence on wavelength and thus on redshift. Consequently, we computed the resolution limit separately for each object in our sample, both for the \ion{Mg}{ii} and [\ion{O}{ii}] positions, and for each NB and continuum images. The resulting scale length resolution limits range from 0.6 to 1.6 kpc for the LRS2 data and, 0.3 to 1.2 kpc for the KCWI data. If the best-fit scale length fell below this resolution threshold, we considered the value as an upper limit. The spatial measurements and fitting results can be found in Table~\ref{tab:fitting}. 

The fourth column of Figs.~\ref{fig:profiles} and~\ref{fig:profiles_2} (panels d) shows the data and best-fit model radial SB profiles obtained for the \ion{Mg}{ii} and [\ion{O}{ii}] emission as well as their respective continuum (see Appendix~\ref{ap:2} for all objects). These profiles were computed by azimuthally averaging the flux over the continuum-subtracted NB and continuum images (Sects.~\ref{sec:31} and~\ref{sec:32}) in concentric 2-pixel wide annuli (or 0\farcs8 and 0\farcs3 for LRS2 and KCWI data, respectively) centered on the \ion{Mg}{ii} best-fit centroids.

For most of the objects, the modeled radial SB profiles are a good representation of the observed profiles. We note that object J0130-0014 shows a SB excess at outer radii because of a close neighbor (Fig.~\ref{fig:profiles2}). In this case, our one-component model only describes the central object.

\subsection{Spatial extent of the \ion{Mg}{ii} and {\rm [\ion{O}{ii}]} emission}
\label{sec:34}

Of the 22 galaxies, two objects do not show any \ion{Mg}{ii} radiation (J0130-0014 and J0804+4726) and one (J1503+3644) lacks [\ion{O}{ii}] observations (Sect.~\ref{sec:222}).
From our analysis of the PSF and detection limits (Appendix~\ref{ap:1} and Sect.~\ref{sec:33}), we have identified four objects (J0130-0014, J0804+4726, J1154+2443, and J1256+4509) that have all of their measured scale lengths below their respective detection threshold. These objects are considered as unresolved in our study.
Fourteen objects are resolved in \ion{Mg}{ii}, 13 in [\ion{O}{ii}] and 12 are resolved both in \ion{Mg}{ii} and [\ion{O}{ii}]. The average PSF-corrected \ion{Mg}{ii} and [\ion{O}{ii}] scale lengths  (without accounting for upper limits) are 1.6~kpc and 1.4~kpc, respectively.

\begin{figure*}
\centering
   \resizebox{\hsize}{!}{\includegraphics{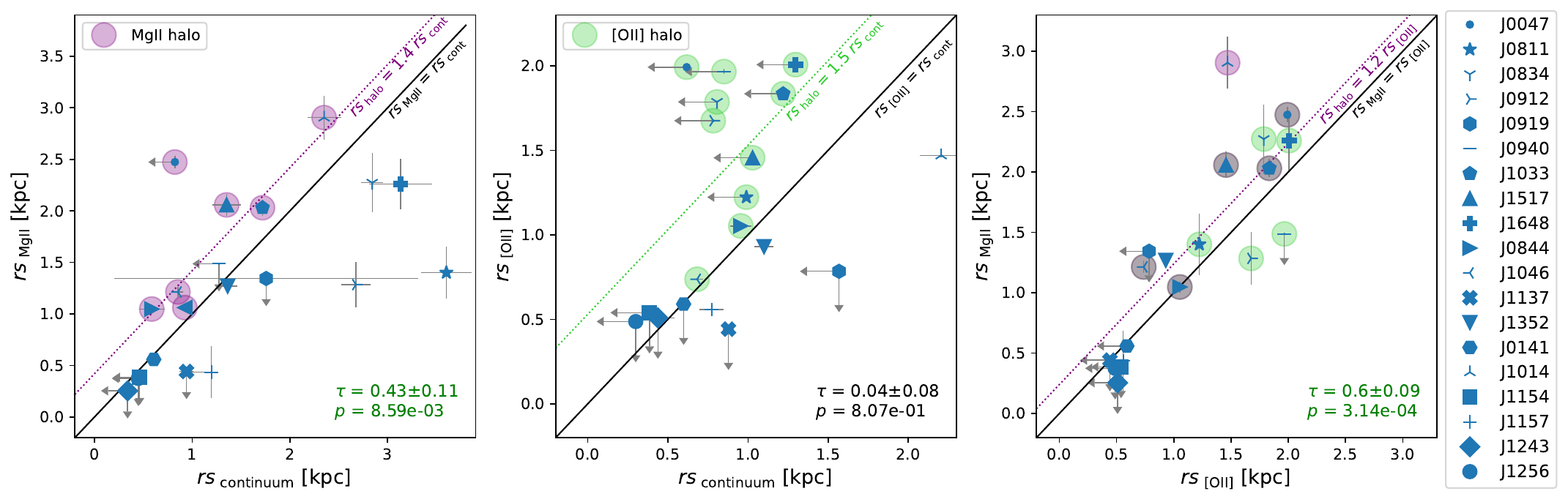}}
    \caption{Comparisons of the spatial scale lenghts ($rs$) as measured in Sect.~\ref{sec:33}: \ion{Mg}{ii} and continuum (Left), [\ion{O}{ii}] and continuum (Middle), and \ion{Mg}{ii} and [\ion{O}{ii}] (Right). Objects with statistically significant \ion{Mg}{ii} and [\ion{O}{ii}] extended emission compared to continuum are indicated by large purple and green symbols, respectively (Sect.~\ref{sec:34}). Grey symbols in the right panel result in the overlap of the purple and green symbols and thus indicate that both \ion{Mg}{ii} and [\ion{O}{ii}] halos are detected. The dotted lines show by how much on average the emission is statistically more extended compared to the continuum (or emission) scale lengths: median $\sim$1.4 times for \ion{Mg}{ii} and $\sim$1.5 times for [\ion{O}{ii}] compared to the continuum, and 1.2 for the \ion{Mg}{ii}/[\ion{O}{ii}] ratio. The black line shows the 1:1 relation (i.e., no extended emission). Upper/lower limit values are shown with arrows. The Kendall correlation coefficient ($\tau$) for every pair of variables and the corresponding false-positive probability that the correlation is real ($p$) are given and colored in green if the correlation is >$2\sigma$ statistically significant (\citealt{Akritas96,Flury22b}, see Sect.~\ref{sec:34}).}
    \label{fig:rs}
\end{figure*}

In order to evaluate whether the resolved emission lines are more spatially extended than the continuum, also refered to as "an emission line halo", we calculated the probability $p_0$ of the two scale lengths to be identical by running a t-test for the objects without upper limits on their emission and continuum sizes. In case of an upper limit on the continuum scale length, we calculated the probability that the emission scale length is less than or equal to the continuum upper limit by considering a normal distribution.
Out of the 14 objects with reliable \ion{Mg}{ii} scale length measurement, 7 have a significant \ion{Mg}{ii} halo with $p_0 < 10^{-5}$ (J0047+0154, J0844+5312, J1014+5501, J1033+6353, J1046+5827, J1503+3644, and J1517+3705; see left panel of Fig~\ref{fig:rs}). Among the 13 sources with robust [\ion{O}{ii}] scale length, 10 have their [\ion{O}{ii}] emission more extended than the continuum with $p_0 < 10^{-5}$ (J0047+0154, J0811+4141, J0834+4805, J0844+5312, J0912+5050, J0940+5932, J1033+6353, J1046+5827, J1517+3705, and J1648+4957; Fig~\ref{fig:rs} middle panel). 
On average, the \ion{Mg}{ii} and [\ion{O}{ii}] extended emissions are at least $\sim$1.6 and $\sim$1.8 times more spread out than the continuum, respectively (the median ratios are $\sim$1.5).
Interestingly, most of the objects with extended \ion{Mg}{ii} emission also show extended [\ion{O}{ii}] emission (except J1014+5501; we cannot conclude for J1503+3644 as [\ion{O}{ii}] observations are not available). Out of this sample of 5 galaxies that display both \ion{Mg}{ii} and [\ion{O}{ii}] extended emission, the \ion{Mg}{ii} emission is always more extended than [\ion{O}{ii}] ($p_0 < 10^{-5}$) by a factor of 1.3 on average (median of 1.2, right panel of Fig~\ref{fig:rs}).
The sole exception is potentially observed for J0844+5312, where the probability is lower at $p_0 = 0.1$ but still generally consistent with \ion{Mg}{ii} being more extended than [\ion{O}{ii}].
Conversely, 5 objects with extended [\ion{O}{ii}] emission are not extended in \ion{Mg}{ii}, suggesting that showing extended [\ion{O}{ii}] emission does not necessarily imply extended \ion{Mg}{ii} emission.

In order to quantify potential correlations with \ion{Mg}{ii} and [\ion{O}{ii}] spatial distributions, we computed the Kendall correlation coefficient ($\tau$) following the \cite{Akritas96} prescription for censored data to take into account the upper limits on variables. We used the routine described in \cite{Flury22b}\footnote{https://github.com/sflury/kendall} that also provides uncertainties on $\tau$ estimated by bootstrapping. Following \cite{amorin2024} with a similar sample size, we considered correlations to be (i) significant if the false-positive probability ($p$) that the correlation is real is $\lesssim 2.275$ $\times$ $10^{-2}$ (2$\sigma$ confidence) and (ii) strong if $|\tau|$ $\gtrsim$ 0.261. While there is a 2$\sigma$ correlation between the extent of the \ion{Mg}{ii} emission and continuum, we found that the spatial extent of the [\ion{O}{ii}] halo does not strongly depend on the stellar continuum size. The strongest correlation (>3$\sigma$) is observed between the \ion{Mg}{ii} and [\ion{O}{ii}] scale lengths (Fig.~\ref{fig:rs}).
We also compared our \ion{Mg}{ii} scale length measurements with the UV half light radius ($r^{\rm UV}_{50}$, y-axis) measured in HST/COS data \citep{Flury22, Izotov22} and found a tentative <2$\sigma$ correlation ($\tau=0.35\pm0.09$ and $p=0.05$, see top panel of Fig.~\ref{fig:r50_offset}), suggesting a possible connection between the UV size of the galaxies and their \ion{Mg}{ii} spatial extent.

\subsection{Spatial offset between \ion{Mg}{ii}, [O~II] and the continuum}
\label{sec:35}

Our modeling procedure also provides us with centroid measurements for the emission line maps and their respective continua ($\Delta_{\rm \ion{Mg}{ii}-cont}$ and $\Delta_{\rm [\ion{O}{ii}]-cont}$ for \ion{Mg}{ii} and [\ion{O}{ii}], respectively). 
Our emission line to continuum spatial offset measurements range from no significant (<3$\sigma$) offset to 2.83 kpc (J0919+4906) for \ion{Mg}{ii} and 2.16 kpc (J0804+4726) for [\ion{O}{ii}]. 
Seven objects have a >3$\sigma$ significant \ion{Mg}{ii} offset to the continuum (J0047+0154, J0834+4805, J0912+5050, J1033+6353, J1046+5827, J1137+3605, J1154+2443). Most of our sources show a significant (>3$\sigma$) spatial offsets between the [\ion{O}{ii}] emission and the continuum (all except J0919+4906, J1352+5617, J1014+5501, J1154+2443). Finally, 14 objects have a significant offset between \ion{Mg}{ii} and [\ion{O}{ii}] (all except J0940+5932, J1648+4957, J0141-0304, J1157+5801).
The average $\Delta_{\rm \ion{Mg}{ii}-cont}$, $\Delta_{\rm [\ion{O}{ii}]-cont}$ and $\Delta_{\rm \ion{Mg}{ii}-[\ion{O}{ii}]}$ values (considering significant spatial offsets only) are 1.2, 0.5 and 0.9 kpc, respectively.

We find a tentative (2$\sigma$) correlation between the \ion{Mg}{ii} scale length and $\Delta_{\rm \ion{Mg}{ii}-cont}$ ($\tau$=0.37, $p$=0.025).
We do not observe such a correlation between the [\ion{O}{ii}] size and offset from the continuum. We also report no correlation between the different scale length ratios ($rs_{\rm \ion{Mg}{ii}}$ / $rs_{\rm cont}$, $rs_{\rm [\ion{O}{ii}]}$ / $rs_{\rm cont}$, and $rs_{\rm \ion{Mg}{ii}}$ / $rs_{\rm [\ion{O}{ii}]}$) and their respective spatial offsets ($\Delta_{\rm \ion{Mg}{ii}-cont}$, $\Delta_{\rm [\ion{O}{ii}]-cont}$ and $\Delta_{\rm \ion{Mg}{ii}-[\ion{O}{ii}]}$). The corresponding figures can be found in Fig.~\ref{fig:rs_offset}.

\begin{figure}
\centering
   \resizebox{\hsize}{!}{\includegraphics{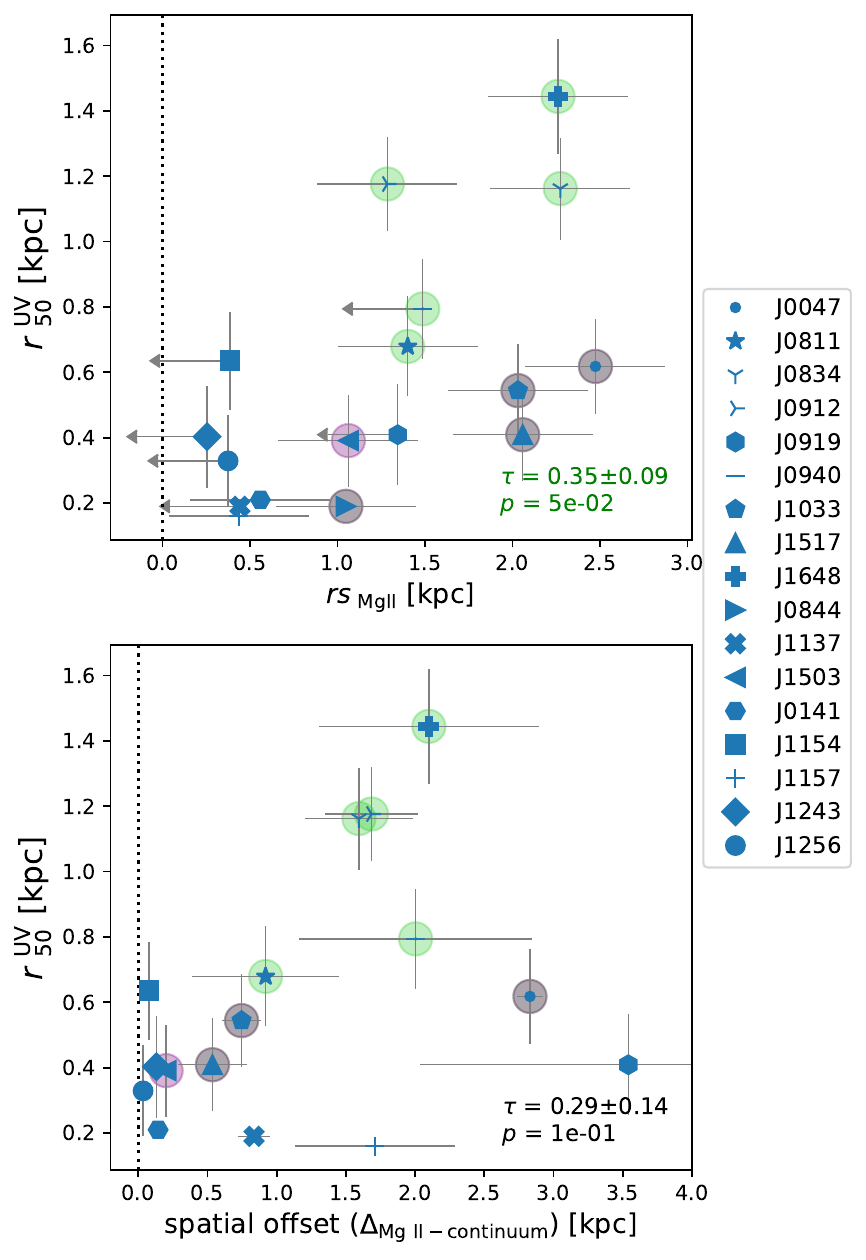}}
    \caption{Relation between the UV half light radius ($r^{\rm UV}_{50}$, y-axis) measured in HST/COS data \citep{Flury22, Izotov22} and the spatial extent (top) and offset (bottom) between the \ion{Mg}{ii} emission and continuum centroids (x-axis). The Kendall correlation coefficient ($\tau$) for every pair of variables and the corresponding false-positive probability that the correlation is real ($p$) are given and colored in green if the correlation is statistically significant at $\approx$2$\sigma$ (\citealt{Akritas96,Flury22b}, see Sect.~\ref{sec:34}). Objects with statistically significant \ion{Mg}{ii} and [\ion{O}{ii}] extended emission are indicated by large purple and green symbols, respectively (Sect.~\ref{sec:34}). Grey symbols result in the overlap of the purple and green symbols and thus indicate that both \ion{Mg}{ii} and [\ion{O}{ii}] halos are detected. Relations between the spatial offsets and other size measurements are shown in Fig.~\ref{fig:offset_rsall}.}
    \label{fig:r50_offset}
\end{figure}

We find that the $\Delta_{\rm \ion{Mg}{ii}-cont}$ offset correlates with most of the galaxy size measurements (\ion{Mg}{ii}, [\ion{O}{ii}], and continuum scale lengths, see Fig.~\ref{fig:offset_rsall}),
including the half light radius as measured in \cite{Flury22} and \cite{Izotov22} (bottom panel of Fig.~\ref{fig:r50_offset}). This last correlation is weaker but holds ($\tau$ = 0.3, $p$ = 0.1), suggesting a tentative relationship between the size of the galaxies and their \ion{Mg}{ii} spatial offsets.

\begin{table*}
\centering
\captionof{table}{Results from our two-dimensional exponential fitting procedure (Sect.~\ref{sec:33}).}
\def\arraystretch{1.5}
\begin{tabular}{lccccccc}
\hline
ID & $rs_{\rm \ion{Mg}{ii}}$ & $rs_{\rm cont@\ion{Mg}{ii}}$ & $rs_{\rm [\ion{O}{ii}]}$ & $rs_{\rm cont@[\ion{O}{ii}]}$ & $\Delta_{\rm \ion{Mg}{ii}-[\ion{O}{ii}]}$ & $\Delta_{\rm \ion{Mg}{ii}-cont}$ & $\Delta_{\rm [\ion{O}{ii}]-cont}$ \\
 & [kpc] & [kpc] & [kpc] & [kpc] & [kpc] & [kpc] & [kpc] \\
\hline
\hline
J0047+0154       & 2.47±0.06 & <0.8 & 1.99±0.01 & <0.6 & 2.49±0.1 & 2.83±0.1 & 0.3±0.05 \\
J0130-0014       & -- & -- & <0.4 & <0.5 & -- & -- & 1.09±0.14 \\
J0141-0304       & 0.56±0.02 & 0.61±0.06 & <0.6 & 0.6±0.03 & 0.01±0.03 & 0.14±0.07 & 0.21±0.03 \\
J0804+4726$^{*}$ & -- & -- & <1.2 & <1.2 & -- & -- & 2.16±0.49 \\
J0811+4141       & 1.4±0.25 & 3.6±0.26 & 1.22±0.02 & <1.0 & 1.36±0.27 & 0.92±0.53 & 0.69±0.07 \\
J0834+4805       & 2.27±0.29 & 2.84±0.11 & 1.79±0.01 & <0.8 & 1.27±0.25 & 1.59±0.39 & 0.12±0.02 \\
J0844+5312       & 1.05±0.05 & 0.59±0.13 & 1.05±0.01 & 0.96±0.06 & -- & 0.29±0.13 & 0.22±0.07 \\
J0912+5050       & 1.28±0.22 & 2.68±0.15 & 1.67±0.01 & <0.8 & 1.76±0.19 & 1.68±0.34 & 0.43±0.02 \\
J0919+4906       & <1.3 & 1.76±1.55 & <0.8 & <1.6 & 2.25±0.19 & 3.54±1.5 & 1.73±0.67 \\
J0940+5932       & <1.5 & <1.3 & 1.97±0.02 & <0.9 & 2.18±0.79 & 2.0±0.84 & 0.4±0.03 \\
J1014+5501       & 2.9±0.21 & 2.35±0.17 & 1.47±0.01 & 2.21±0.13 & 0.96±0.28 & 1.08±0.45 & 0.33±0.13 \\
J1033+6353$^{*}$ & 2.03±0.08 & 1.72±0.07 & 1.83±0.01 & <1.2 & 0.47±0.09 & 0.75±0.14 & 0.09±0.02 \\
J1046+5827       & 1.21±0.05 & 0.85±0.05 & 0.74±0.01 & 0.68±0.04 & 0.53±0.05 & 0.36±0.09 & 0.31±0.04 \\
J1137+3605       & <0.4 & 0.94±0.07 & <0.4 & 0.88±0.03 & 0.39±0.08 & 0.84±0.12 & 0.31±0.03 \\
J1154+2443$^{*}$ & <0.4 & <0.5 & <0.5 & <0.4 & 0.07±0.02 & 0.08±0.02 & 0.01±0.01 \\
J1157+5801       & 0.43±0.25 & <1.2 & 0.56±0.01 & 0.77±0.07 & 1.02±0.37 & 1.71±0.58 & 0.85±0.09 \\
J1243+4646$^{*}$ & <0.3 & <0.3 & <0.5 & 0.44±0.1 & 0.08±0.01 & 0.13±0.04 & 0.48±0.01 \\
J1256+4509$^{*}$ & <0.4 & <0.5 & <0.5 & <0.3 & 0.11±0.02 & 0.03±0.02 & 0.19±0.06 \\
J1352+5617       & 1.27±0.04 & 1.36±0.1 & 0.93±0.01 & 1.1±0.06 & 0.28±0.05 & 0.38±0.14 & 0.07±0.06 \\
J1503+3644       & 1.06±0.03 & 0.92±0.05 & -- & -- & -- & 0.2±0.07 & -- \\
J1517+3705       & 2.06±0.11 & 1.35±0.14 & 1.46±0.01 & <1.0 & 0.39±0.11 & 0.54±0.25 & 0.26±0.04 \\
J1648+4957       & 2.26±0.25 & 3.13±0.32 & 2.01±0.02 & <1.3 & 0.77±0.28 & 2.1±0.8 & 0.79±0.08 \\
\hline
\end{tabular}
\smallskip\\
\small{\textbf{Notes.} ID: source identifier. Objects identified with $^{*}$ are considered as strong LyC leakers following the criteria of \cite{Flury22b}: $>5\sigma$ detection and \fesc>5\%. $rs_{\rm \ion{Mg}{ii} ([\ion{O}{ii}])}$: exponential scale length of the \ion{Mg}{ii} ([\ion{O}{ii}]) emission in physical kpc. $rs_{\rm cont@\ion{Mg}{ii} ([\ion{O}{ii}])}$: exponential scale length of the continuum at \ion{Mg}{ii} ([\ion{O}{ii}]) wavelenghts in kpc. $\Delta_{\rm \ion{Mg}{ii}-[\ion{O}{ii}]}$: spatial offset between the \ion{Mg}{ii} and [\ion{O}{ii}] emission best-fit centroids in kpc. $\Delta_{\rm \ion{Mg}{ii}([\ion{O}{ii}])-cont}$: spatial offset between the \ion{Mg}{ii} ([\ion{O}{ii}]) emission and the continuum best-fit centroids in kpc.}
\label{tab:fitting}
\end{table*}

\section{Connecting LyC leakage to gas distribution}
\label{sec:4}

Our characterization of the spatial distributions of the neutral and ionized gas in our galaxy sample unveils diverse gaseous configurations. Our targets indeed show from very compact (unresolved) to extended \ion{Mg}{ii} and [\ion{O}{ii}] emission (up to $\sim$10~kpc), as well as from significant (up to few kpc) to zero spatial offsets between the nebular gas and the stellar continuum.
We now explore the connection between the spatial properties of the emitting gas and the escape of ionizing photons, derived in the literature from fitting the COS UV spectrum \citep{Flury22,Saldana-Lopez2022,Izotov22}, using both individual (Sects.~\ref{sec:41} and ~\ref{sec:42}) and stacking measurements (Sect~\ref{sec:43}).

\subsection{\ion{Mg}{ii} and {\rm [\ion{O}{ii}]} spatial extents versus \fesc}
\label{sec:41}

\begin{figure*}
\centering
   \resizebox{\hsize}{!}{\includegraphics{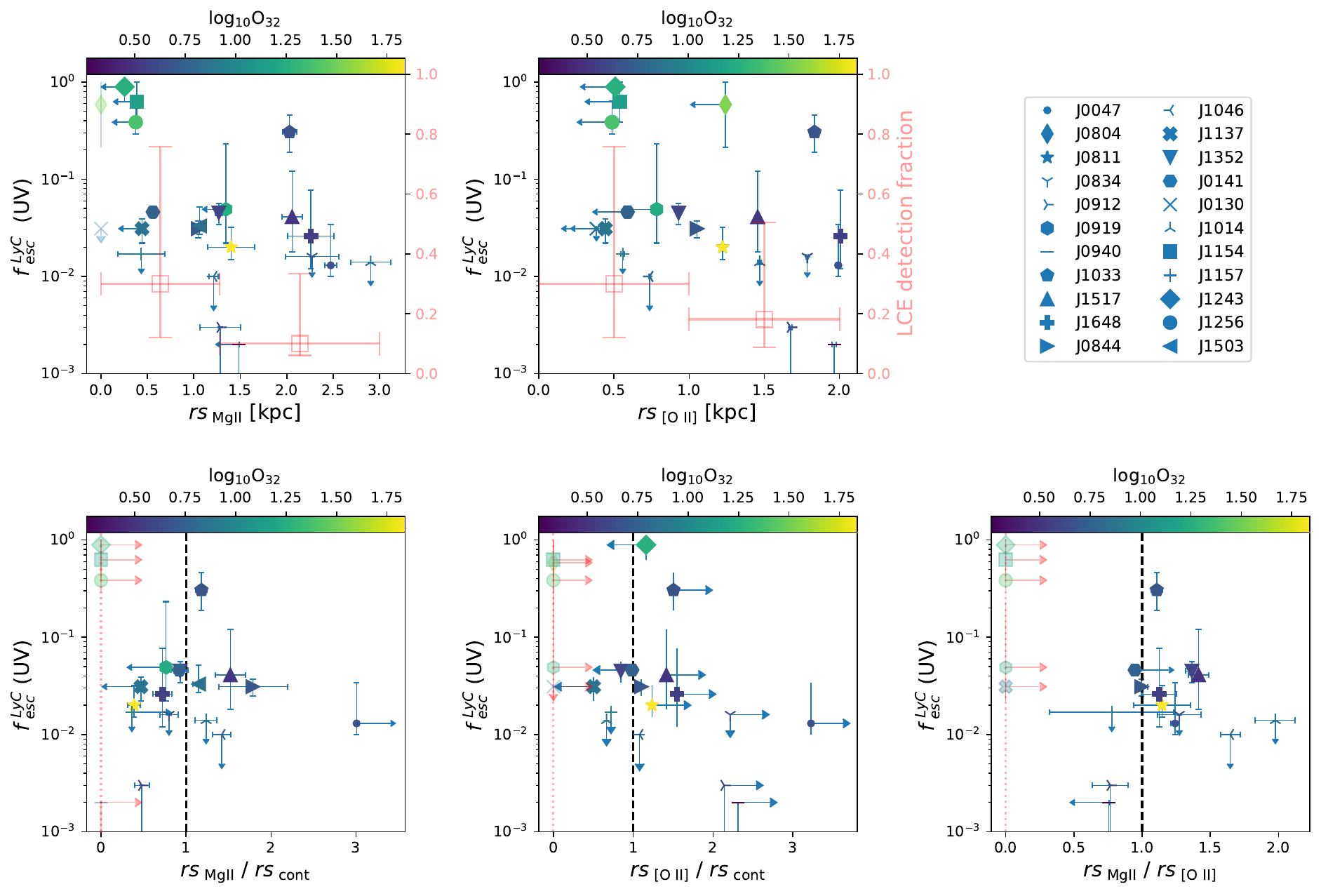}}
    \caption{Relation between the LyC escape fraction and the spatial extent of the neutral and ionized gas as traced by \ion{Mg}{ii} and [\ion{O}{ii}], respectively. The points are color-coded by the O$_{32}$ values measured in \cite{Flury22}.
    \textit{Top panels:} \fesc\ as a function of the \ion{Mg}{ii} (left) and [\ion{O}{ii}] (right) scale length. Objects with undetected \ion{Mg}{ii} are shown with higher transparency at $x=0$. The red squares indicate the fraction of strong LyC leakers (5$\sigma$ detection and \fesc>5\% following \citealt{Flury22}) detected in the bin delimited by the horizontal error bars. The errors on the fractions correspond to a 1$\sigma$ confidence interval (see Sect.~\ref{sec:51}).
    \textit{Bottom panels:} \fesc\ as a function of the ratio of the \ion{Mg}{ii} ([\ion{O}{ii}]) scale length to continuum scale length is shown in the left (middle) panel. The right panel shows \fesc\ versus the \ion{Mg}{ii} and [\ion{O}{ii}] scale length ratio. Objects with upper limits for both scale lengths are shown at $x=0$ with red lower limits and higher transparency. While weak and non LyC emitters show a wide diversity of \ion{Mg}{ii} and [\ion{O}{ii}] spatial extents, the neutral and ionized gas of strong leakers is compact (see Sect.~\ref{sec:41} and~\ref{sec:53}).}
    \label{fig:fesc_rs}
\end{figure*}

Figure~\ref{fig:fesc_rs} shows the connection between the escape of ionizing photons measured using the FUV continuum fits in \cite{Flury22} and the different exponential scale lengths and scale length ratios from our analysis (Sect.~\ref{sec:34}). We found that the strongest LyC leakers (\fesc > 30\%) are unresolved, and therefore compact ($rs\lesssim$ 0.5 kpc), in both \ion{Mg}{ii} and [\ion{O}{ii}] emission (top first and second panels),  whereas the weaker or non leakers show a wider diversity with scale lengths ranging from upper limits (i.e., unresolved) to 3 kpc and 2 kpc for \ion{Mg}{ii} and [\ion{O}{ii}], respectively. 
We compute the fraction of strong LyC leakers (LCE detection fraction) detected in equal size bins of small and large scale lengths. Following \cite{Flury22}, an object is considered as a strong leaker if \fesc > 5\% with $>5\sigma$ significance. The statistical uncertainties for fractions (i.e., Bernoulli trials) are given by a binomial proportion confidence interval using the Wilson approximation formula (Wilson 1927). Our 1$\sigma$ uncertainties correspond to 68.3\% confidence intervals. 
The trends with both scale lengths suggest that the LCE detection fraction decreases with increasing \ion{Mg}{ii} and [\ion{O}{ii}] spatial extents. We note that such a trend is more significant for \ion{Mg}{ii}.
Interestingly, J1033+6353 appears like an exception because it is a strong leaker (\fesc $\sim$ 30\%) with large \ion{Mg}{ii} and [\ion{O}{ii}] spatial scale lengths ($\approx$2 kpc). We also note that J0804+4726 is a strong leaker (\fesc $\sim$ 60\%) with no \ion{Mg}{ii} detection and only an upper limit on its [\ion{O}{ii}] scale length. J0804+4726 has large uncertainties on its stellar continuum fit leading to a poorly constrained LyC escape fraction (\fesc $\simeq$ 0.58$\pm$0.40). This object is also our lowest metallicity source (12+log$_{10}$(O/H)=7.5), which might explain its weak \ion{Mg}{ii} emission. The LyC measurement for J1033+6353 is more reliable. We discuss this last object in Sect.~\ref{sec:53}. 

When comparing the ratio between the emission and continuum scale lengths -- excluding objects with upper limits on both their emission and continuum size measurements for which we cannot conclude (i.e., most of the strong leakers) -- we do not find any strong correlation with \fesc\ (Fig.~\ref{fig:fesc_rs} bottom). \ion{Mg}{ii} and [\ion{O}{ii}] halos are indeed detected both in weak and non leakers. While J1033+6353 shows significant extended neutral and ionized gas, both \ion{Mg}{ii} and [\ion{O}{ii}] halos are less than 1.5 times more extended than the continuum. Objects with larger halos (>1.5 times the continuum extent) are all weak (\fesc < 3\%) or non-leakers. This is also valid when comparing the neutral and ionized gas extents; sources with \ion{Mg}{ii} more extended than [\ion{O}{ii}] are non-leakers (bottom right panel). We note that larger samples and higher spatial resolution are required to confirm these trends.

\subsection{Spatial offsets versus \fesc}
\label{sec:42}

\begin{figure*}
\centering
   \resizebox{\hsize}{!}{\includegraphics{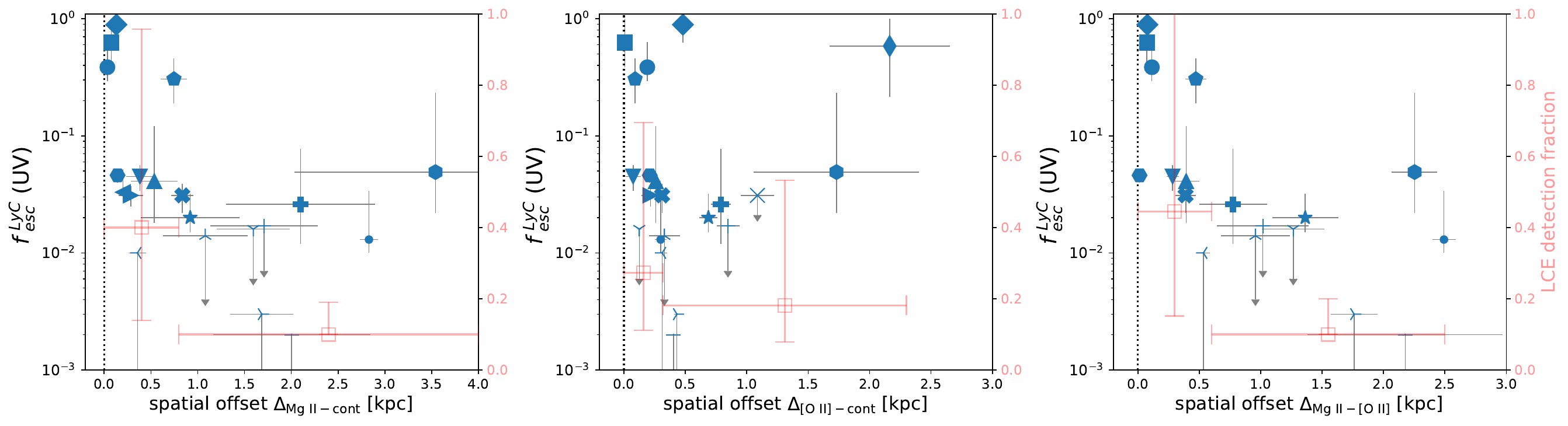}}
    \caption{Relation between the LyC escape fraction and the spatial offset of the neutral (as traced by \ion{Mg}{ii}, $\Delta_{\rm \ion{Mg}{ii}-cont}$, left panel) and ionized gas (as traced by [\ion{O}{ii}], $\Delta_{\rm [\ion{O}{ii}]-cont}$, middle panel) from the stellar continuum. The right panel shows \fesc\ as a function of the \ion{Mg}{ii} spatial offset relative to the [\ion{O}{ii}] centroid ($\Delta_{\rm \ion{Mg}{ii}-[\ion{O}{ii}]}$). The red squares indicate the fraction of strong LyC leakers detected in the bin shown by the horizontal error bars. The errors on the fractions correspond to a 1$\sigma$ confidence interval (see Sect.~\ref{sec:41}). Symbols legend is the same as Fig.~\ref{fig:fesc_rs}. While weak and non LyC emitters show a wide diversity of \ion{Mg}{ii} and [\ion{O}{ii}] spatial offset from the continuum, leakers no/weak offset due to their compact configurations (see Sect.~\ref{sec:41}).}
    \label{fig:fesc_offset}
\end{figure*}

We now connect the LyC leakage fraction to the spatial offsets measured between the emission line and continuum centroids (Sect.~\ref{sec:35}). First panel of Figure~\ref{fig:fesc_offset} shows that strong leakers have non or small (<1 kpc) spatial offset between \ion{Mg}{ii} and stellar continuum. We computed the LCE detection fraction similarly as in Sect.~\ref{sec:41} and found that it significantly decreases with increasing $\Delta_{\rm \ion{Mg}{ii}-cont}$. We found a similar trend between \fesc\ and the spatial offset between \ion{Mg}{ii} and [\ion{O}{ii}] centroids ($\Delta_{\rm \ion{Mg}{ii}-[\ion{O}{ii}]}$, right panel).
However, this trend is less significant when considering the spatial offset between [\ion{O}{ii}] and the continuum ($\Delta_{\rm [\ion{O}{ii}]-cont}$, middle panel).
The same results are obtained when we normalize the offsets by the $r_{50}^{\rm UV}$.

\subsection{\ion{Mg}{ii}, {\rm [\ion{O}{ii}]} and continuum stacks}
\label{sec:43}

\begin{figure*}
\centering
   \resizebox{\hsize}{!}{\includegraphics{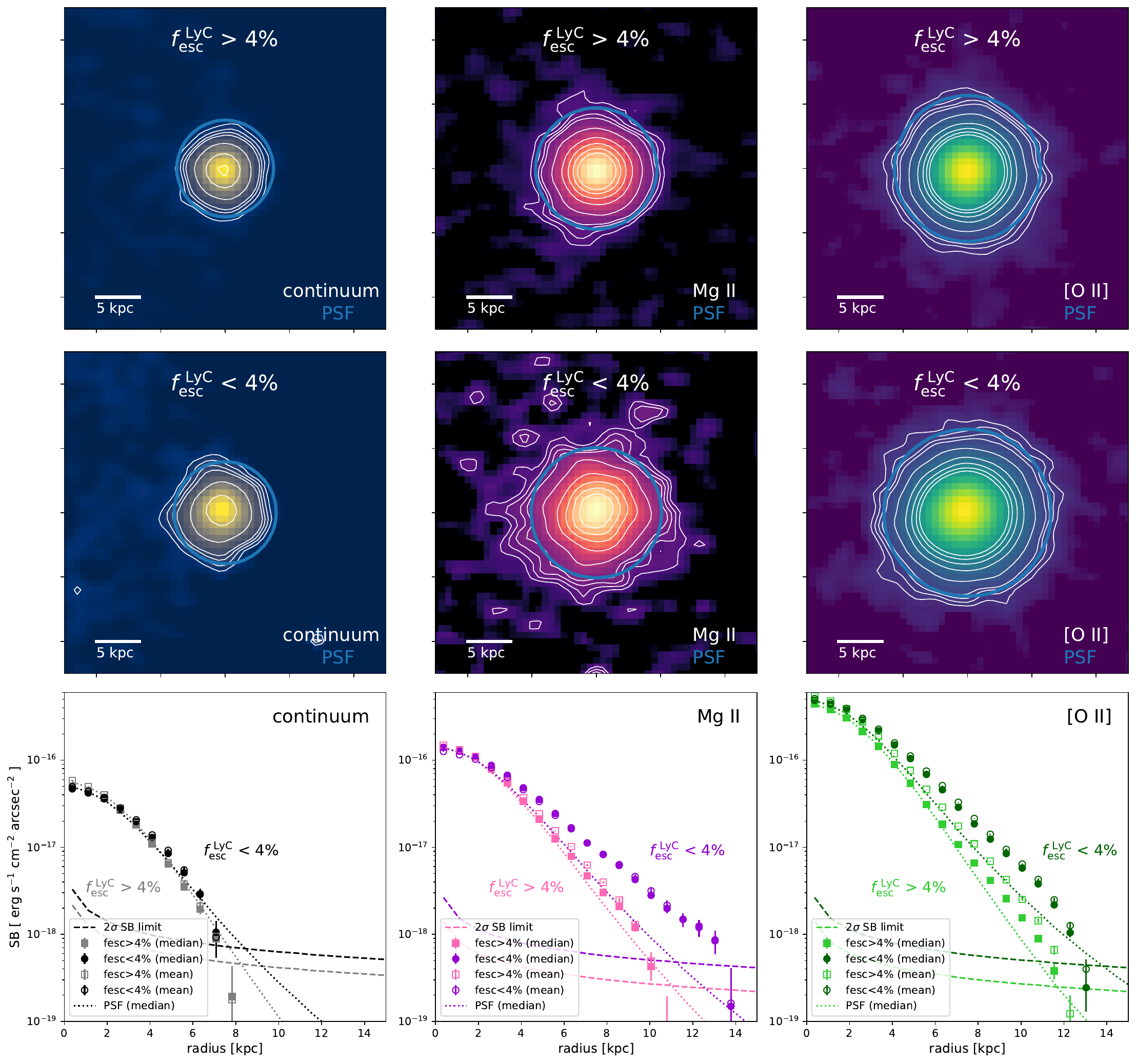}}
    \caption{Composite images of our strong (\fesc$>4\%$, top row) and non/weak (\fesc$<4\%$, middle row) LyC leakers observed with KCWI (see Sect.~\ref{sec:43}). The UV continuum, \ion{Mg}{ii} and [\ion{O}{ii}] mean stacks are shown from left to right, respectively. The contours are spaced logarithmically by 0.2 dex, with the lowest contour level always at $4\times10^{-18}~\sbl$.
    The bottom row compares the continuum (left), \ion{Mg}{ii} (middle) and [\ion{O}{ii}] (right) radial SB profiles of the two subsets composites images. In each bottom panels, the lighter (darker) colored profiles with square (circle) symbols correspond to the strong (non/weak) LyC leaker subsample. The full and empty markers show the profiles of the median and mean composite images, respectively. The dotted lines correspond to the SB profiles of the median stacked PSF images of the different subsamples. The dashed lines show the 2$\sigma$ significance levels.}
    \label{fig:stacks}
\end{figure*}

To increase the signal-to-noise ratio and determine the average SB profile of LyC leakers and weak or non leakers, we adopt a 2D stacking procedure. We only consider the objects with KCWI data because they were observed in very similar conditions (PSF FWHM $\approx$ 1$\arcsec$) compared to the LRS2 observations (see Table~\ref{tab:psf}). We exclude J0130-0014 because it has a larger PSF (FWHM$\approx$1\farcs3) and a bright neighboring source (see Fig.~\ref{fig:profiles2}). J1014+5501 is also excluded because it has a large PSF (FWHM$\approx$1\farcs5) and a different pixel scale because it was observed with a different slicer (see Table~\ref{tab:obs}). J0844+5312 has also been observed under less good conditions (FWHM$\approx$1\farcs2) but is included in the stacks. The inclusion of J0844+5312 does not change our conclusions.
Our KCWI targets have very similar redshifts ($z = 0.32-0.43$). We therefore do not rescale the flux of each individual image to correct for the impact of cosmological dimming. We also do not normalize the flux of the images before stacking to retain physical units. However, we verify that the brightest sources are not dominating the composite images by ensuring that the normalization does not affect our results.
We split our KCWI in two equal size subsamples resulting in 5 objects with \fesc$>4\%$ and 5 objects with \fesc$<4\%$.
The continuum subtracted \ion{Mg}{ii} and [\ion{O}{ii}] images, and the continuum images were extracted as detailed in Sects.~\ref{sec:31} and~\ref{sec:32} (spatially centered on the continuum peak). Each subsample image was averaged using both median and mean functions for comparison. Figure.~\ref{fig:stacks} shows the resulting averaged stacks.

\begin{figure}
\centering
   \resizebox{\hsize}{!}{\includegraphics{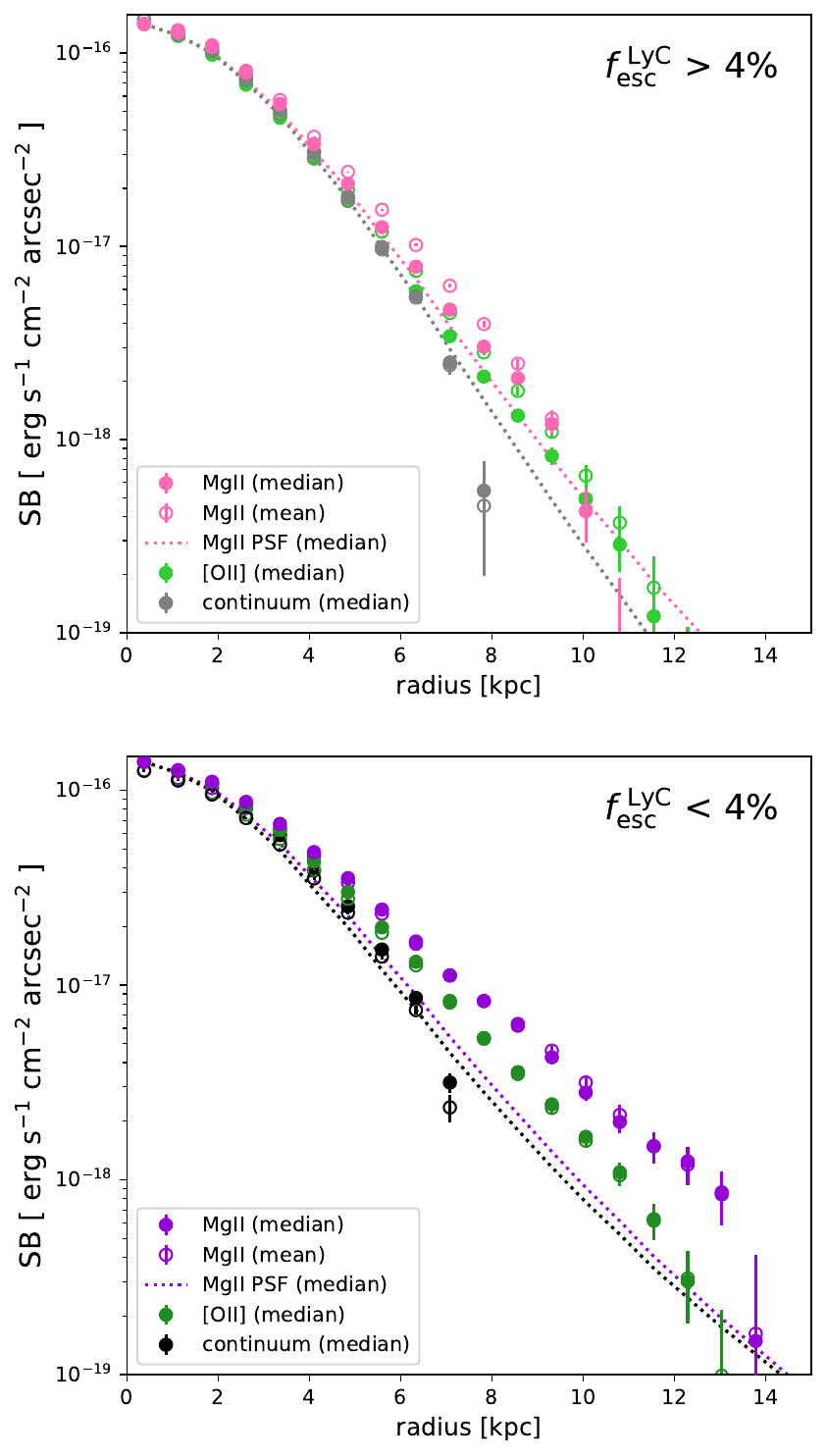}}
    \caption{Direct comparison between the radial SB profiles of the strong (\fesc$>4\%$, top) and weak/non (\fesc$<4\%$, bottom) LyC leakers composite images. The legend is the same as in Fig.~\ref{fig:stacks}.}
    \label{fig:stacks_prof}
\end{figure}

Although our statistics are small, we gain a factor of more than 2.5 in terms of limiting SB levels compared to individual images, reaching levels of $\approx$1$\times$10$^{-18}$ and 3$\times$10$^{-18}$ \sbl\ for the [\ion{O}{ii}] and \ion{Mg}{ii} composite images, respectively. We see a trend that both \ion{Mg}{ii} and [\ion{O}{ii}] are more extended than the continuum in weak/non LyC leakers (\fesc$<4\%$) compared to stronger leakers (\fesc$>4\%$). This is even clearer when comparing the radial SB profiles of the two subsamples (Fig.~\ref{fig:stacks} bottom). While the emission lines are more extended, the continuum sizes are not significantly different, indicating that strong and weak LyC emitters have different nebular gas configurations.
Besides highlighting the compact/extended nature of the strong/weak leakers, the profiles reveal that the \ion{Mg}{ii} emission around weak/non leakers is more extended than the [\ion{O}{ii}] emission (Fig.~\ref{fig:stacks_prof}). These stacking results are in good agreement with the trends from our individual analysis (Sects.~\ref{sec:41} and~\ref{sec:42}). We however remind the reader that the individual measurements reveal a large diversity of spatial properties in the weak/non leakers sample. 

\section{Discussion} 
\label{sec:5}

\subsection{Previous results on extended \ion{Mg}{ii} and {\rm [\ion{O}{ii}]} emission}
\label{sec:51}

Thanks to absorption studies \citep[see][for a review]{T17}, we know that galaxies are surrounded by metal-enriched gaseous halos. However, such pencil-beam surveys do not provide information about the morphology, porosity or global kinematics of the CGM.
Detecting gaseous halos directly in emission would be ideal but is challenging because of their low surface brightness ($\lesssim10^{-18} ~\sbl$). Very recent work by \cite{Guo2023Natur} reported the detection of Mg-enriched gas by stacking MUSE observations of $z = 0.7-2.3$ galaxies. This work revealed that the \ion{Mg}{ii} emission is statistically more extended than the stellar light and preferentially extends along the minor axis of massive galaxies ($M_*>10^{9.5}$ M$_\odot$) following a biconical geometry tracing outflowing gas. Lower mass galaxies, resembling our galaxy sample, show rather circular \ion{Mg}{ii} distribution extending up to 10 kpc. Similarly, \cite{Dutta2023} stacked $\approx$600 galaxies of median mass $M_*\sim2\times10^{9}$ M$_\odot$ at $z=0.7-1.5$ observed with MUSE and found \ion{Mg}{ii} emission extending up to 25 kpc at a SB level of 10$^{-20}$ \sbl. The ubiquity of \ion{Mg}{ii} halos is also observed in simulated star-forming galaxies, regardless of the the stellar mass or redshift \citep{Nelson2021}. 
Although we expect \ion{Mg}{ii} halos to be universal, individual detections are still not very numerous in the literature.
The few recent maps of extended \ion{Mg}{ii} emission around individual galaxies have been built due to sensitive IFUs like MUSE or KCWI.
\cite{Burchett2021} and \cite{Zabl2021} reported \ion{Mg}{ii} halos around two $z\simeq0.7$ star forming galaxies, extending out from the galaxy center to a radius of $\approx$20 and 25 kpc at 1$\sigma$ SB limits of $7\times10^{-19}$ and $5\times10^{-19}$ \sbl, respectively. 
\cite{Shaban2022} recently found patchy extended \ion{Mg}{ii} emission extending out to a radial distance of 27 kpc around a gravitationally lensed star-forming galaxy at $z=1.7$ (1$\sigma$ detection limit of $\approx1\times10^{-18}$ \sbl). While these galaxies from the literature are at higher redshift and are more massive than our sample ($z\sim0.35$, $M_*\sim10^{10}$ M$_\odot$), our 7 \ion{Mg}{ii} halos have similar maximal spatial extents ($\approx 10 - 20$ kpc at a $1\sigma$ SB limit of $8\times10^{-18}$ \sbl). This suggests that the \ion{Mg}{ii} spatial extent does not strongly depend on stellar masses or redshift, in contrast to the \ion{Mg}{ii} spectral shapes. \cite{Finley17} and \cite{F18} indeed show that massive galaxies present high column density \ion{Mg}{ii} absorption, whereas low mass galaxies typically show \ion{Mg}{ii} emission and reside in the low column density regime (implying little to no scattering, i.e., most likely the leaking regime). 
The lack of observed trends between the \ion{Mg}{ii} spatial extent and the stellar masses is in agreement with the simulation work of \cite{Nelson2021}.
The observed spatial extent values are however more than ten times larger than the simulated spatial extents of \cite{Nelson2021} at a fixed SB of $1\times10^{-18}$ \sbl\ in the $z\sim0.3$ redshift bin and the smallest ($M_*\sim10^{9}$ M$_\odot$) mass bin (their Fig.~3). Given that these simulations do not take into account radiative transfer effects, this difference suggests that \ion{Mg}{ii} resonant scattering has a strong effect on the \ion{Mg}{ii} halos spatial extents. 

Similarly to our results, most of the studies cited above also report [\ion{O}{ii}] emission more extended than the stellar continuum when \ion{Mg}{ii} is extended, both in stacks and individual objects. 
In particular, our stack experiment reveals for the first time that on average \ion{Mg}{ii} is statistically more extended than both the continuum and [\ion{O}{ii}] for objects with \fesc< 4\%. This is expected when there are numerous resonant scatterings that propagate \ion{Mg}{ii} photons out spatially.
In individual objects, \ion{Mg}{ii} also appears more extended than [\ion{O}{ii}] \citep[see][and some objects of our study]{Zabl2021}, but we also find a large diversity of configurations with sources showing similar \ion{Mg}{ii} and [\ion{O}{ii}] extents or larger [\ion{O}{ii}] scale lengths. 
\cite{Dutta2023} found [\ion{O}{ii}] to be more extended in their composite images of galaxies within large scale groups where environmental effects could be at play. \cite{Rupke2019} also discovered a nebula that extended 100 kpc [\ion{O}{ii}] around a massive galaxy ($M_*\sim10^{11}$ M$_\odot$) at $z=0.46$ ($1\sigma$ SB limit of $1\times10^{-18}$ \sbl), with \ion{Mg}{ii} emission a lot more compact than [\ion{O}{ii}]. We find similar objects in our sample, with 5 galaxies showing ionized [\ion{O}{ii}] gas but no extended neutral gas, as traced by \ion{Mg}{ii}. We also note that our ground-based data do not resolve some of the most extreme LyC leakers, leaving the full picture for their extent largely unclear. Overall, our \ion{Mg}{ii} and [\ion{O}{ii}] halo properties are comparable to previous studies reported in the literature. 

Finally, while our metal-enriched emitting halos are $\approx$10 times less extended than their continuum compared to the high-$z$ Ly$\alpha$ halos \citep[$z>3$, e.g.,][]{W16,L17,W18,Ku20,Claeyssens2022}, we find similar emission to continuum size ratios ($rs_{\rm \ion{Mg}{ii}}$/$rs_{\rm cont}$ and $rs_{\rm [\ion{O}{ii}]}$/$rs_{\rm cont}$ ranging from 1 to 3) as for the Ly$\alpha$ and H$\alpha$ halos reported in nearby galaxies \citep{H13,Rasekh2022}. In contrast with the \ion{Mg}{ii} halos, \citet{Rasekh2022} report a correlation between the size of the Ly$\alpha$ halos and the stellar masses. Larger samples of \ion{Mg}{ii} halos will be needed to understand this difference.

\subsection{Physical properties for compact \ion{Mg}{ii} and [\ion{O}{ii}] gas configurations}
\label{sec:52}

\begin{figure*}
\centering
   \resizebox{\hsize}{!}{\includegraphics{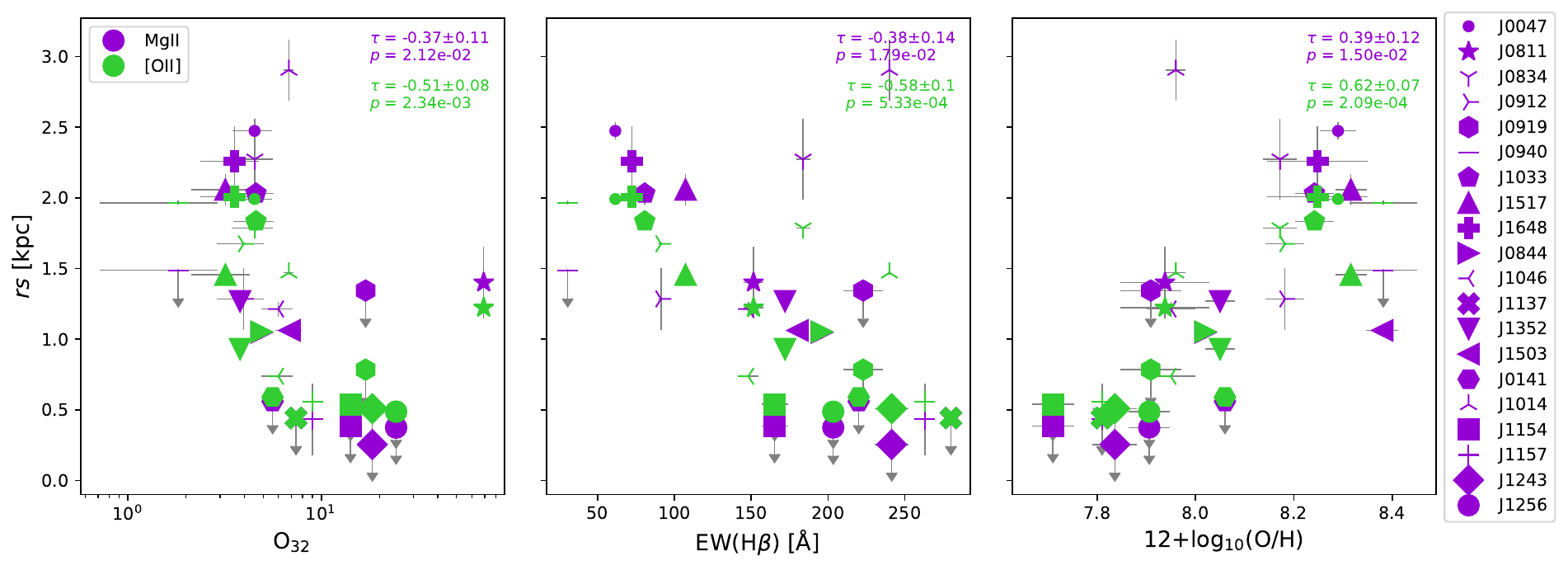}}
    \caption{Comparisons of the emission scale lengths (\ion{Mg}{ii} in purple and [\ion{O}{ii}] in green) and the O$_{32}$ ratios (left), H$\beta$ equivalent widths (middle), and metallicities (right). Upper limit values are shown with arrows. The Kendall correlation coefficient ($\tau$) for every pair of variables and the corresponding false-positive probability that the correlation is real ($p$) are given for both emission lines in the top right corners following the same color coding as the data points (\citealt{Akritas96,Flury22b}, see Sect.~\ref{sec:34}).}
    \label{fig:rs_o3hbz}
\end{figure*}

As mentioned in the previous section, the first few simulations and observations of \ion{Mg}{ii} halos indicate that the spatial extent of the neutral gas is independent of redshift and stellar mass (Sect.~\ref{sec:51}). Here we investigate whether any other physical parameters impact the neutral and ionized gas extent. We compare the \ion{Mg}{ii} exponential scale lengths measured on our statistical sample of 22 objects with their global properties as measured in \cite{Flury22} and \cite{Izotov22} in Fig.~\ref{fig:rs_o3hbz}. We find significant correlations with the O$_{32}$ ratio ($\tau=-0.37\pm0.12$ and $p=0.02$), H$\beta$ equivalent width ($\tau=-0.38\pm0.14$ and $p=0.02$), metallicity ($\tau=0.39\pm0.12$ and $p=0.015$), and marginally with $r^{\rm UV}_{50}$ ($\tau=0.35\pm0.12$ and $p=0.05$, see top panel of Fig.~\ref{fig:r50_offset}). 
In other words, compact configurations of the neutral gas are preferentially found in galaxies with high ionization parameter, large H$\beta$ equivalent width, low metallicity, and rather compact in UV sizes.
The spatial extent of the ionized gas, as traced by [\ion{O}{ii}], is even more strongly correlated (>3$\sigma$) with these exact same quantities. The fact that we observe weaker correlations for \ion{Mg}{ii} could be due to the fact that (i) \ion{Mg}{ii} is intrinsically fainter than [\ion{O}{ii}] and (ii) \ion{Mg}{ii} resonant scattering increases the chances for \ion{Mg}{ii} photons to be absorbed by dust, adding scatter in some correlations. We find no correlations with $E$(B$-$V), UV $\beta^{1500}$ slope, stellar mass, star formation rate (SFR), SFR surface density, specific SFR, or stellar age. We also report a lack of strong correlation between the \ion{Mg}{ii} spatial extent and the Ly$\alpha$ equivalent width or escape fraction.

The inverse correlation between the ionization of the galaxy and the total extent of the neutral gas suggests that galaxies that have more compact neutral gas sizes are more highly ionized. This could arise from the stellar populations within the galaxies ionizing most of the neutral gas within the galaxy and halo, such that galaxies with high O$_{32}$ and large H$\beta$ equivalent widths ionize all of the neutral gas within 1~kpc. The literature has often called this a "density-bounded" ISM configuration \citep[e.g.,][]{JaskotOey2013,NakajimaOuchi2014,Gazagnes2018,Gazagnes2020}.
In other words, such compact objects might have a ISM/CGM with very low \ion{H}{i} column density. This is consistent with the detection of double-peaked Ly$\alpha$ profile with narrow peak separation observed in strong LyC emitting galaxies \cite[e.g.,][]{Verhamme2017,Izotov18b, Izotov21}.
This is also consistent with the results of \cite{Kanekar2021} that report a low \ion{H}{i} 21 cm detection rate in local compact galaxies with the high (>10) O$_{32}$ ratios.
We discuss the relationship between the compactness and the LyC escape in the next section (Sect.~\ref{sec:53}).

\subsection{LyC escape and spatial compactness}
\label{sec:53}

\cite{Flury22b} found a scattered but significant correlation between the UV half-light radius measured on the COS acquisition images and \fesc, indicating that strong LyC emitters have compact stellar cores. In this work, we looked at the nebular gas distribution and similarly found that strong leakers have very compact gas distributions. The LCE detection fraction is indeed higher at small \ion{Mg}{ii} and [\ion{O}{ii}] spatial extents (Sect.~\ref{sec:4} and red points in Fig.~\ref{fig:fesc_rs}). This is in agreement with the recent work of \cite{Choustikov2024} where they found that galaxies with high \fesc\ tend to have less extended neutral gas (Ly$\alpha$) halos than non-leaking sources.
These results indicate that the CGM of strong leakers has a low \ion{H}{i} column density, which would allow the LyC photons that escape the ISM to go through the CGM without being absorbed.
These low column densities in the galaxy surroundings can be caused by stellar populations within galaxies ionizing most of the neutral gas in the CGM (see Sect.~\ref{sec:52}) and/or powerful outflows and feedback effects capable of clearing the galaxy surroundings \citep[but see \citealt{Jaskot2017}]{Chisholm2017,Trebitsch2017,Carr2021}. Indeed, \cite{amorin2024} recently found that the strongest LyC emitters show strong indication of ionized outflows and more complex kinematics. We note that we plan to analyze the resolved kinematics of our KCWI sources in future work (Leclercq et al. in prep.)
Such effects might be even more important at high redshift with the increase in ionization of galaxies \citep[e.g.,][]{Endsley2021,Rinaldi2023,Cameron2023}, and the presence of higher velocity \citep[$z\sim5-6$]{Sugahara2019} and more frequent spherical outflows \citep[$z\sim3-9$]{XuYi2023}. 
Moreover, in the nearby reionization-era analog galaxy Haro 11, \cite{LeReste2023} showed that merger-driven interactions can induce a significant shift of neutral gas by several kpc from the galaxy center, facilitating LyC escape in our line of sight.
The compactness of the nebular gas in strong LyC leakers, coupled with its link to highly ionized and density-bounded interstellar medium configurations (Sect.\ref{sec:52}), may elucidate the mechanism behind the escape of ionizing photons from these galaxies. 

The J1033+6353 galaxy however behaves differently compared to the other strong LyC emitters which might indicate a different leakage mechanism. It is indeed the only strong leaker of the sample with \fesc $\approx$30\% to show \ion{Mg}{ii} emission, i.e., neutral gas, beyond its stellar continuum. Although \ion{Mg}{ii} is only more extended than the continuum by a factor 1.5, this result suggests that the presence of neutral gas in the CGM is not incompatible with strong LyC leakage, which might thus escape anisotropically through ionized channels that go through both the ISM and CGM, as proposed in the literature \citep{Borthakur14, Trebitsch2017, Gazagnes2018, Wang19, Mauerhofer21, Gazagnes2020, Flury22b}. This scenario is reinforced by the detection of extended [\ion{O}{ii}] emission that traces the ionized gas. Moreover, \cite{Saldana-Lopez2022} measured a relatively low \ion{H}{i} covering fraction of $C_{\rm f}(\ion{H}{I})\approx$0.6 for J1033+6353 compared to the rest of the sample ($0.4<C_{\rm f}(\ion{H}{I})<1$) which favors a patchy configuration of the ISM/CGM. Interestingly, J1033+6353 has a lower O$_{32}$ ratio ($\approx$5) compared to the other strong leakers (>14). This suggests that LyC photons may escape J1033+6353 through ionized channels that cross through otherwise lower ionization gas  \citep[e.g., ][]{Flury22b}.

Finally, it is important to highlight that our study has revealed that weak and non LyC leakers show diverse \ion{Mg}{ii} and [\ion{O}{ii}] spatial extents spanning from compact to >3 times more extended than the stellar continuum (Sect.~\ref{sec:41}). The compact nebular configurations of weak and non LyC leakers indicates that the LyC photons are absorbed within the ISM. Higher spatial resolution data is needed to get better understanding of the physical state of the ISM/CGM and the inner propagation of the LyC photons.

\subsection{Using nebular gas extent to infer \fesc\ at high redshift}
\label{sec:54}

Recent JWST observations reveal that galaxies are compact at $z>6$ with effective radii $\lesssim$1 kpc \citep[e.g.,][]{Naidu2022-jwst,Yang2022,Ono2023,Robertson2023,Ormerod2024}. 
Moreover, \ion{Mg}{ii} has been detected at very high redshift \citep[e.g., GNz11][and Gazagnes in prep.]{Bunker2023}.
Although only few spatial measurements of the nebular gas at such high redshift exist \citep[e.g.,][]{Arribas2023}, we discuss the possibility of using the \ion{Mg}{ii} and [\ion{O}{ii}] compactness as an indicator of leakage in the EoR. Our study has revealed that the neutral and ionized gas distributions of most LyC emitting galaxies have compact configurations. However, we also found that weak LyC emitters or non leakers are diverse, with some also showing compact configurations. As a consequence, a compactness criterion is not enough to disentangle the population of strong and weak leakers. Fig.~\ref{fig:fesc_rs} shows the relation between the spatial extent of the gas and \fesc with data points color coded by the O$_{32}$ ratios. Combined with other diagnostics like strong O$_{32}$ ratio, high H$\beta$ equivalent width, or low metallicity, \ion{Mg}{ii} and [\ion{O}{ii}] spatial compactness (although [\ion{O}{ii}] is slightly less statistically reliable) appear like good indicators of LyC escape. As an example, we can expect high redshift \ion{Mg}{ii} and [\ion{O}{ii}] emitters unresolved in NIRSpec/IFU ($\lesssim$0.5 kpc at $z=7$ considering the spatial scale of 0\farcs1) and with high O$_{32}$ ratios to be leaking ionizing photons.

\section{Summary \& conclusions}
\label{sec:6}

Thanks to our IFU campaigns targeting stringently confirmed LyC emitters and non-emitters with KCWI and LRS2, we characterized the gas distributions of 22 z$\approx$0.35 galaxies and connected their spatial properties to the escape of ionizing photons. We used \ion{Mg}{ii} as a tracer of neutral gas and [\ion{O}{ii}] as a tracer of ionized gas. The spatial distributions were parameterized individually using a 2D exponential model. In order to confirm the trends observed for individual objects between the neutral and ionized gas distribution properties and the escape of ionizing photons, we performed stacking experiments. 
Our results can be summarized as follows:

\begin{enumerate} 
    \item The radial surface brightness profiles of the \ion{Mg}{ii}, [\ion{O}{ii}], and continuum emission are reasonably well-fit by an single component exponential model (Sect.~\ref{sec:33}). The average \ion{Mg}{ii} and [\ion{O}{ii}] exponential scale lengths of our sample (without accounting for upper limits) are 1.6 kpc and 1.4 kpc, respectively. Out of the 14 (13) objects with reliable \ion{Mg}{ii} ([\ion{O}{ii}]) scale length measurement, 7 (10) have a significant \ion{Mg}{ii} ([\ion{O}{ii}]) halo extending up to >10 kpc (Sect.~\ref{sec:34}). 
    \item On average, the \ion{Mg}{ii} and [\ion{O}{ii}] extended emission are at least $\sim$1.6 and $\sim$1.8 times more spread out than the continuum respectively (the median ratios are $\sim$1.5). Most of the objects with extended \ion{Mg}{ii} emission also show extended [\ion{O}{ii}] emission (all but one object); in this case, \ion{Mg}{ii} emission is always more extended than [\ion{O}{ii}] by a factor 1.3 on average. Five objects with extended [\ion{O}{ii}] emission are not extended in \ion{Mg}{ii}, suggesting that extended [\ion{O}{ii}] emission does not necessarily imply \ion{Mg}{ii} extended emission (Sect.~\ref{sec:34}).
    \item We observed a 2$\sigma$ correlation between the \ion{Mg}{ii} emission and continuum extents and a >3$\sigma$ correlation between the \ion{Mg}{ii} and [\ion{O}{ii}] scale lengths (Sect.~\ref{sec:34}).
    \item Our emission line to continuum spatial offset measurements range from no significant offset to 3.54 kpc and 2.16 kpc for \ion{Mg}{ii} and [\ion{O}{ii}], respectively. The average $\Delta_{\rm \ion{Mg}{ii}-cont}$, $\Delta_{\rm [\ion{O}{ii}]-cont}$ and $\Delta_{\rm \ion{Mg}{ii}-[\ion{O}{ii}]}$ values are 1.5, 0.5 and 1.0 kpc, respectively (Sect.~\ref{sec:35}).
    \item We find that the \ion{Mg}{ii} offset correlates with the galaxy size measurements (\ion{Mg}{ii}, [\ion{O}{ii}], and continuum scale) and NUV half-light radius. We do not observe such a correlation for [\ion{O}{ii}]. We also report no correlation between the \ion{Mg}{ii}/continuum, [\ion{O}{ii}]/continuum and \ion{Mg}{ii}/[\ion{O}{ii}] scale length ratios and their spatial offsets (Sect.~\ref{sec:35}).
    \item We found that the strong LyC leakers (\fesc > 30\%) are unresolved, and therefore compact ($rs$ $\lesssim$ 0.5 kpc), in both \ion{Mg}{ii} and [\ion{O}{ii}] emission, whereas the weaker or non leakers show a wider diversity with scale lengths ranging from upper limits (i.e., unresolved) to 3 kpc and 2 kpc for \ion{Mg}{ii} and [\ion{O}{ii}], respectively. The LCE detection fraction decreases with increasing \ion{Mg}{ii} and [\ion{O}{ii}] spatial extents (Sect.~\ref{sec:41}). The same results are obtained when we normalize the offsets by the $r_{50}^{UV}$.
    \item Strong leakers have non or small (< 1 kpc) spatial offset between \ion{Mg}{ii} and stellar continuum. The LCE detection fraction significantly decreases with increasing \ion{Mg}{ii}-continuum offset. We found a similar trend between \fesc\ and the spatial offset between \ion{Mg}{ii} and [\ion{O}{ii}] centroids (Sect.~\ref{sec:42}).
    \item Stacking experiments of our KCWI reinforce the trends observed for individual objects that the \ion{Mg}{ii} and [\ion{O}{ii}] emission are more extended in non/weak LyC leakers than in strong leakers. Independently of \fesc, \ion{Mg}{ii} is found to be more extended than [\ion{O}{ii}] on average (Sect.~\ref{sec:43}).
    \item Our \ion{Mg}{ii} and [\ion{O}{ii}] halos properties are comparable to previous studies. Our results reinforce the trend that the spatial extent of the neutral gas is independent of redshift and stellar mass (Sect.~\ref{sec:51}). In addition, we find significant anti-correlations between the spatial extent of the neutral gas and the O$_{32}$ ratio, and H$\beta$ equivalent width, as well as positive correlations with metallicity and UV size, suggesting that galaxies that have more compact neutral gas sizes are more highly ionized (Sect.~\ref{sec:52}). We find no correlations with $E$(B$-$V), UV $\beta^{1500}$ slope, stellar mass, SFR, SFR surface density, specific SFR, stellar age, EW(Ly$\alpha$), and Ly$\alpha$ escape fraction.
    \item The fact that the LCE detection fraction is higher at small \ion{Mg}{ii} scale lengths indicates that the CGM of strong leakers has a low \ion{H}{i} column density caused by stellar populations ionizing most of the surrounding neutral gas (Sect.~\ref{sec:53}). J1033+6353 is a strong leaker with a lower O$_{32}$ ratio ($\approx$5) than other strong leakers (>14) surrounded by a \ion{Mg}{ii} halo, suggesting a different leakage mechanism where LyC photons escape through ionized channels that go through both the ISM and the CGM (Sect.~\ref{sec:53}).
    \item Combined with high ionization diagnostics like strong O$_{32}$ ratio, high H$\beta$ equivalent width or low metallicity, \ion{Mg}{ii} and [\ion{O}{ii}] spatial compactness  appear like good indicators of LyC escape in the EoR (Sect.~\ref{sec:54}).
\end{enumerate}

Our IFU study of the spatial distribution of the neutral and ionized gas around a statistical and robust sample of 22 LyC emitting and non emitting galaxies has revealed a large diversity of gas configurations within the sample, with a trend for strong leakers to be compact. 
Our findings suggest that highly ionized galaxies are not surrounded by extended gaseous envelopes, implying that LyC photons can escape in many directions due to a widespread ionization and/or due to an ISM with numerous porous lines of sight distributed fairly isotropically.
We therefore propose to use the compactness of the neutral and ionized gas around galaxies as an additional LyC leakage indicator. 
A future paper will delve into the kinematics and resolved properties maps derived from our IFU data, complementing the spatial insights presented in this paper (Leclercq et al. in prep.).
Larger samples with deeper and higher resolution IFU observations will be needed to get a deeper understanding of the mechanisms governing ionizing photon escape in galaxies. The Lyman-alpha and Continuum Origins Survey (LaCOS) is a 119-orbit HST program (PID 17069) that will obtain spatially resolved images of 41 LzLCS+ galaxies. 
Simultaneously, we will obtain high resolution Ly$\alpha$ spectra for 15 LzLCS+ galaxies using HST/COS (PID 17153).
The Lyman $\alpha$ images and spectra will unveil the physical state of the ISM/CGM and its role in regulating LyC escape. 
Connecting the Ly$\alpha$ properties with the \ion{Mg}{ii} will help understand how \ion{Mg}{ii} can be used at the EoR (when Ly$\alpha$ is not observable) to constrain the LyC transmission in the JWST era.

\begin{acknowledgements}
This research is based on observations made with the NASA/ESA Hubble Space Telescope obtained from the Space Telescope Science Institute, which is operated by the Association of Universities for Research in Astronomy, Inc., under NASA contract NAS 5–26555. These observations are associated with program 15845. The Low Resolution Spectrograph 2 (LRS2) was developed and funded by the University of Texas at Austin McDonald Observatory and Department of Astronomy and by Pennsylvania State University. We thank the Leibniz-Institut für Astrophysik Potsdam (AIP) and the Institut für Astrophysik Göttingen (IAG) for their contributions to the construction of the integral field units. -- We would like to acknowledge that the HET is built on Indigenous land. Moreover, we would like to acknowledge and pay our respects to the Carrizo \& Comecrudo, Coahuiltecan, Caddo, Tonkawa, Comanche, Lipan Apache, Alabama-Coushatta, Kickapoo, Tigua Pueblo, and all the American Indian and Indigenous Peoples and communities who have been or have become a part of these lands and territories in Texas, here on Turtle Island. -- We acknowledge the Texas Advanced Computing Center (TACC) at The University of Texas at Austin for providing high performance computing, visualization, and storage resources that have contributed to the results reported within this paper. -- The authors wish to recognize and acknowledge the very significant cultural role and reverence that the summit of Maunakea has always had within the indigenous Hawaiian community.  We are most fortunate to have the opportunity to conduct observations from this mountain.
Y.I. acknowledges support from the National Academy of Sciences of Ukraine (Project No.0121U109612).
R.A. acknowledges support from ANID/Fondecyt 1202007.
O.B. is supported by the {\em AstroSignals} Sinergia Project funded by the Swiss National Science Foundation.
M.T. acknowledges support from the NWO grant 016.VIDI.189.162 (“ODIN”).
\end{acknowledgements}

%%%%%%%%%%%%%%%%%%%%%%%%%%%%%%%%%%%%%%%%%%%%%%%%%%%%%%%%%%%%%%%%%%%%%%%%%%%%%%%%%%%%%%%%%%%%%%%%%%%%%%%%%%%%%%%%%%%%%%%%%%%%%%%%%%%%%%%%%%%%%%%%%%%%%%%%%%%%%%%%

\bibliographystyle{aa} 
\bibliography{biblio}

\begin{appendix}

\onecolumn
\section{Point spread function characterization}
\label{ap:1}

Here we provide a detailed description of our procedure to characterize the point spread function (PSF) for both our LRS2 (\ref{ap:1a}) and KCWI (\ref{ap:1b}) observations. The resulting PSF parameters can be found in Table~\ref{tab:psf}.

\begin{table*}[h!]
\centering
\captionof{table}{Moffat PSF model parameters estimated at the observed \ion{Mg}{ii} and [\ion{O}{ii}] wavelength of the science object.}
\def\arraystretch{1.5}
\begin{tabular}{lcccccccc}
\hline
ID & $FWHM_{\rm X}^{\rm Mg~II}$ & $FWHM_{\rm Y}^{\rm Mg~II}$ & $\beta^{\rm Mg~II}$ & $\phi^{\rm Mg~II}$ & $FWHM_{\rm X}^{\rm [O~II]}$ & $FWHM_{\rm Y}^{\rm [O~II]}$ & $\beta^{\rm [O~II]}$ & $\phi^{\rm [O~II]}$ \\
   & [arcsec] & [arcsec] & & [deg] & [arcsec] & [arcsec] & & [deg] \\
\hline
\hline
J0047+0154  & 1.88 & --   & 5.0  & --   & 1.67 & -- & 4.6 & -- \\
J0130-0014  & 1.37 & 1.24 & 18.4 & 9.61 & 1.34 & 1.29 & >20 & 9.59 \\
J0141-0304  & 0.88 & 0.73 & 4.1  & 3.21 & 0.76 & 0.72 & 3.9 & 4.8 \\
J0804+4726  & 1.99 & --   & 4.7  & --   & 1.86 & -- & 3.9 & -- \\
J0811+4141  & 1.94 & --   & 5.0  & --   & 1.74 & -- & 4.6 & -- \\
J0834+4805  & 2.0  & --   & 4.9  & --   & 1.95 & -- & 4.7 & -- \\
J0844+5312  & 1.25 & 1.15 & 3.3  & 179  & 1.14 & 1.04 & 3.3 & 0.01 \\
J0912+5050  & 2.09 & --   & 6.7  & --   & 1.91 & -- & 5.8 & -- \\
J0919+4906  & 1.85 & --   & 3.6  & --   & 1.74 & -- & 3.5 & -- \\
J0940+5932  & 1.99 & --   & 4.0  & --   & 1.85 & -- & 4.1 & -- \\
J1014+5501  & 1.69 & 1.35 & 5.8  & 1.59 & 1.51 & 1.23 & 4.8 & 0.01 \\
J1033+6353  & 1.66 & --   & 4.9  & --   & 1.55 & -- & 4.8 & -- \\
J1046+5827  & 0.99 & 0.97 & 3.5  & 91.  & 0.91 & 0.96 & 3.9 & 91.13 \\
J1137+3605  & 0.99 & 0.99 & 3.6  & 91.1 & 0.9  & 0.96 & 3.9 & 91.13 \\
J1154+2443  & 0.94 & 0.88 & 4.8  & 89.5 & 0.94 & 0.92 & 5.5 & 89.52 \\
J1157+5801  & 1.02 & 0.99 & 3.4  & 3.19 & 0.92 & 1.0 & 3.0 & 3.19 \\
J1243+4646  & 1.02 & 0.89 & 4.4  & 3.19 & 1.04 & 0.93 & 5.0 & 4.79 \\
J1256+4509  & 1.05 & 1.01 & 4.7  & 89.5 & 0.95 & 0.92 & 5.4 & 89.53 \\
J1352+5617  & 0.99 & 0.98 & 3.7  & 91.1 & 0.91 & 0.96 & 3.9 & 91.13 \\
J1503+3644  & 0.89 & 1.01 & 3.9  & 179 & --   & -- & -- & -- \\
J1517+3705  & 1.51 & --   & 3.7  & --   & 1.43 & -- & 4.4 & -- \\
J1648+4957  & 1.66 & --   & 4.5  & --   & 1.52 & -- & 4.4 & -- \\
\end{tabular}
\smallskip\\
\small{\textbf{Notes.} For the LRS2 observations, the PSF is close to circular and therefore only one value of FWHM is given ($FWHM_{\rm X}$). For KCWI observations, we provide the two spatial dimensions $FWHM_{\rm X}$ and $FWHM_{\rm Y}$ as well as the rotation angle $\phi$. $\beta$ is the Moffat shape parameter.  At the redshift of our sample ($z\approx0.35$) the \ion{Mg}{ii} and [\ion{O}{ii}] lines fall at $\approx$3785 $\AA$ and $\approx$5040 $\AA$, respectively.}
\label{tab:psf}
\end{table*}

\subsection{KCWI PSF}
\label{ap:1a}

To characterize the PSF of the KCWI instrument, we fit standard stars that were observed immediately before or/and after the science observations. Given that one of our primary scientific objectives involves the detection of extended emissions, we devoted particular attention to understanding the wings of the PSF. 

\begin{figure*}[h!]
\centering
   \resizebox{\hsize}{!}{\includegraphics{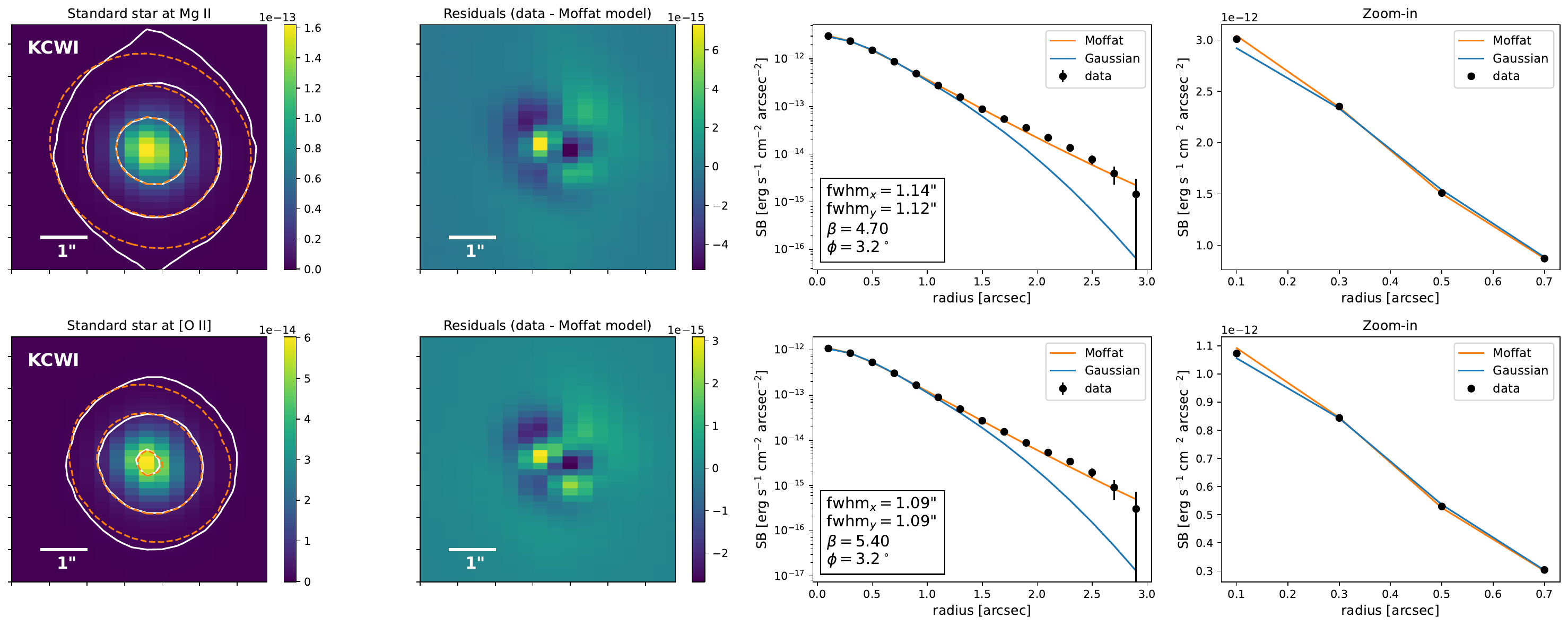}}
    \caption{Characterization of the KCWI point spread function. Example for the standard star (Feige66) where the first and second rows show the size measurement at \ion{Mg}{ii} ($\approx$ 3790 \AA) and [\ion{O}{ii}] ($\approx$ 5045 \AA) wavelengths of the corresponding science object observed (J1256+4509), respectively. \textit{First panel}: narrow-band image at the wavelengths of interest ($\approx$ 1000 \kms\ wide or 13 \AA\ at $z=0.35$). The white and orange contours corresponds to SB levels of $10^{-14}, 10^{-13}$, and $10^{-12}$ \sbl\ for the data and Moffat best-fit model, respectively. \textit{Second panel}: Fit residuals of the Moffat best-fit model. \textit{Third panel}: Radial SB profiles of the standard star data (black), Gaussian best-fit model (blue) and Moffat best-fit model (orange). The best-fit model parameters are shown in the bottom left. \textit{Forth panel}: Zoom-in on the inner profiles. A Moffat model better describes both the core and wings of the PSF for our KCWI observations.}
    \label{fig:psf_kcwi}
\end{figure*}

The comparison between the Gaussian and Moffat best-fit models, as illustrated in Fig.~\ref{fig:psf_kcwi} indeed reveals that a Gaussian PSF model does not capture the PSF wings. Consequently, utilizing a Gaussian model would artificially lead to the detection of extended emissions. We therefore adopted a Moffat PSF model and fit the standard star observations at the wavelengths pertinent to this study, i.e., \ion{Mg}{ii} ($\approx$ 3790 \AA) and [O~II] ($\approx$ 5045\AA) and within a window of $\approx$ 1000 \kms\ (or 13 \AA\ at $z=0.35$).
We note that some objects of our sample have two standard stars observations (before and after the science observations). In this case, we consider the average of the two sets of Moffat parameters set as PSF parameter values.
The resulting elliptical Moffat model parameters are compiled in table~\ref{tab:psf}.

\subsection{LRS2 PSF}
\label{ap:1b}

\begin{figure*}[h!]
\centering
   \resizebox{\hsize}{!}{\includegraphics{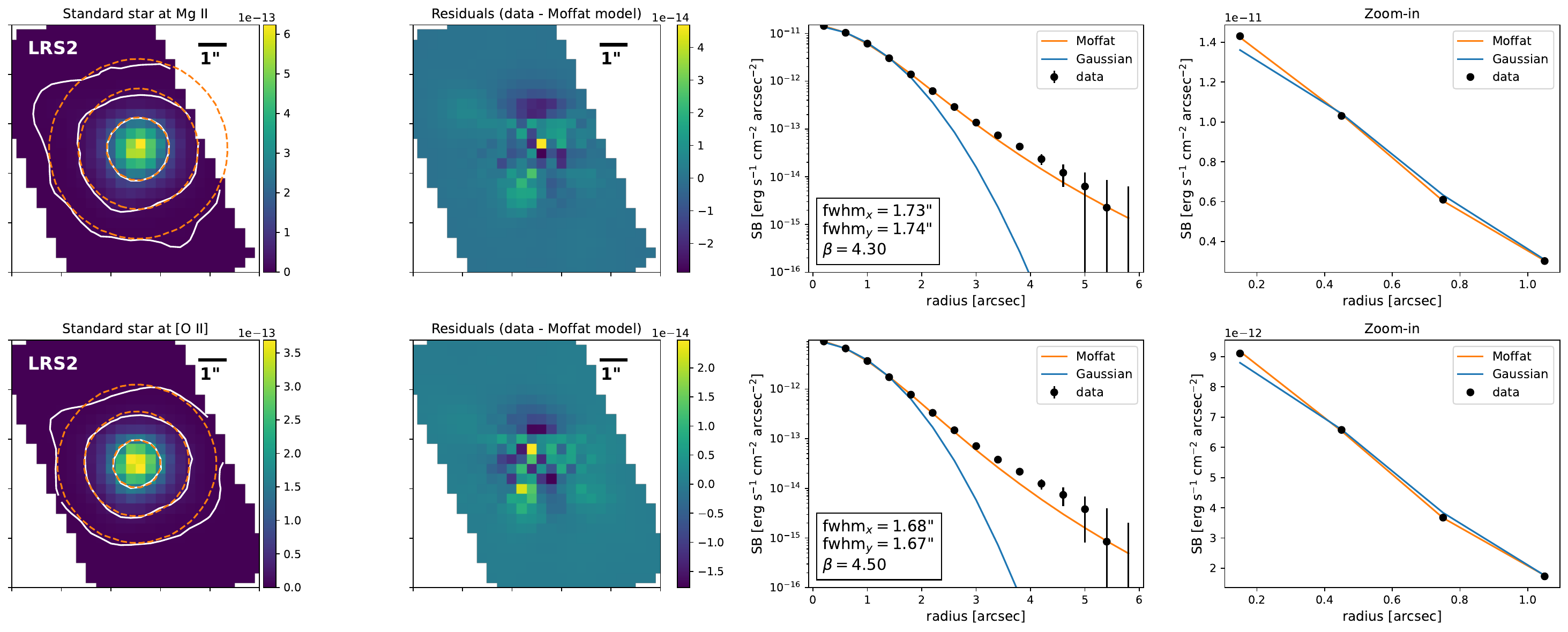}}
    \caption{Same as Fig.~\ref{ap:1a} but for the LRS2 PSF characterization. We show the size measurements of the standard star BD+210607 observed at the end of the night of the J0804+4726 observations. We do not include a rotation measurement because most of the standards stars can be considered as circular (see Fig.~\ref{fig:psf_lrs2_dimm}). The size of the stars as observed with LRS2 and therefore the LRS2 PSF is better described with a Moffat model.}
    \label{fig:psf_lrs2}
\end{figure*}

\begin{figure*}[h!]
\centering
   \resizebox{\hsize}{!}{\includegraphics{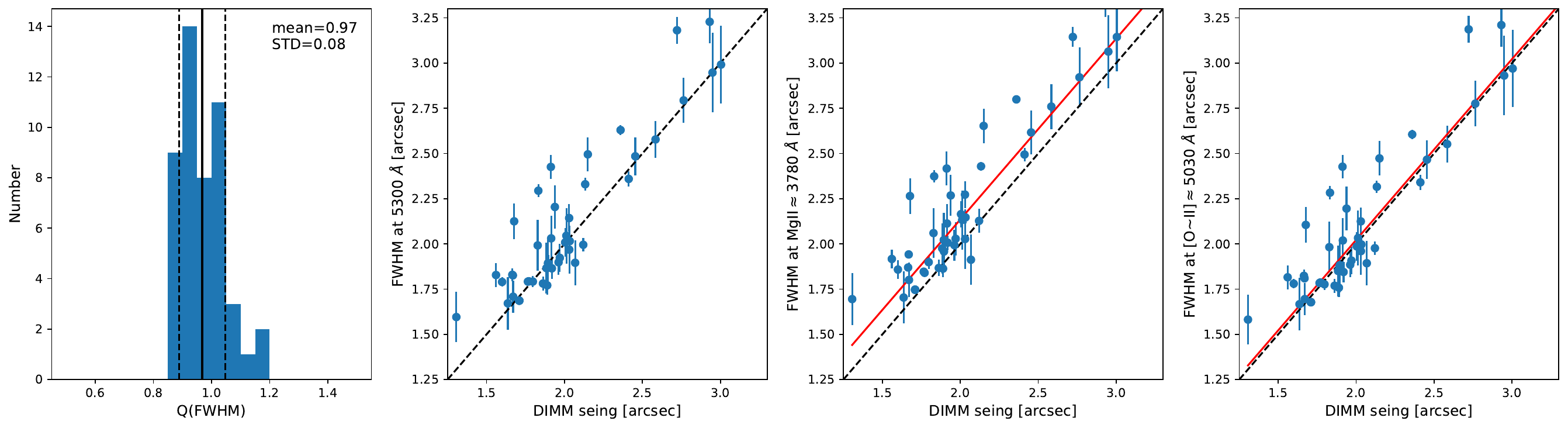}}
    \caption{\textit{Left panel}: Distribution of the FWHM PSF ratio, Q(FWHM) = $FWHM_{\rm X}$/$FWHM_{\rm Y}$, measured on the standard stars observed with LRS2. The PSF is on average consistent with being circular with a mean FWHM ratio of 0.97±0.08. \textit{Second left to right}: comparison between the DIMM seeing at the telescope and the FWHM (mean between the x- and y-axis of the elliptical Moffat function, see Fig.~\ref{ap:1b}) measured on the standard stars (Sect.~\ref{ap:1b}) at the middle of the LRS2-B wavelength window ($\sim$5300 \AA), at the \ion{Mg}{ii} and at the [\ion{O}{ii}] wavelengths, respectively. The dashed line shows the one-to-one relation. The DIMM seeing provides a relatively good estimation of the PSF FWHM near $\sim$ 5300 \AA\ but overestimates it at \ion{Mg}{ii} wavelengths ($\approx$ 3780 \AA). The red line shows the estimated PSF FWHM based on the DIMM seeing calculated using Eq.~\ref{eq:lrs2_psf} (see Sect.~\ref{ap:1b}).} 
    \label{fig:psf_lrs2_dimm}
\end{figure*}

We applied the same fitting process as for the KCWI standard stars (Sect.~\ref{ap:1a}). We selected standard stars observed on the same night as the science observations and fit them with an elliptical Moffat model. This fitting was performed within a $\approx$ 1000 \kms\ spectral window around both the \ion{Mg}{ii} and [\ion{O}{ii}] wavelengths of the science object, which correspond approximately to 3780 \AA\ and 5030 \AA, respectively. Similarly to the KCWI standard stars observations, the core and wings of the LRS2 PSF are better described by a Moffat distribution than a Gaussian model (see Fig.~\ref{fig:psf_lrs2} for the example of BD+210607). Moreover, our measurements show that the LRS2 PSF is consistent with being circular with a mean axis ratio of $0.97\pm0.08$ (Fig.~\ref{fig:psf_lrs2_dimm}, first panel).
Characterizing the PSF of our LRS2 observations is however more challenging than for the KCWI observations. This is because standard stars are not always observed right after our main observations, but at the end of the night when the seeing conditions are not as good as during our main observations. As a result, our PSF size estimates would end up being larger than what it actually is, which would hamper a precise spatial analysis.

\begin{figure*}[h!]
\centering
   \resizebox{\hsize}{!}{\includegraphics{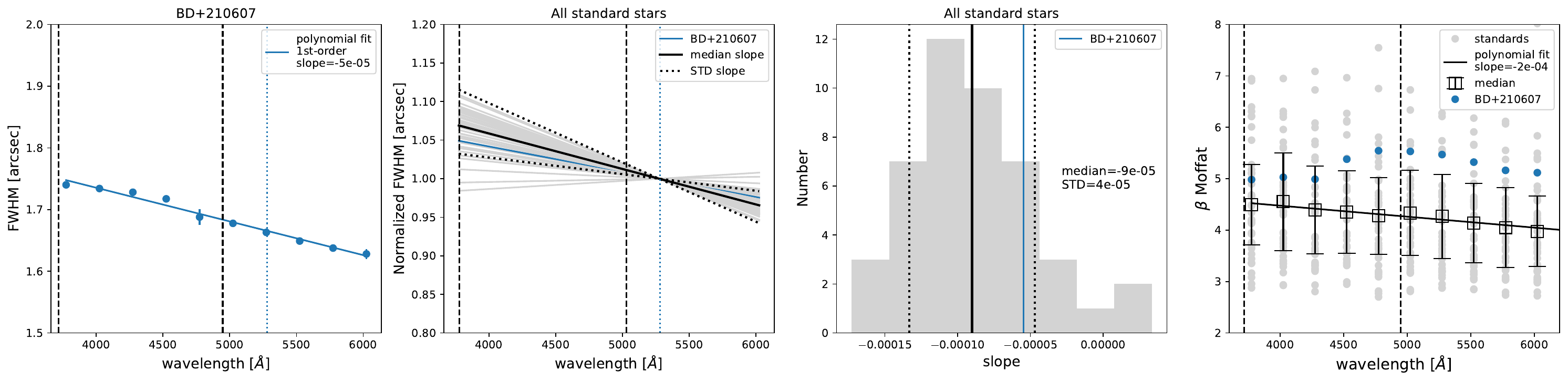}}
    \caption{\textit{First panel}: Evolution of the PSF FWHM with wavelength for the standard star BD+210607 observed with LRS2. The vertical dashed line indicate the \ion{Mg}{ii} and [\ion{O}{ii}] wavelengths of the science object J080425+472607. The vertical dotted blue line corresponds to the central wavelength of LRS2B ($\approx$ 5300 \AA). The solid blue line shows the best first-order polynomial fit to the data. The resulting slope is $-5 \times 10^{-5}$ which is consistent with the average value obtained from all the standard stars (middle and right panels). \textit{Second panel}: Best first-order polynomial fits of the PSF FWHM - wavelength relation of all the standard stars (grey). The relation resulting from the median slope and the standard deviation is shown with solid and dotted black line, respectively. \textit{Third panel}: Distribution of the slope measured on the standard stars. The solid and dotted vertical lines indicate the median and standard deviation of the distribution (values on middle right), respectively. \textit{Fourth panel}: Evolution of the Moffat $\beta$ parameter with wavelength for all the standard stars (grey). The empty squares and corresponding errorbars give the median value and median absolute deviation in each wavelength bin. The blue dots show the BD+210607 values.}
    \label{fig:psf_lrs2_all}
\end{figure*}

In order to best characterize the PSF of our observations and its evolution with wavelength, we adopt the following approach. The only information that we have about the seeing conditions during the science observations is the DIMM seeing at the telescope which is reported in the observation log. We show in Fig.~\ref{fig:psf_lrs2_dimm} (second to fourth panels) that the DIMM seeing size reported during the standard stars observations provides a reasonably good estimation of the PSF FWHM near the central wavelength of LRS2 ($\approx$ 5300 \AA), but underestimates the PSF size at \ion{Mg}{ii} wavelengths ($\approx$ 3780 \AA). This discrepancy arises because the PSF shape evolves as a function of wavelength. We measured this evolution for each standard star observation by dividing the datacubes into 14 wavelength slices of 250 \AA. The first and second panels of Figure~\ref{fig:psf_lrs2_all} shows the PSF FWHM as a function of wavelength for the standard star BD+210607 (blue) and all the standard stars observed on the same nights of our science observations, respectively. As expected, we observe a decrease in FWHM with increasing wavelengths. The median slope of this decrease is $-9\pm4\times10^5$ (third panel). 
The PSF size at the wavelength of interest can then be estimated using this slope and the DIMM seeing of the observations -- under the assumption that the DIMM seeing is a good approximation of the PSF FWHM at 5300 \AA\ which is the case for our data (see second panel of Fig.~\ref{fig:psf_lrs2_dimm}) -- using this equation:
\begin{equation}
FWHM_{\text{PSF,LRS2}}(\lambda) = -9 \times 10^{-5} \left(\lambda - 5300~\AA\right) + \text{DIMMseeing,}
\label{eq:lrs2_psf}
\end{equation}
where the wavelength $\lambda$ is in units of \AA\ and the FWHM and DIMMseeing in arcsec. This relation applied at the observed \ion{Mg}{ii} and [\ion{O}{ii}] wavelengths ($\approx$3780 \AA) is showed in red in the third and fourth panels of Fig.~\ref{fig:psf_lrs2_dimm}. The estimated PSF based on the DIMM seeing (Eq.~\ref{eq:lrs2_psf}) is now closer to the observed values. We performed the same exercice for the Moffat shape parameter $\beta$. The fourth panel of Fig~\ref{fig:psf_lrs2_all} shows the individual standard star measurements along with the overall trend of $\beta$ evolution with wavelength. We highlighted the measurements of BD+210607 as example (blue). While doing this analysis, we found that the individual FWHM($\lambda$) slope and beta values measured for each standard star can be quite different from the average values across all the standard stars. We therefore decided to use the individual slope and $\beta$ measurements, rather than the average values. These individual measurements are reported in Table~\ref{tab:psf}. As for KCWI, we note that some objects of our sample have several standard stars observations. In this case, we consider the average of the different sets of Moffat parameters set as PSF parameter values.

\newpage
\section{\ion{Mg}{ii} and [\ion{O}{ii}] maps and SB profiles}
\label{ap:2}

\begin{figure*}[h!]
    \centering
    \begin{minipage}{0.87\textwidth}
        \begin{subfigure}[t]{\linewidth}
            \vspace{0.1cm}
            \includegraphics[width=\textwidth]{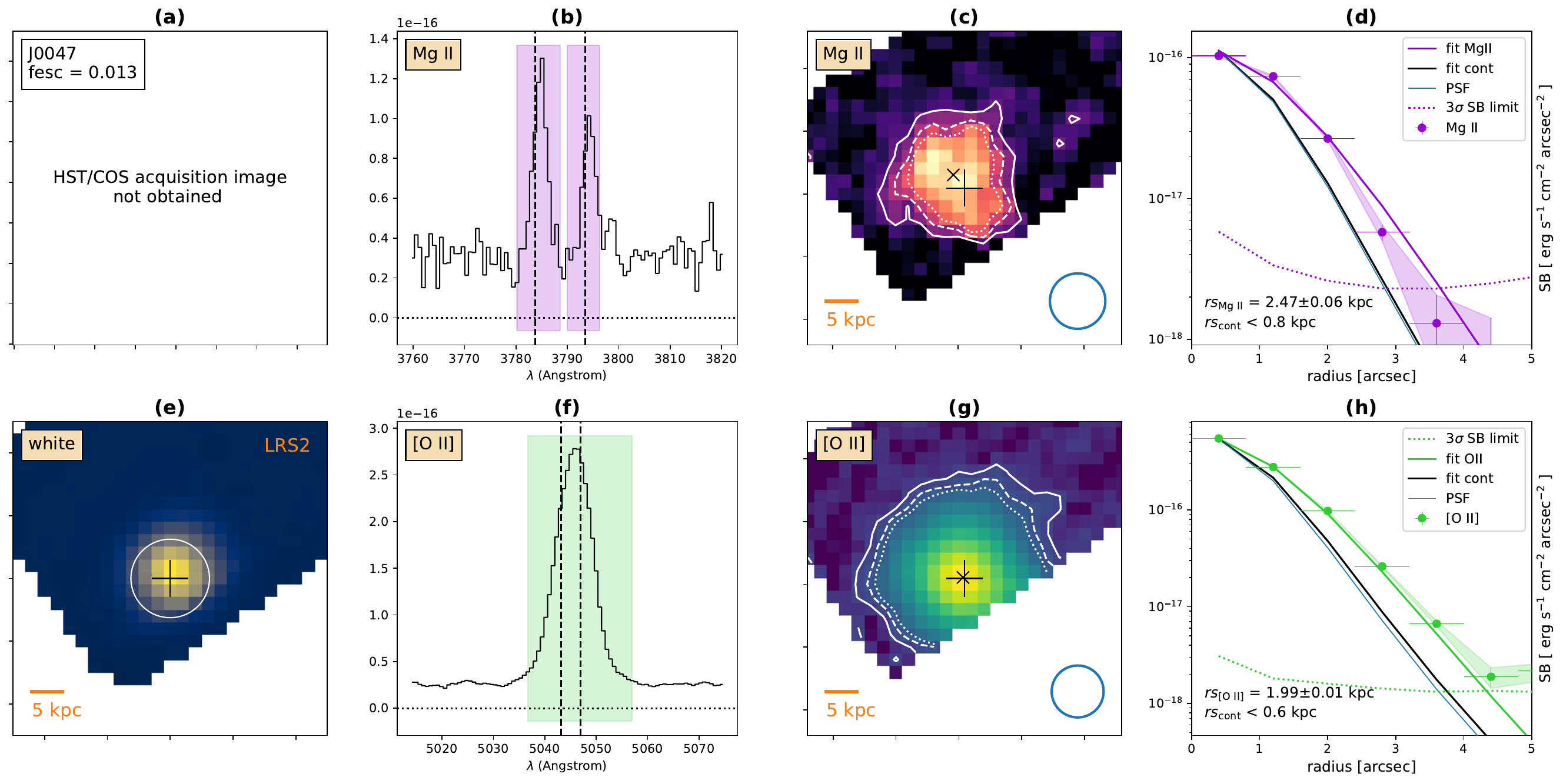}
        \end{subfigure}    
    \end{minipage}
    \caption{Same as Fig.~\ref{fig:profiles} (Part 1/8)}
    \label{fig:profiles1}
\end{figure*}

\begin{figure*}[p]
    \centering
    \begin{minipage}{0.87\textwidth}
        \begin{subfigure}[b]{\linewidth}            
            \includegraphics[width=\textwidth]{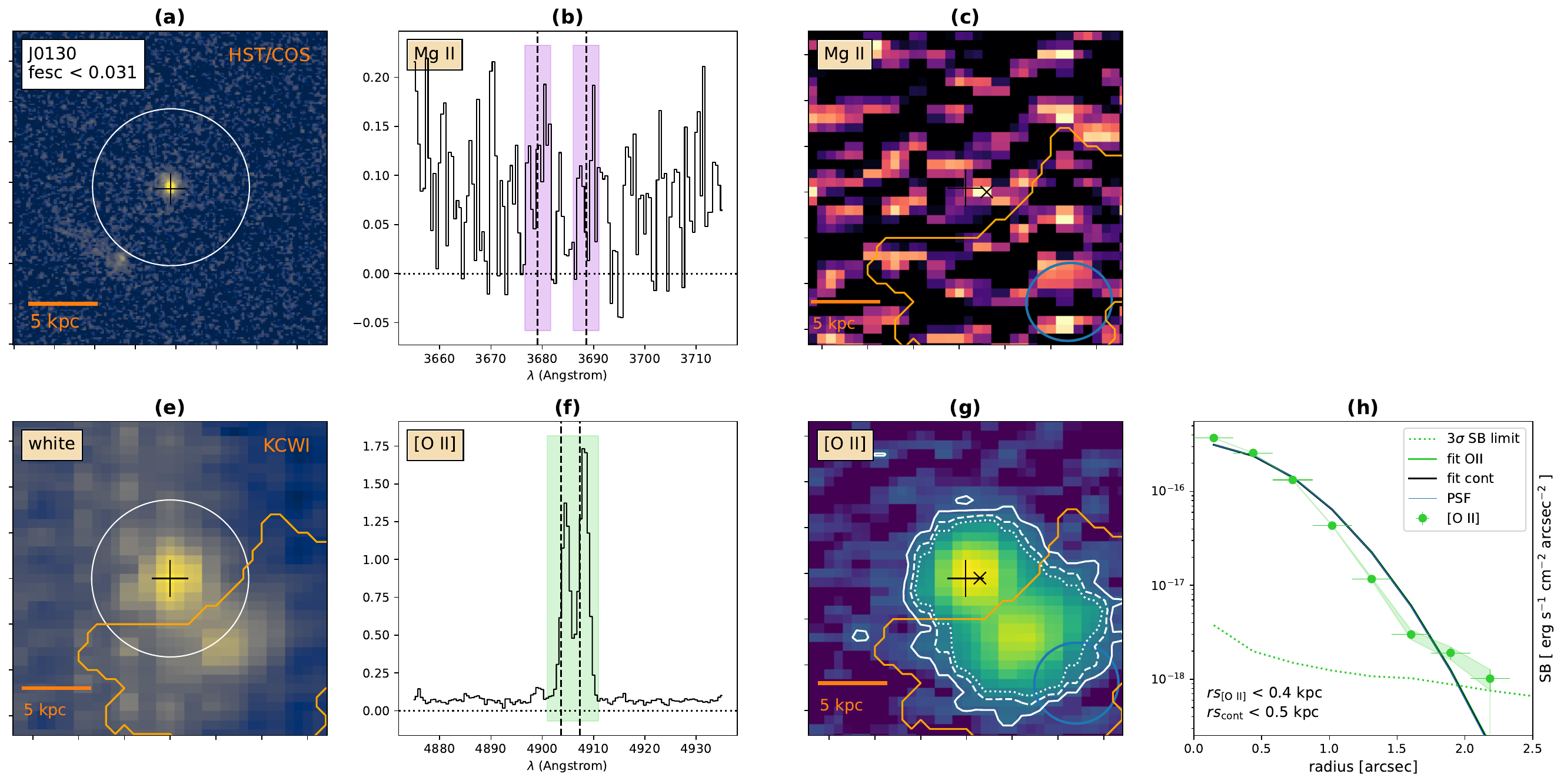}            
        \end{subfigure} 
        \begin{subfigure}[b]{\linewidth}            
            \includegraphics[width=\textwidth]{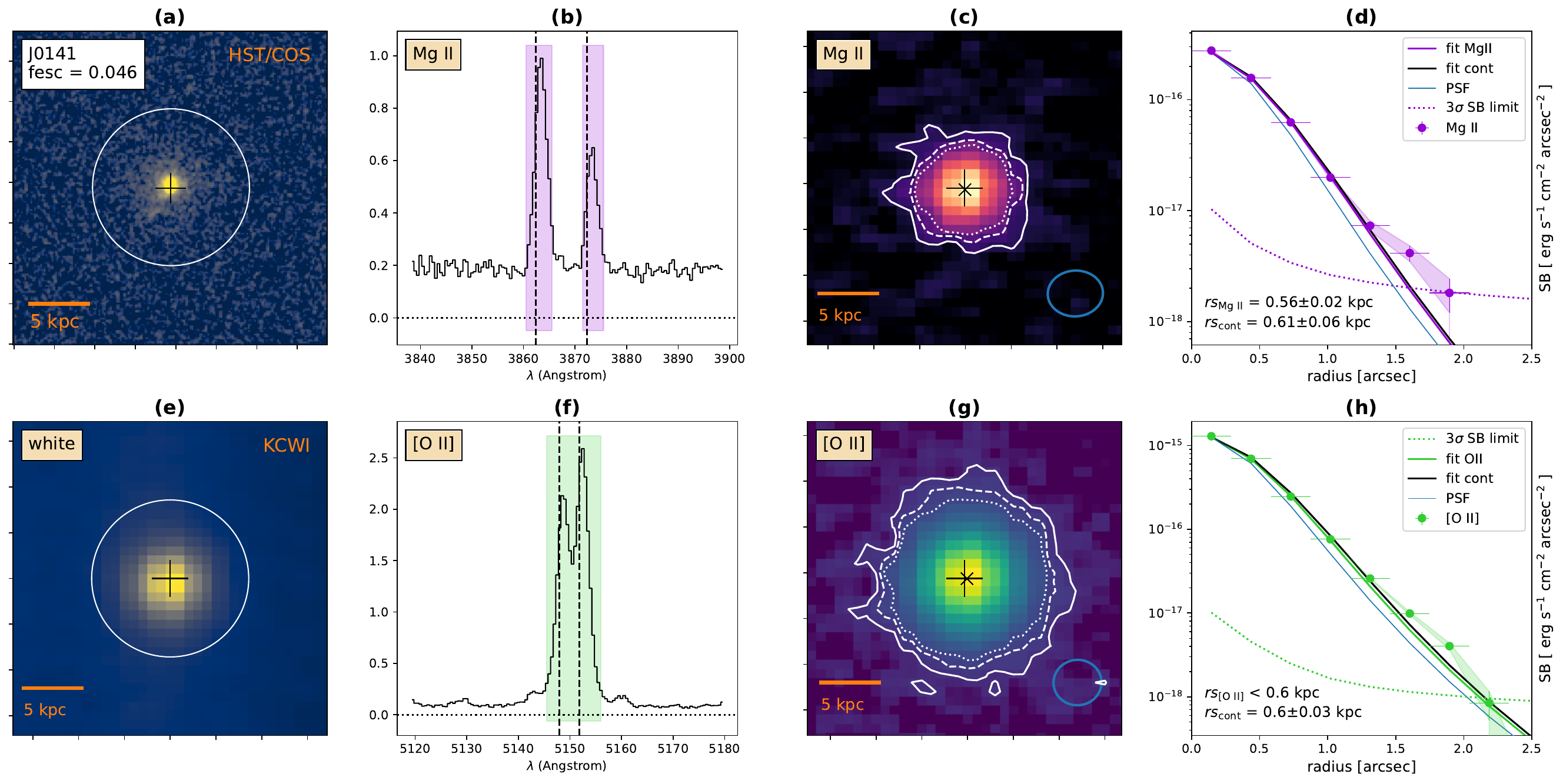}
        \end{subfigure} 
    
        \begin{subfigure}[t]{\linewidth}
            \vspace{0.1cm}
            \includegraphics[width=\textwidth]{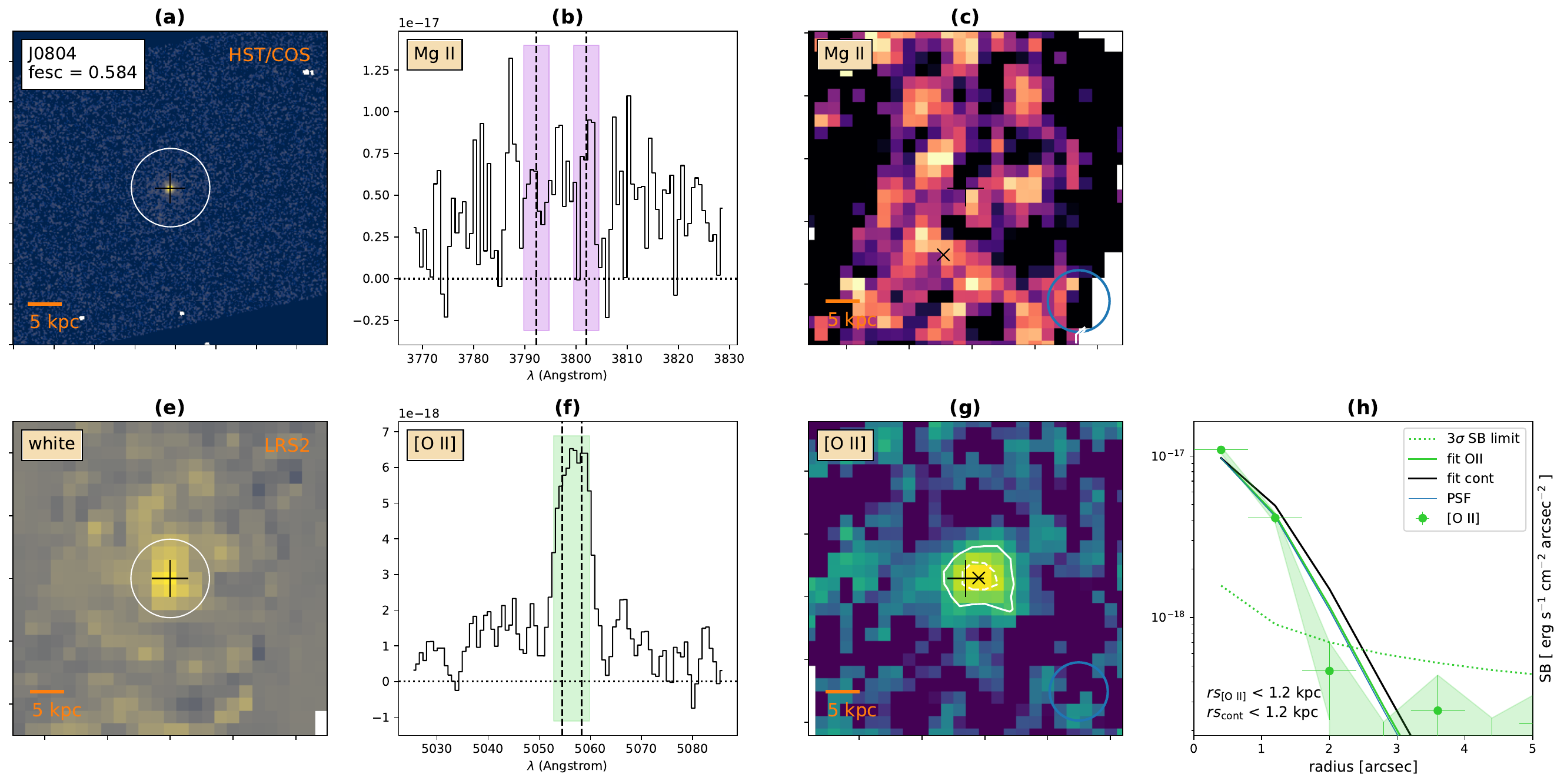}
        \end{subfigure}

    \end{minipage}
    \caption{Same as Fig.~\ref{fig:profiles} (Part 2/8)}
    \label{fig:profiles2}
\end{figure*}

\begin{figure*}[p]
    \centering
    \begin{minipage}{0.87\textwidth}     
        \begin{subfigure}[b]{\linewidth}
            \includegraphics[width=\textwidth]{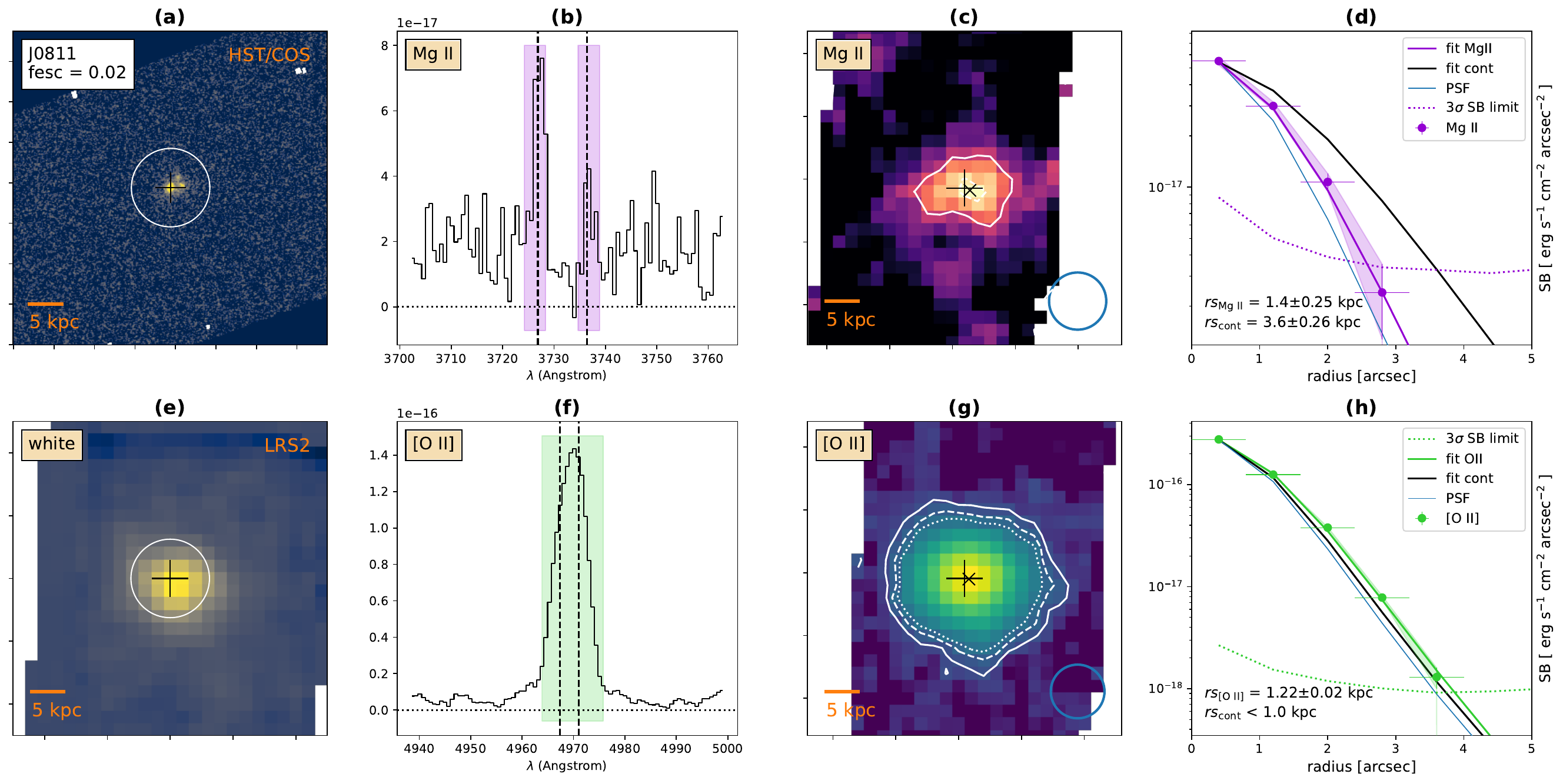}
        \end{subfigure} 

        \begin{subfigure}[b]{\linewidth}
            \includegraphics[width=\textwidth]{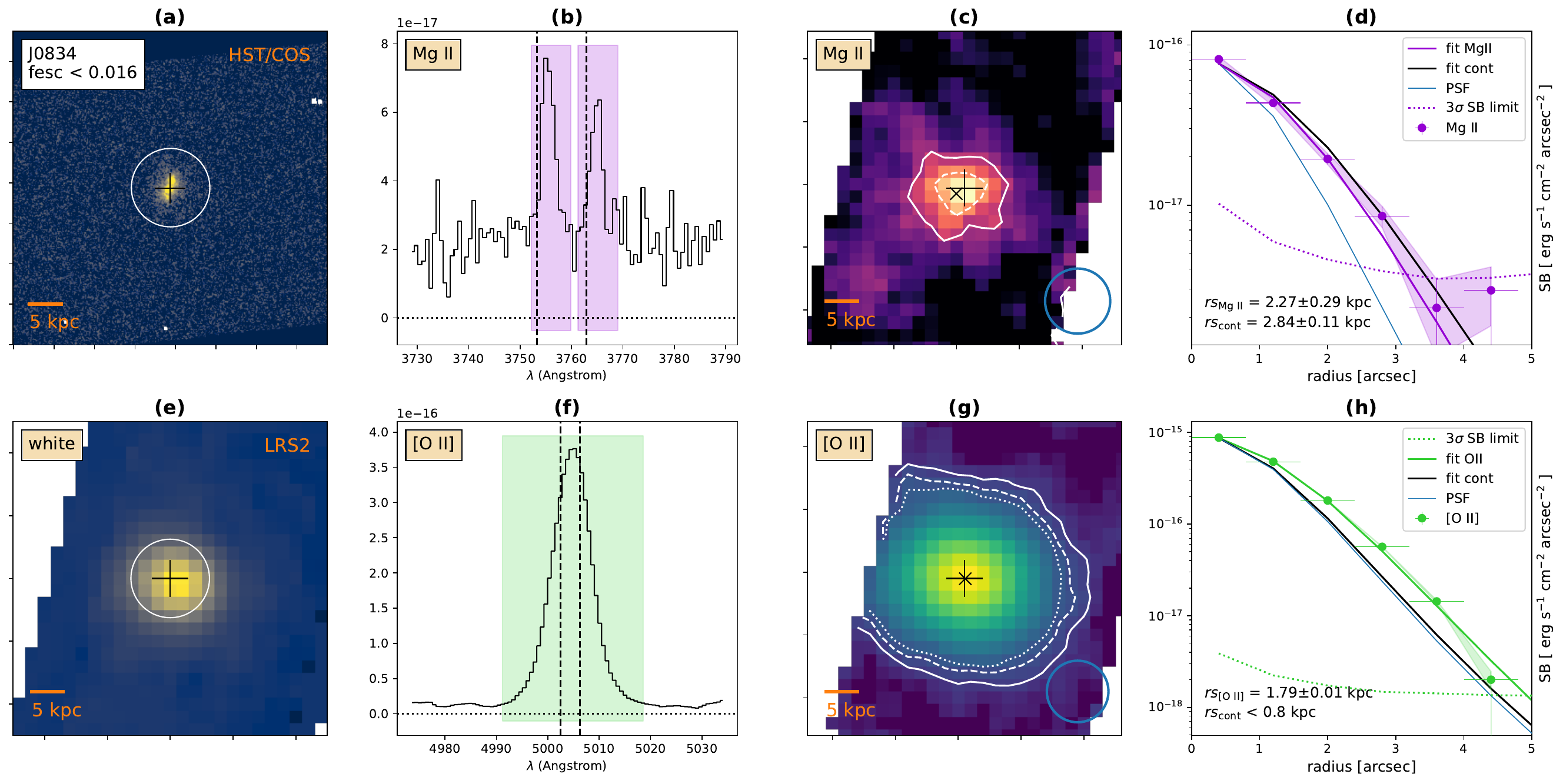}
        \end{subfigure}

        \begin{subfigure}[t]{\linewidth}
            \vspace{0.1cm}
            \includegraphics[width=\textwidth]{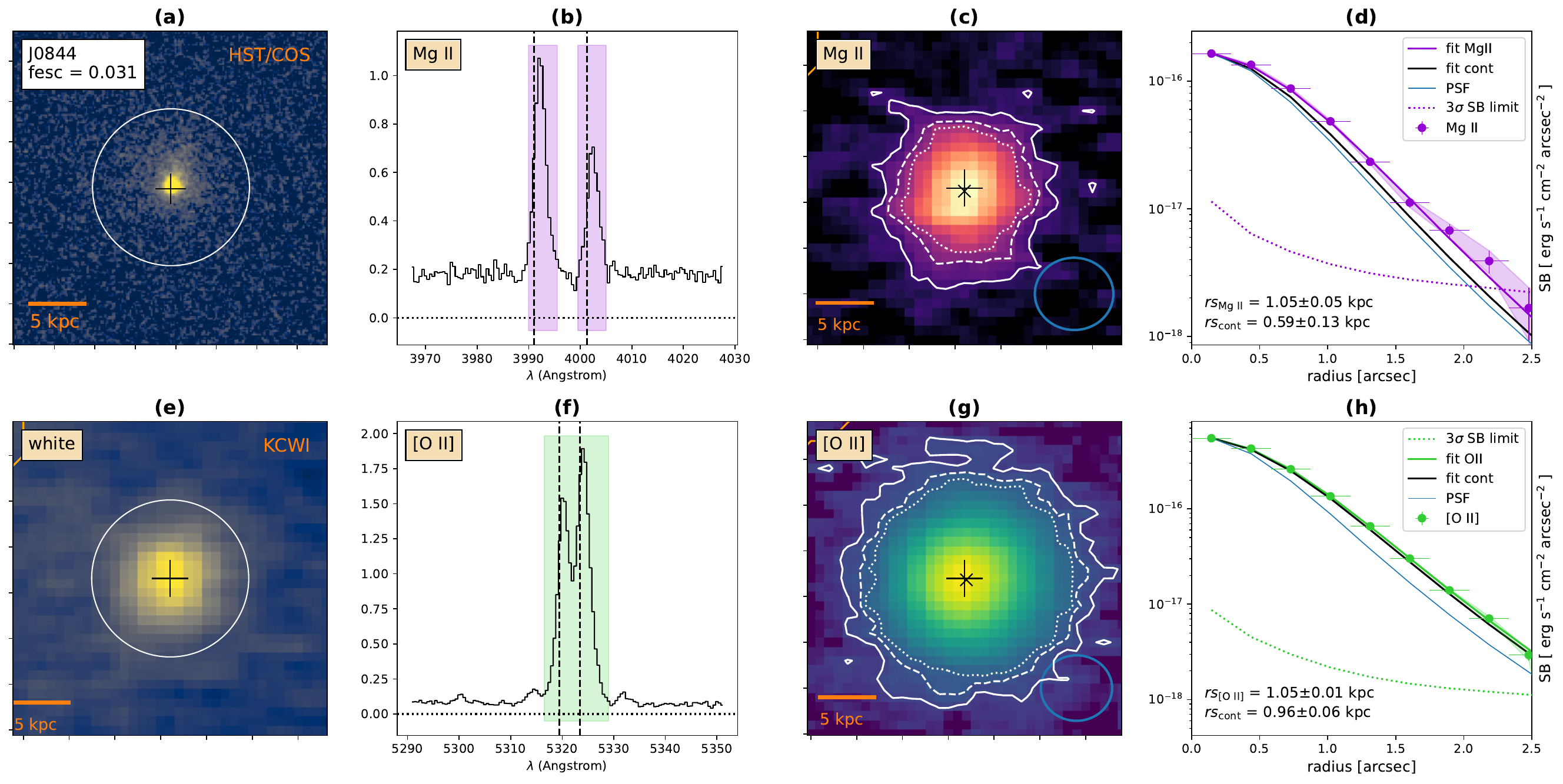}
        \end{subfigure}
        
    \end{minipage}
    \caption{Same as Fig.~\ref{fig:profiles} (Part 3/8)}
    \label{fig:profiles3}
\end{figure*}

\begin{figure*}[p]
    \centering
    \begin{minipage}{0.87\textwidth}

         \begin{subfigure}[b]{\linewidth}
            \includegraphics[width=\textwidth]{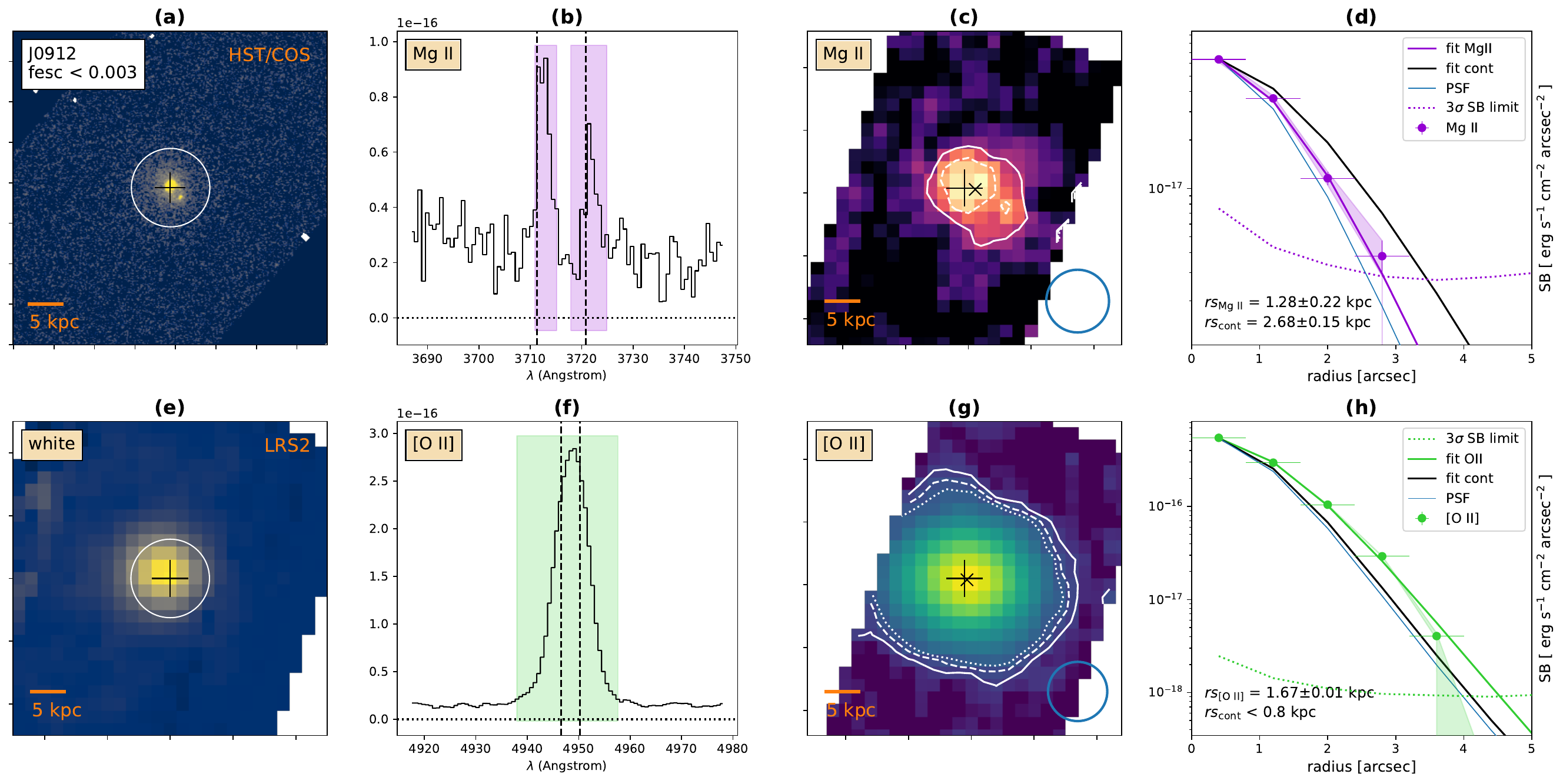}
        \end{subfigure} 

        \begin{subfigure}[b]{\linewidth}
            \includegraphics[width=\textwidth]{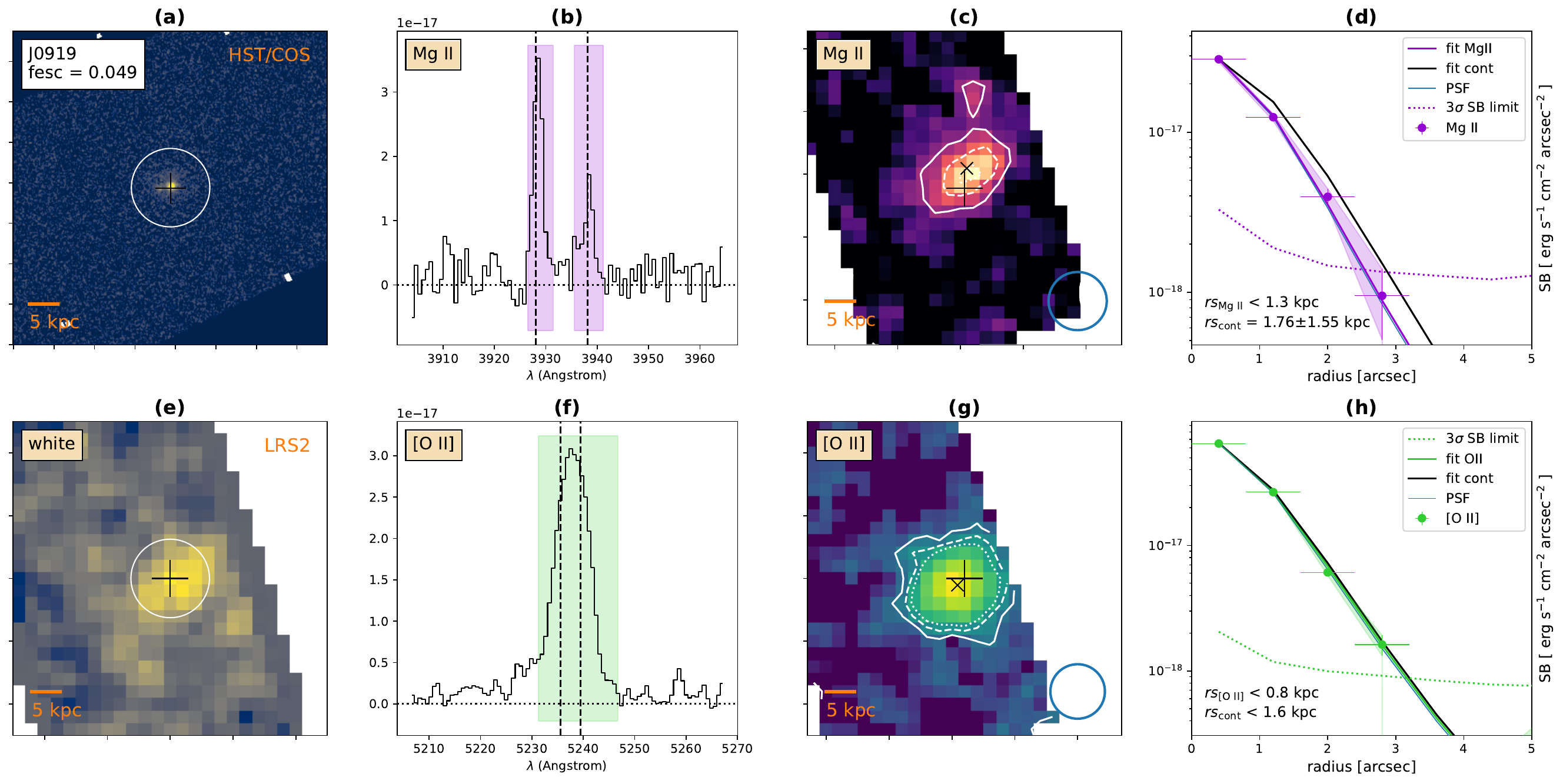}
        \end{subfigure} 
    
        \begin{subfigure}[t]{\linewidth}
            \vspace{0.1cm}
            \includegraphics[width=\textwidth]{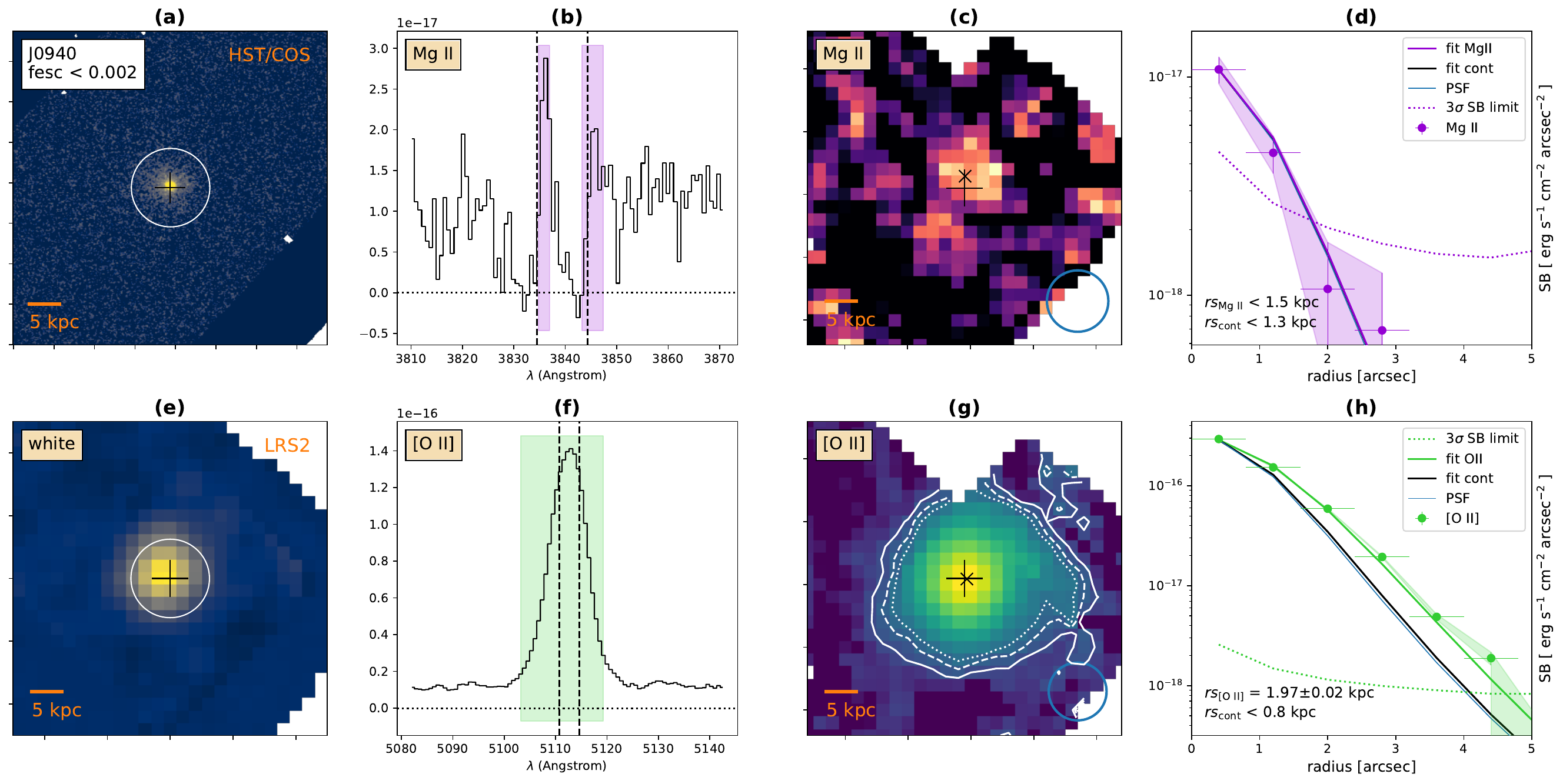}
        \end{subfigure}
        
    \end{minipage}
    \caption{Same as Fig.~\ref{fig:profiles} (Part 4/8)}
    \label{fig:profiles4}
\end{figure*}

\begin{figure*}[p]
    \centering
    \begin{minipage}{0.87\textwidth}

        \begin{subfigure}[b]{\linewidth}
            \includegraphics[width=\textwidth]{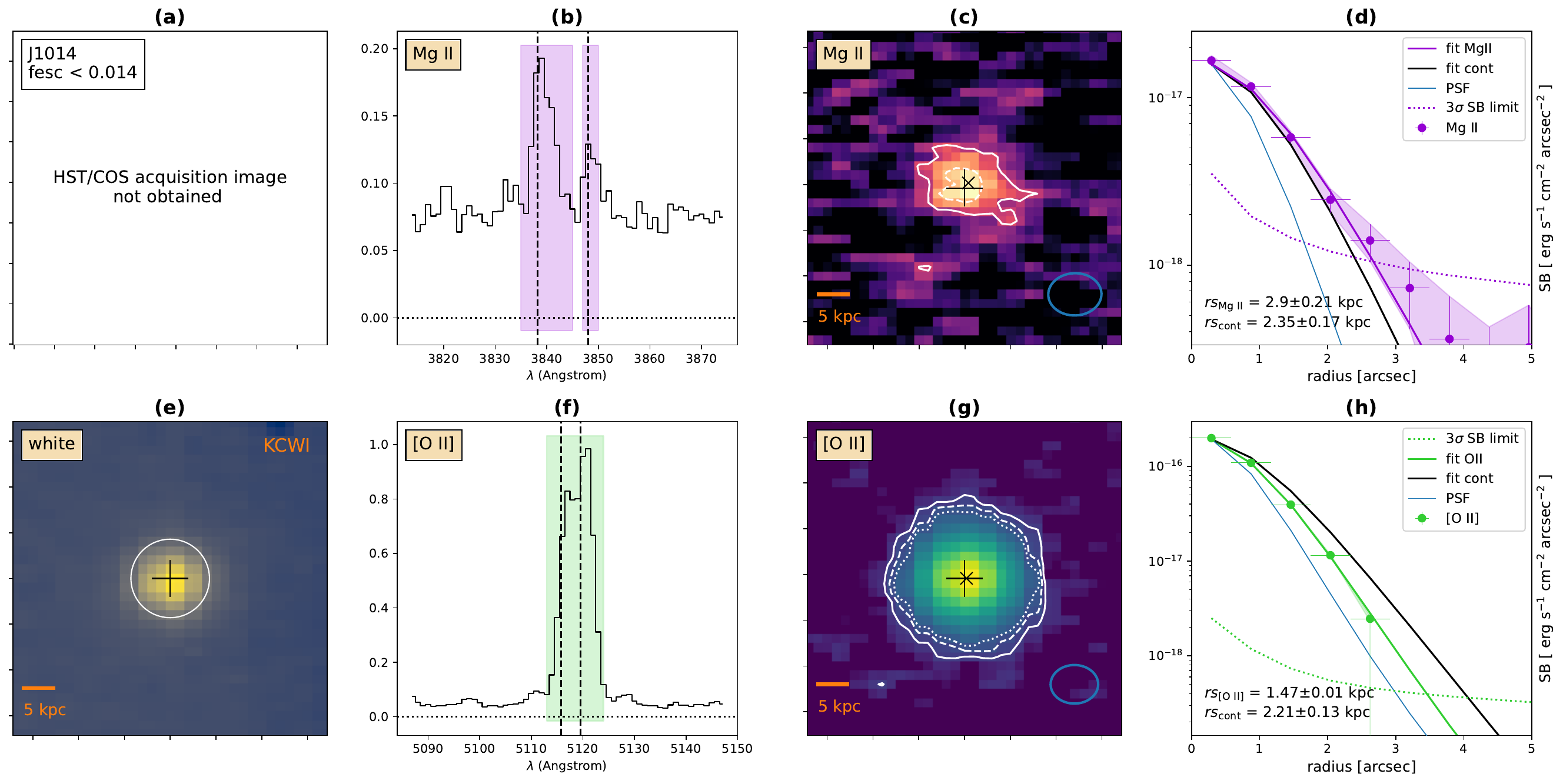}
        \end{subfigure} 

        \begin{subfigure}[b]{\linewidth}
            \includegraphics[width=\textwidth]{plot_w_lines-J1033_sca_FINAL_int.pdf}
        \end{subfigure} 
        
        \begin{subfigure}[t]{\linewidth}
            \vspace{0.1cm}
            \includegraphics[width=\textwidth]{plot_w_lines-J1046_sca_FINAL_int.pdf}
        \end{subfigure}
        
    \end{minipage}
    \caption{Same as Fig.~\ref{fig:profiles} (Part 5/8)}
    \label{fig:profiles5}
\end{figure*}

\begin{figure*}[p]
    \centering
    \begin{minipage}{0.87\textwidth}

        \begin{subfigure}[b]{\linewidth}
            \includegraphics[width=\textwidth]{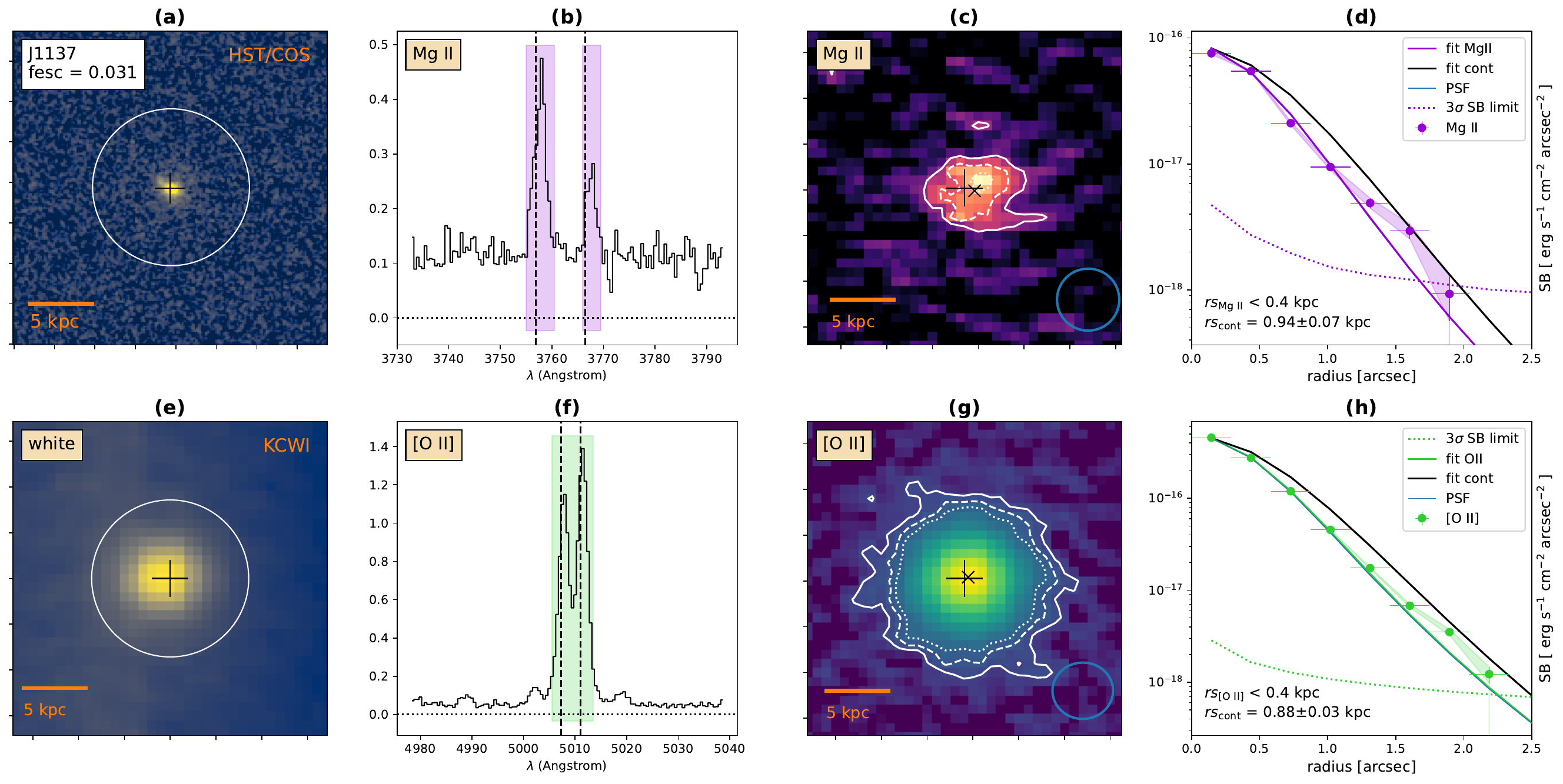}
        \end{subfigure} 

        \begin{subfigure}[b]{\linewidth}
            \includegraphics[width=\textwidth]{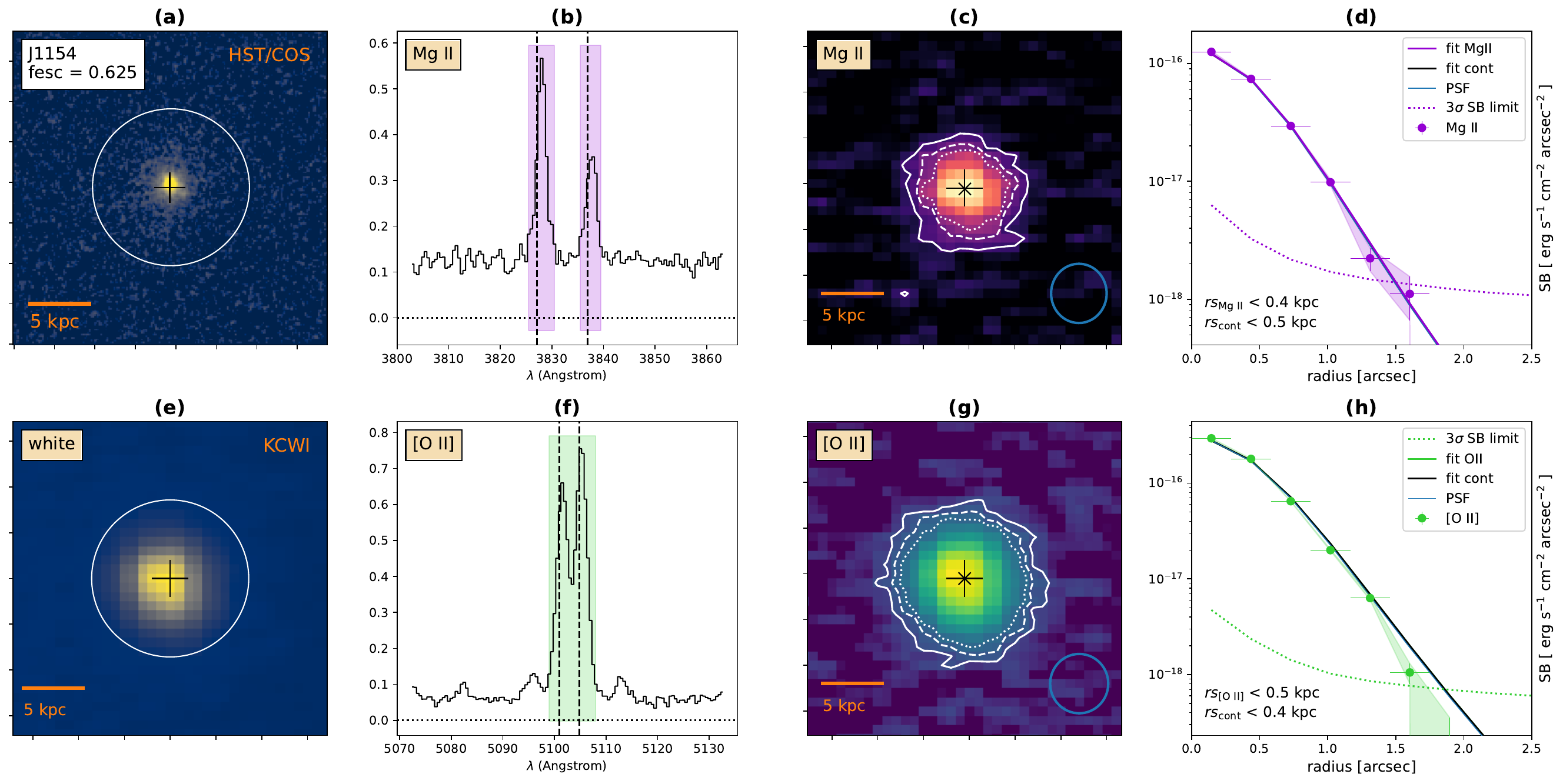}
        \end{subfigure} 
    
        \begin{subfigure}[t]{\linewidth}
            \vspace{0.1cm}
            \includegraphics[width=\textwidth]{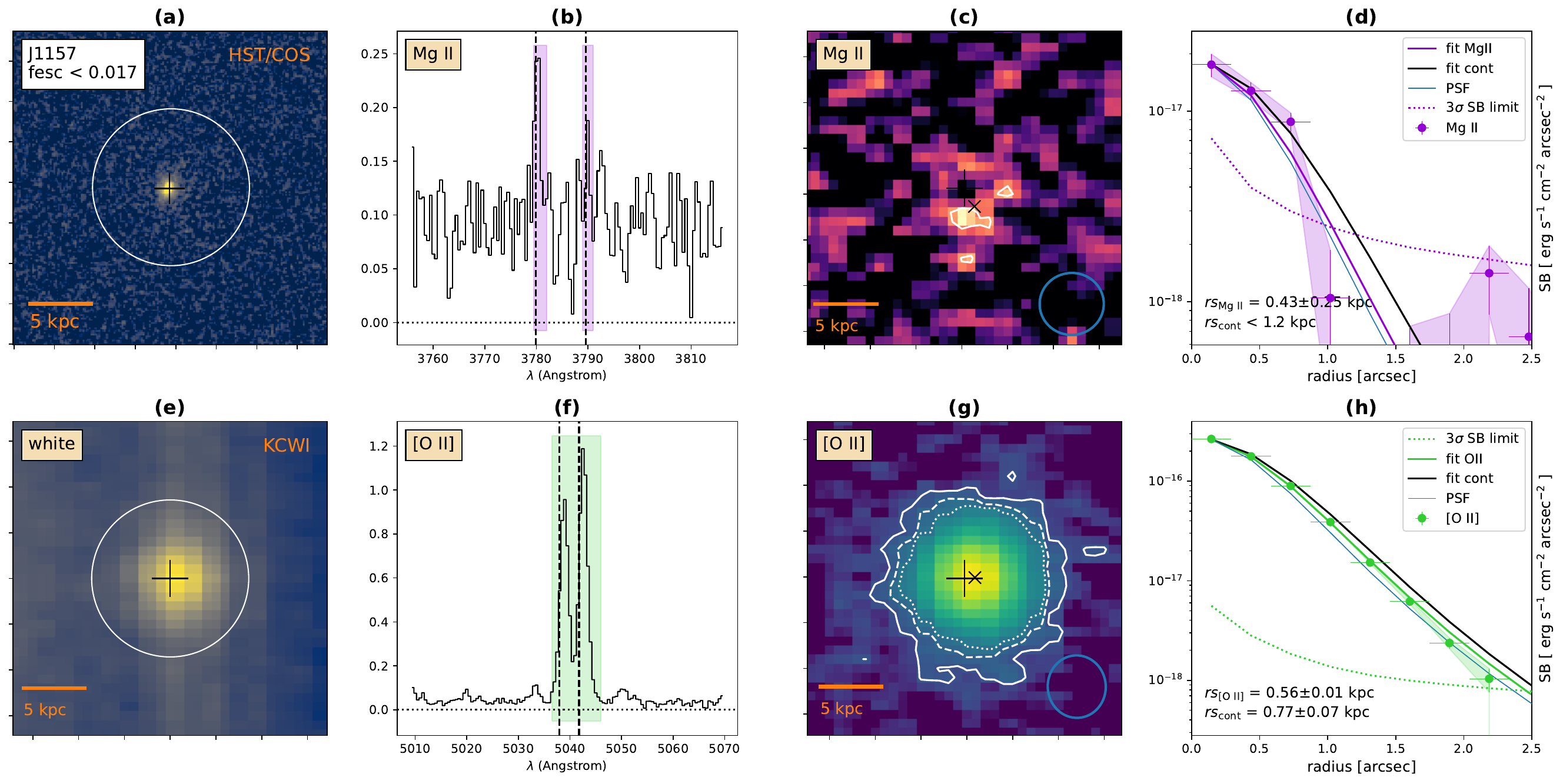}
        \end{subfigure}
        
    \end{minipage}
    \caption{Same as Fig.~\ref{fig:profiles} (Part 6/8)}
    \label{fig:profiles6}
\end{figure*}

\begin{figure*}[p]
    \centering
    \begin{minipage}{0.87\textwidth}

        \begin{subfigure}[b]{\linewidth}
            \includegraphics[width=\textwidth]{plot_w_lines-J1243_sca_FINAL_int.pdf}
        \end{subfigure} 

        \begin{subfigure}[b]{\linewidth}
            \includegraphics[width=\textwidth]{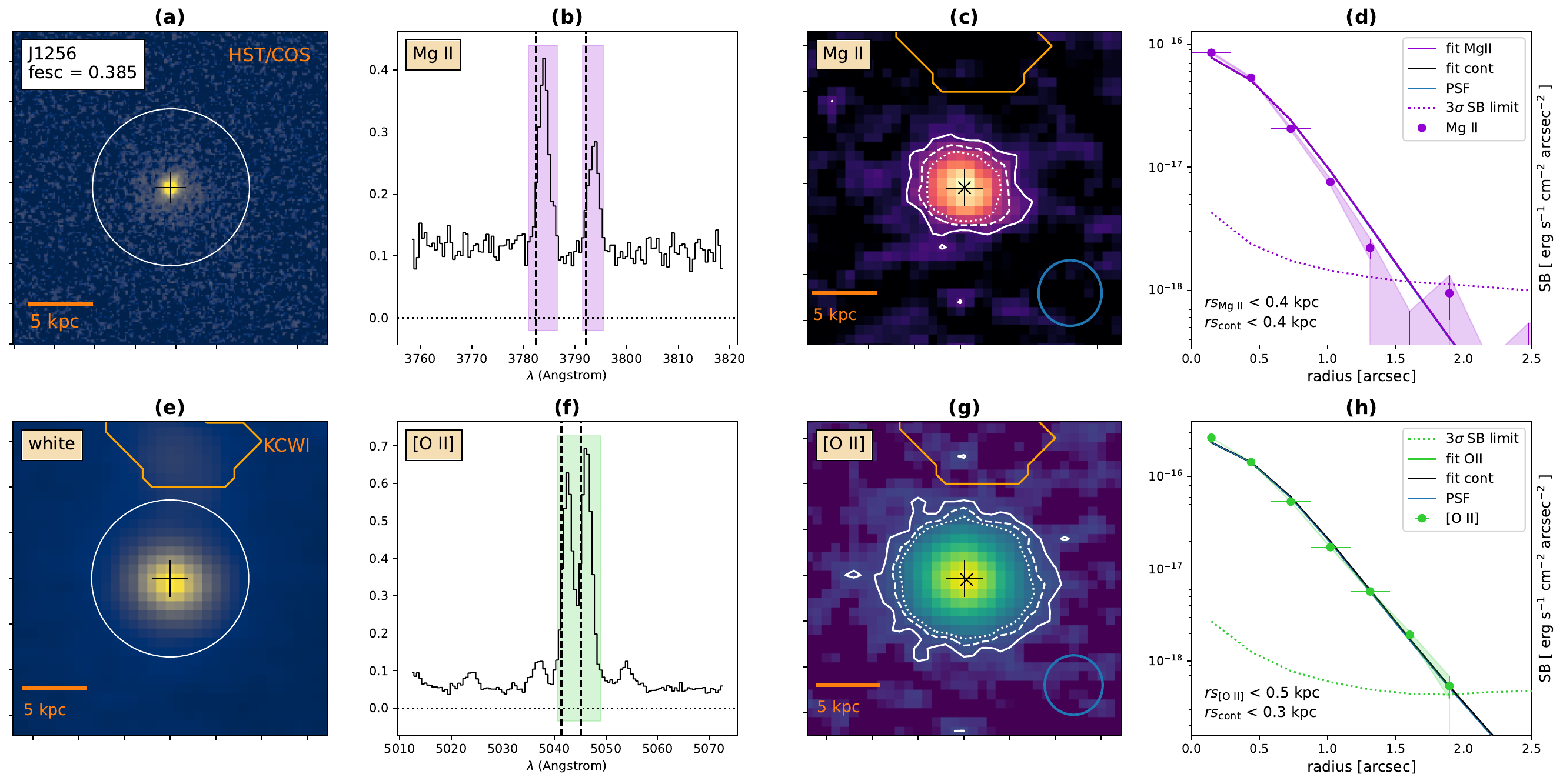}
        \end{subfigure} 
    
        \begin{subfigure}[t]{\linewidth}
            \vspace{0.1cm}
            \includegraphics[width=\textwidth]{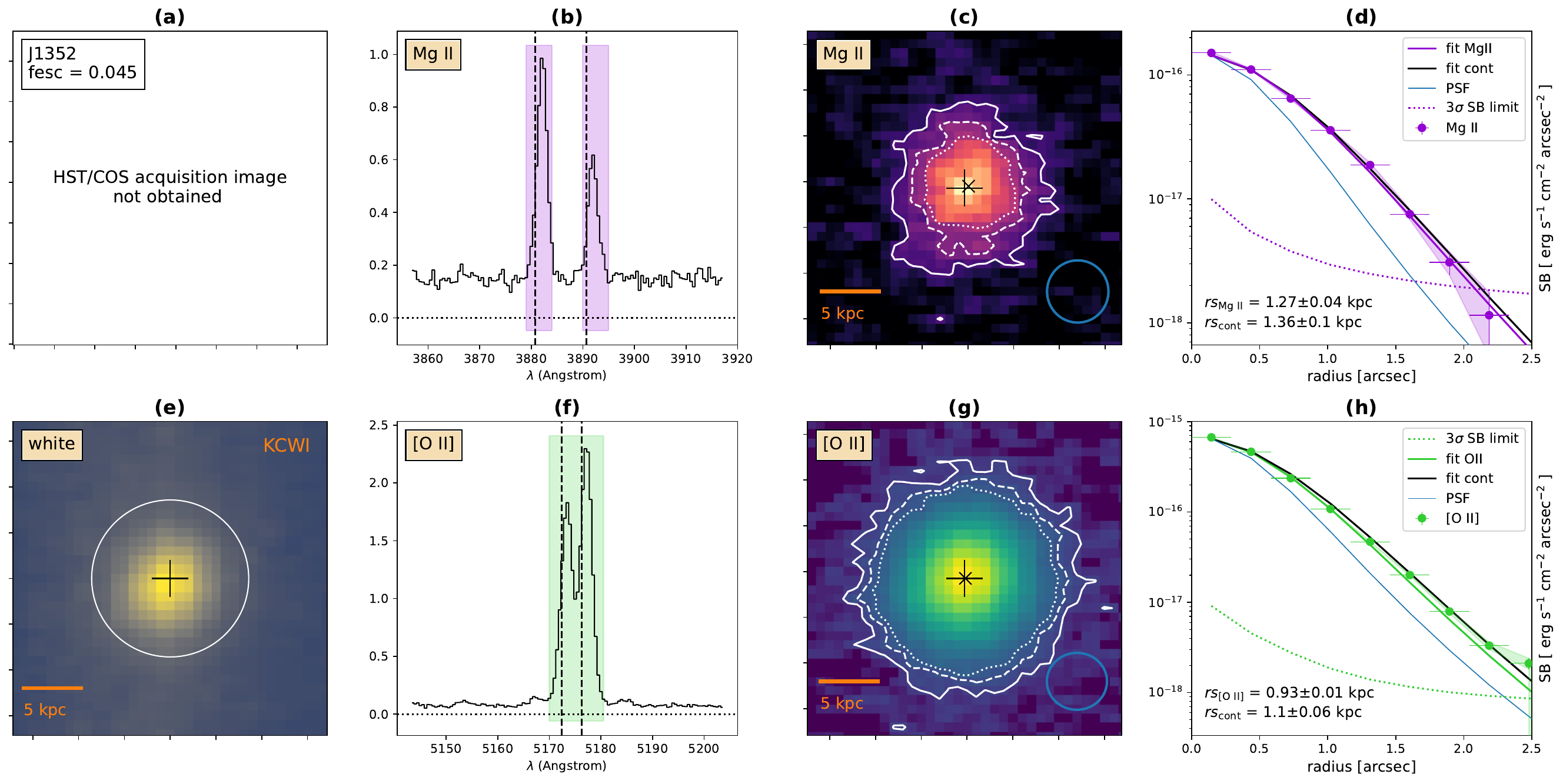}
        \end{subfigure}
        
    \end{minipage}
    \caption{Same as Fig.~\ref{fig:profiles} (Part 7/8)}
    \label{fig:profiles7}
\end{figure*}

\begin{figure*}[p]
    \centering
    \begin{minipage}{0.87\textwidth}

        \begin{subfigure}[b]{\linewidth}
            \includegraphics[width=\textwidth]{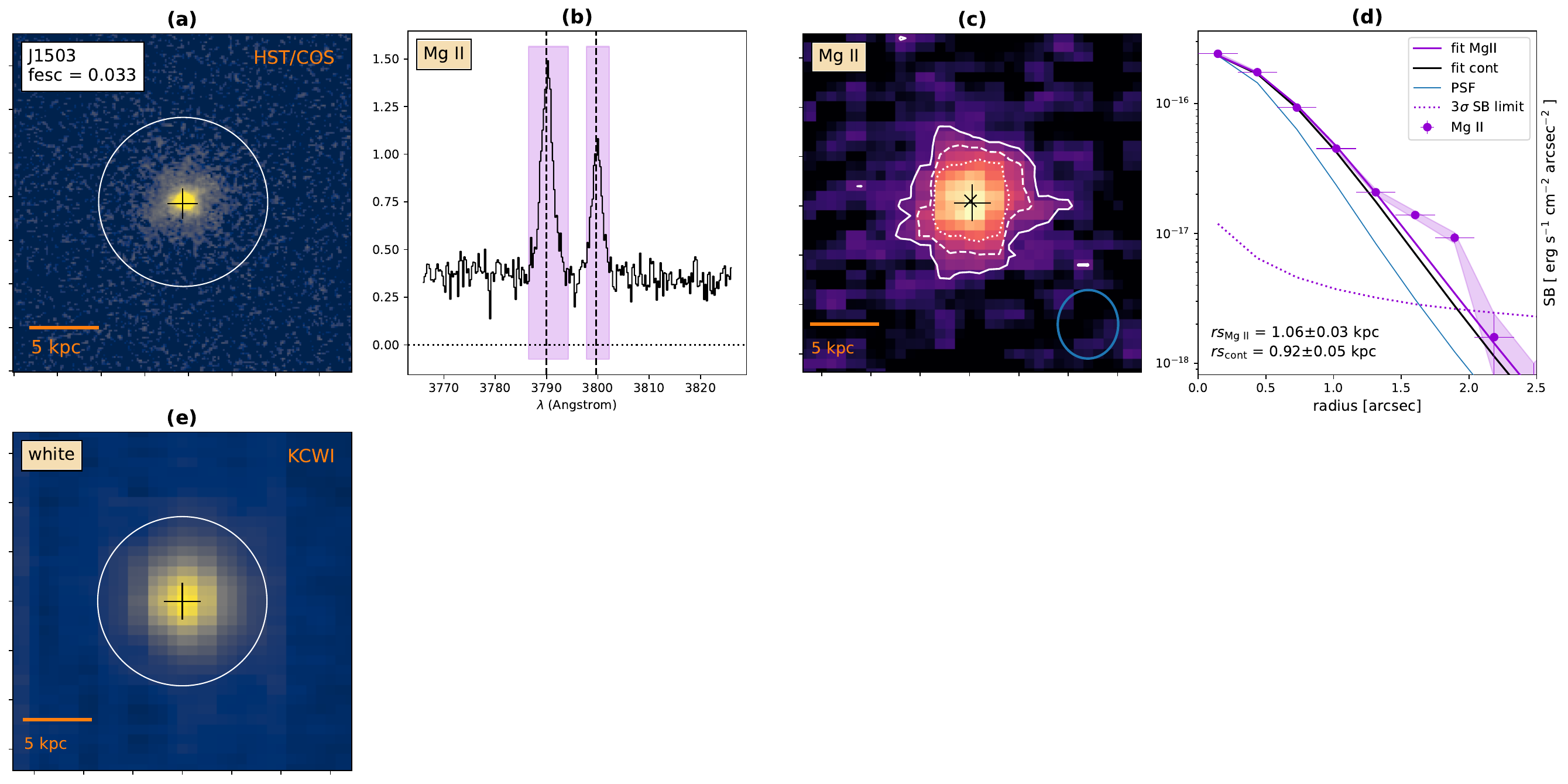}
        \end{subfigure} 

        \begin{subfigure}[b]{\linewidth}
            \includegraphics[width=\textwidth]{plot_w_lines-J1517_sca_FINAL_int.pdf}
        \end{subfigure} 
    
        \begin{subfigure}[t]{\linewidth}
            \vspace{0.1cm}
            \includegraphics[width=\textwidth]{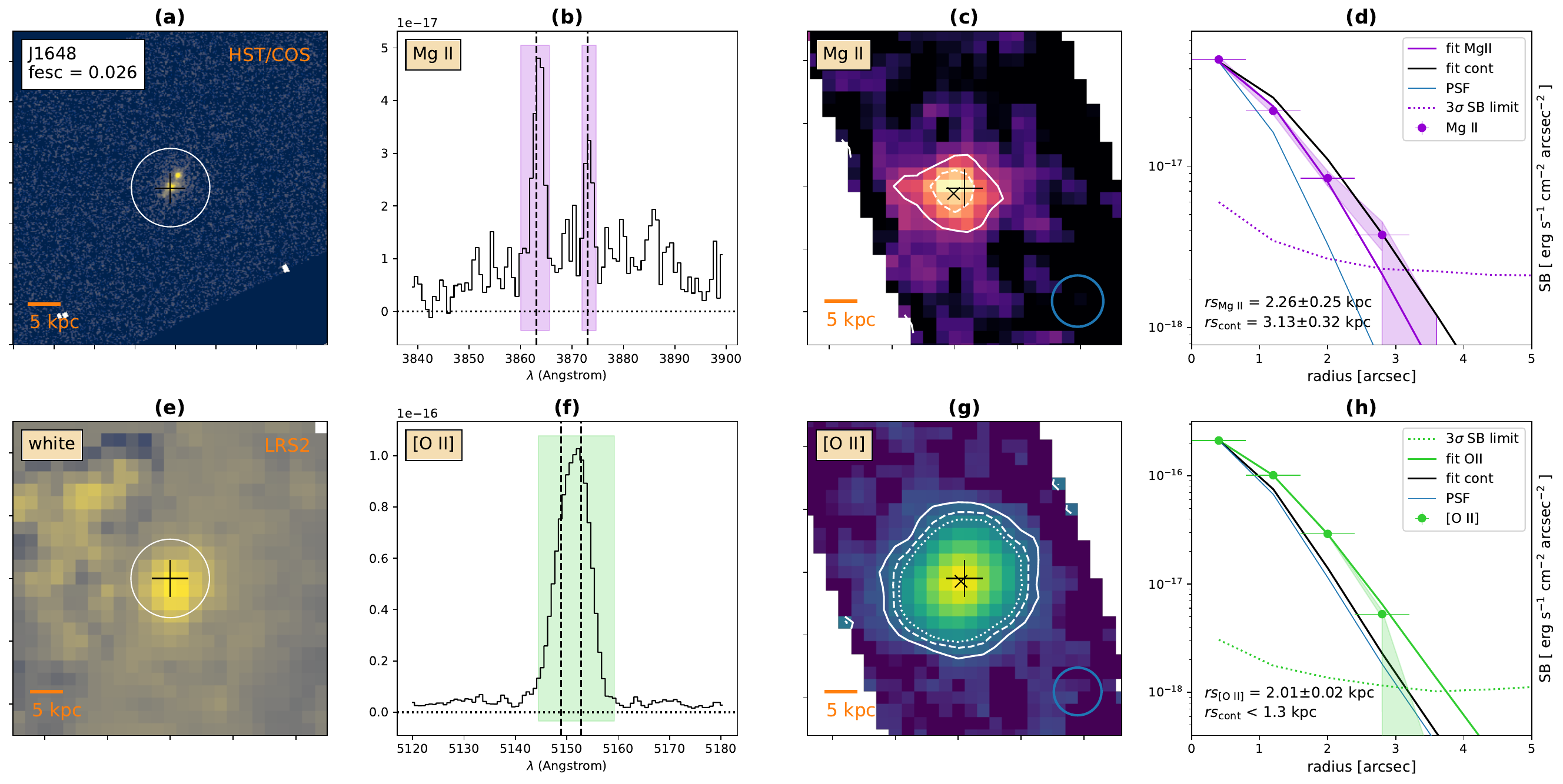}
        \end{subfigure}
    \end{minipage}
    \caption{Same as Fig.~\ref{fig:profiles} (Part 8/8)}
    \label{fig:profiles8}
\end{figure*}

\newpage
\section{Correlations between spatial offset and size}
\label{ap:3}

\begin{figure*}[h!]
\centering
   \resizebox{\hsize}{!}{\includegraphics{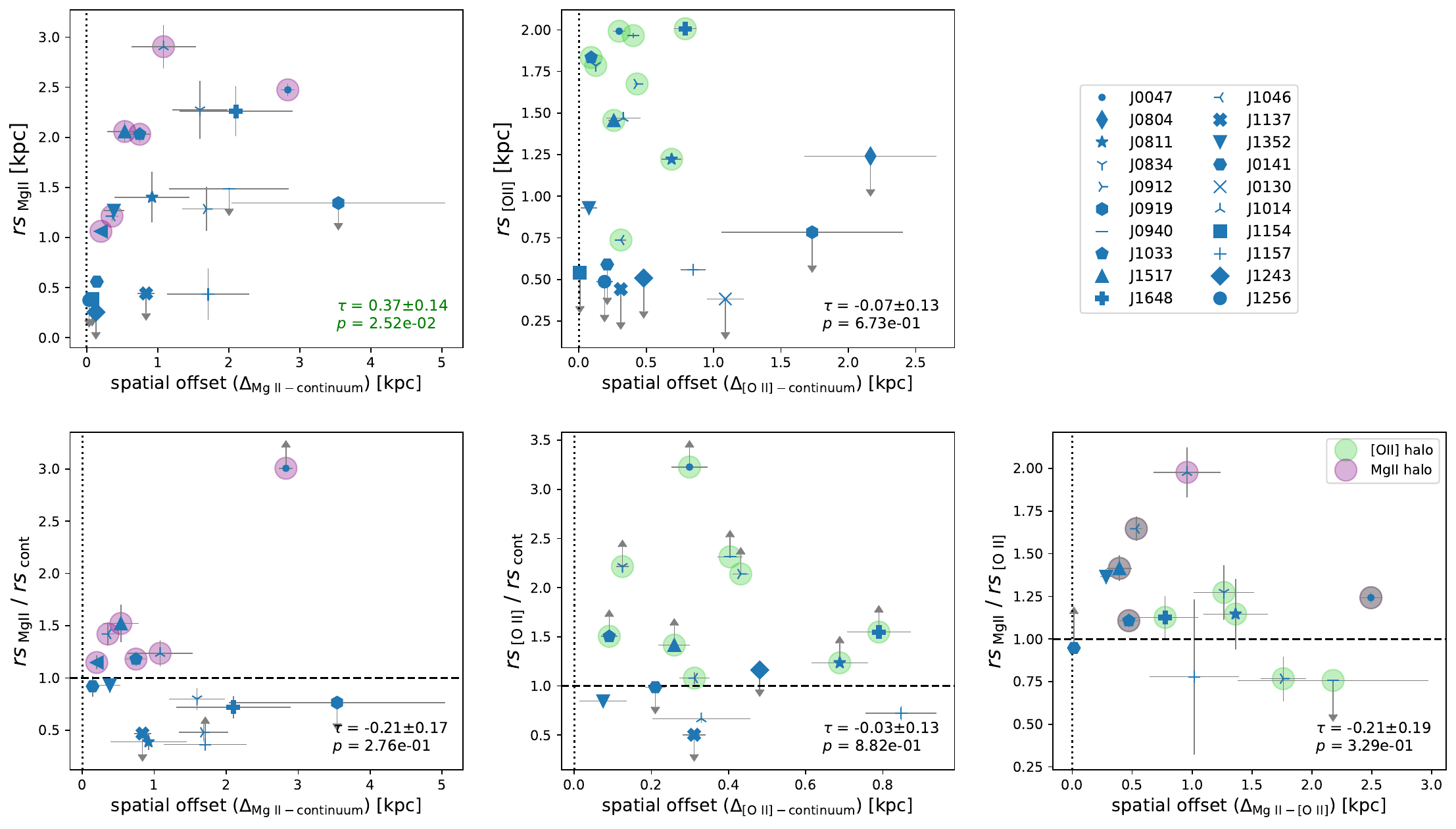}}
    \caption{Relation between the \ion{Mg}{ii} and [\ion{O}{ii}] spatial extent and offset measured for our sample. \textit{Top left (right)}: \ion{Mg}{ii} ([\ion{O}{ii}]) exponential scale length plotted as a function of the \ion{Mg}{ii} ([\ion{O}{ii}]) spatial offset relative to the continuum. \textit{Bottom left (middle):} Ratio between the \ion{Mg}{ii} ([\ion{O}{ii}]) and continuum scale length versus the \ion{Mg}{ii} ([\ion{O}{ii}]) spatial offset relative to the continuum. \textit{Bottom right:} Ratio between the \ion{Mg}{ii} and [\ion{O}{ii}] scale length versus the \ion{Mg}{ii} spatial offset relative to the [\ion{O}{ii}] emission. The Kendall correlation coefficient ($\tau$) and the corresponding probability that the correlation is real ($p$) are given and colored in green if the correlation is statistically significant (\citealt{Akritas96,Flury22b}, see Sect.~\ref{sec:34}). Objects with statistically significant \ion{Mg}{ii} and [\ion{O}{ii}] extended emission are indicated by large purple and green symbols, respectively (Sect.~\ref{sec:34}).}
    \label{fig:rs_offset}
\end{figure*}

\begin{figure*}[h!]
\centering
   \resizebox{.7\hsize}{!}{\includegraphics{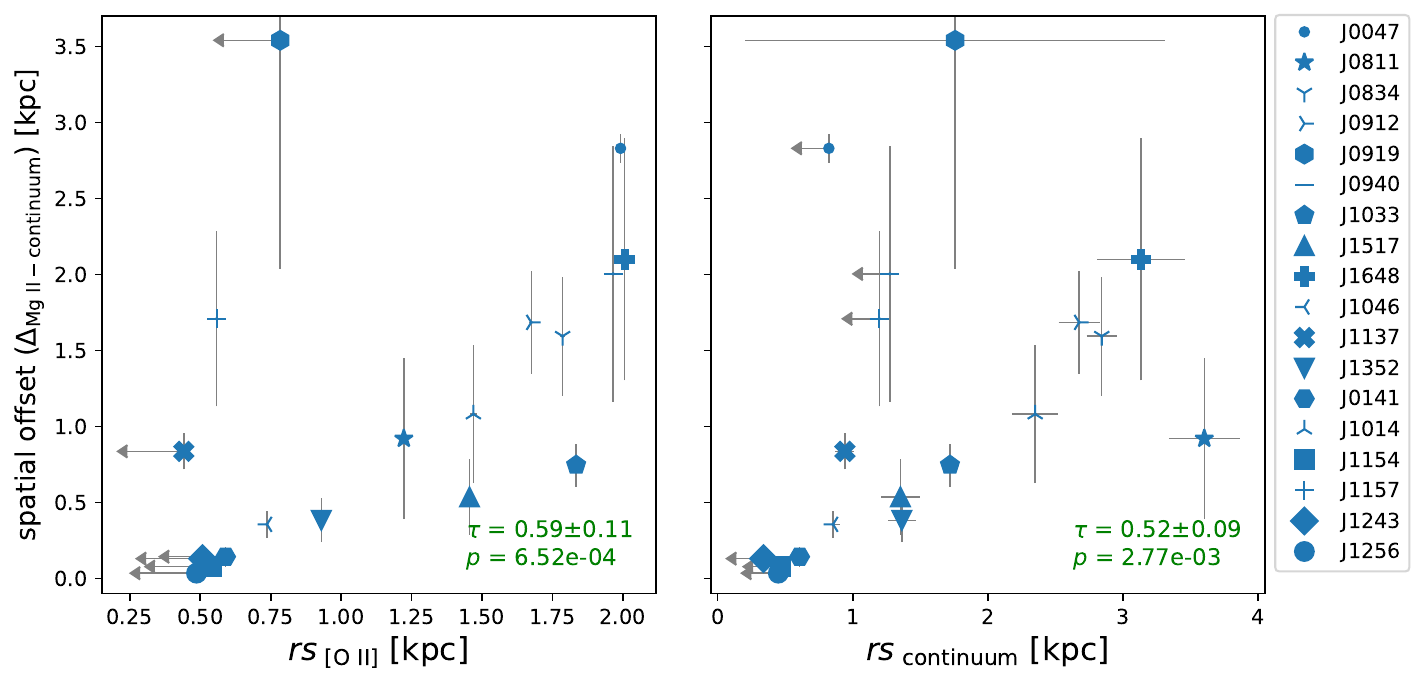}}
    \caption{Relation between the \ion{Mg}{ii} - continuum spatial offsets and the [\ion{O}{ii}] (Left) and continuum (Right) exponential scale lengths. The Kendall correlation coefficient ($\tau$) and the corresponding probability that the correlation is real ($p$) are given and colored in green if the correlation is statistically significant (\citealt{Akritas96,Flury22b}, see Sect.~\ref{sec:34}).}
    \label{fig:offset_rsall}
\end{figure*}

\end{appendix}

\end{document}